\documentstyle[preprint,prc,aps,eqsecnum,epsf]{revtex}
\tighten
\begin{document}
\draft

\title{Parity-Violating Interaction Effects in the $n$$p$ System}
\author{R.\ Schiavilla}
\address{Jefferson Lab, Newport News, Virginia 23606 \\
         and \\
         Department of Physics, Old Dominion University, Norfolk, Virginia 23529}
\author{J.\ Carlson and M.\ Paris}
\address{Theoretical Division, Los Alamos National Laboratory, Los Alamos, New Mexico 87545}
\date{\today}
\maketitle

\begin{abstract}
We investigate parity-violating observables in the $n$$p$ system, including
the longitudinal asymmetry and neutron-spin rotation in $\vec n$$p$ elastic scattering,
the photon asymmetry in $\vec n$$p$ radiative capture, and the asymmetries in deuteron
photo-disintegration $d(\vec{\gamma},n)p$ in the threshold region and electro-disintegration
$d(\vec{e},e^\prime)np$ in quasi-elastic kinematics.  To have an estimate of the
model dependence for the various predictions, a number of different, latest-generation
strong-interaction potentials---Argonne $v_{18}$, Bonn 2000, and Nijmegen I---are used
in combination with a weak-interaction potential consisting of $\pi$-, $\rho$-, and
$\omega$-meson exchanges---the model known as DDH.  The complete bound and scattering
problems in the presence of parity-conserving, including electromagnetic, and parity-violating
potentials is solved in both configuration and momentum space.  The issue of electromagnetic
current conservation is examined carefully.  We find large cancellations between
the asymmetries induced by the parity-violating interactions and those
arising from the associated pion-exchange currents.  In the $\vec n$$p$
capture, the model dependence is nevertheless quite small, because of constraints
arising through the Siegert evaluation of the relevant $E_1$ matrix elements.  In
quasi-elastic electron scattering these processes are found to be insignificant compared to the
asymmetry produced by $\gamma$-$Z$ interference on individual nucleons.  These two
experiments, then, provide clean probes of different aspects of weak-interaction
physics associated with parity violation in the $n$$p$ system.  Finally,
we find that the neutron spin rotation in $\vec{n}$$p$ elastic scattering
and asymmetry in deuteron disintegration by circularly-polarized photons
exhibit significant sensitivity both to the values used for the weak
vector-meson couplings in the DDH model and to the input strong-interaction
potential adopted in the calculation.
This reinforces the conclusion that these short-ranged
meson couplings are not in themselves physical observables,
rather the parity-violating mixings are the physically
relevant parameters.
\end{abstract}

\section{Introduction}
\label{sec:intro}

A new generation of experiments have recently been completed, or are
presently under way or in their planning phase to study the effects of
parity-violating (PV) interactions in $p$$p$ elastic scattering~\cite{Berdoz01},
$n$$p$ radiative capture~\cite{Snow00} and deuteron electro-disintegration~\cite{JLab01}
at low energies.  There is also considerable interest in determining the extent to
which hadronic weak interactions can affect the longitudinal asymmetry measured by the
SAMPLE collaboration in quasi-elastic scattering of polarized electrons off the
deuteron~\cite{Hasty00}, and therefore influence the extraction from these data
(and those on the proton~\cite{Spayde00}) of the nucleon's strange magnetic and
axial-vector form factors at four-momentum transfers squared of $0.04$ and $0.09$
(GeV/c)$^2$.
 
The present is the third in a series of papers dealing with the theoretical
investigation of PV interaction effects in two-nucleon systems.  The first~\cite{Carlson02}
was devoted to $\vec{p}$$p$ elastic scattering, and presented a calculation
of the longitudinal asymmetry induced by PV interactions in the lab-energy
range 0--350 MeV.  The second~\cite{Schiavilla03a} provided a rather cursory 
account of a study of the PV asymmetries in $\vec{n}$$p$ radiative capture
at thermal neutron energies and in deuteron electro-disintegration at quasi-elastic
kinematics.  This work further extends that of Ref.~\cite{Schiavilla03a} by
investigating the neutron spin rotation at zero energy and the longitudinal
asymmetry in $\vec{n}$$p$ elastic scattering at lab energies between 0 and
350 MeV, and the photon-helicity dependence of the $d(\vec{\gamma},n)p$
cross-section from threshold up to energies of 20 MeV.  It also provides a
thorough analysis of the results already presented in Ref.~\cite{Schiavilla03a}.

We adopt the PV potential developed by Desplanques {\it et al.}~\cite{Desplanques80}
over twenty years ago, the so-called DDH model.  In the $n$$p$ sector, it is conveniently
parameterized in terms of $\pi$-, $\rho$-, and $\omega$-meson exchanges.  In
Ref.~\cite{Desplanques80} the pion and vector-meson weak coupling constants
were estimated within a quark model approach incorporating symmetry techniques
like SU(6)$_W$ and current algebra requirements.  Due to limitations inherent
to such an analysis, however, the coupling constants so determined had rather
wide ranges of allowed values.

Our prime motivations are to develop a systematic and consistent framework
for studying PV observables in the few-nucleon systems, where accurate
microscopic calculations are feasible, and to use available and forthcoming
experimental data on these observables to constrain the strengths of the
short- and long-range parts of the two-nucleon weak interaction.  Indeed,
in Ref.~\cite{Carlson02} we showed, for the case of the longitudinal asymmetry
measured in $\vec{p}$$p$ elastic scattering, how available experimental 
data provide strong constraints on allowable combinations of $\rho$- and
$\omega$-meson weak coupling constants.

The remainder of the present paper is organized as follows.  In Sec.~\ref{sec:pot}
the PV potential as well as the parity-conserving strong-interaction
potentials used in this work are briefly discussed, while in
Sec.~\ref{sec:cnt} the model for the nuclear electro-weak currents is
described, including the electromagnetic two-body terms induced by the presence
of PV interactions.  In Sec.~\ref{sec:app} a self-consistent treatment of
the $n$$p$ bound- and scattering-state problems in both configuration
and momentum spaces is provided, patterned after that of Ref.~\cite{Carlson02},
and in Sec.~\ref{sec:obs} explicit expressions are derived for the longitudinal
asymmetry and spin rotation in $\vec{n}$$p$ elastic scattering, the photon
asymmetry in $\vec{n}$$p$ radiative capture, and the asymmetries in deuteron
photo- and electro-disintegration.  In Sec.~\ref{sec:cal} the techniques
used to calculate the PV observables are briefly reviewed, while in
Sec.~\ref{sec:res} a fairly detailed analysis of the results is offered.
Finally, Sec.~\ref{sec:cons} contains some concluding remarks.

\section{Parity-Conserving and Parity-Violating Potentials}
\label{sec:pot}

The parity-conserving (PC), strong-interaction potentials
used in the present work are the Argonne $v_{18}$ (AV18)~\cite{Wiringa95},
Nijmegen I (NIJM-I)~\cite{Stoks94}, and CD-Bonn (BONN)~\cite{Machleidt01}
models.  They were discussed in Ref.~\cite{Carlson02} in connection
with the calculation of the longitudinal asymmetry in $\vec p$$p$ elastic
scattering.  Here, we briefly summarize a few salient points.

The AV18 and NIJM-I potentials were fitted to the Nijmegen
database of 1992~\cite{Bergervoet90,Stoks93}, consisting of 1787 $p$$p$ and
2514 $n$$p$ scattering data, and both produced $\chi^2$ per datum close to one.
The latest version of the charge-dependent Bonn potential, however, has been fit to the
1999 database, consisting of 2932 $p$$p$ and 3058 $n$$p$ data, for which it gives $\chi^2$
per datum of 1.01 and 1.02, respectively~\cite{Machleidt01}.  The substantial increase in
the number of $p$$p$ data is due to the development of novel
experimental techniques---internally polarized gas targets and stored, cooled
beams.  Indeed, using this technology, IUCF has produced a large
number of $p$$p$ spin-correlation parameters of very high precision,
see for example Ref.~\cite{Przewoski98}.  It is worth noting that
the AV18 potential, as an example, fits the post-1992 and both pre- and
post-1992 $p$$p$ ($n$$p$) data with $\chi^2$'s of 1.74 (1.02) and 1.35
(1.07), respectively~\cite{Machleidt01}.  Therefore, while the quality of
their fits (to the $p$$p$ data) has deteriorated somewhat in regard to the
extended 1999-database, the AV18 and NIJM-I models can still be
considered \lq\lq realistic\rq\rq.

As already mentioned in Sec.~\ref{sec:intro}, the form of
the parity-violating (PV) weak-interaction potential was derived
in Ref.~\cite{Desplanques80}---the DDH model.  In the isospin space
of the $n$$p$ pair it is expressed as 

\begin{equation}
v^{\rm PV}_{T^\prime,T}=\langle T^\prime,M_T^\prime=0\mid v^{\rm PV}\mid T,M_T=0\rangle \ ,
\end{equation}
where $T,T^\prime$=0 or 1.  The diagonal and off-diagonal terms are then obtained as

\begin{eqnarray}
v^{\rm PV}_{T,T}=\sum_{\alpha=\rho,\omega}&-&
\frac{g_\alpha \, h^{np}_\alpha}{4\pi} \frac{m_\alpha}{m}
\Bigg[ m_\alpha (1+\kappa_\alpha)Y^\prime (m_\alpha r)
({\bbox \sigma}_1\times {\bbox \sigma}_2) \cdot \hat{\bf r} \nonumber \\
&+&({\bbox \sigma}_1 - {\bbox \sigma}_2) \cdot \left[ {\bf p} \, ,
\, Y(m_\alpha r)\right]_{+} \Bigg] \ ,
\label{eq:DDH1}
\end{eqnarray}
\begin{eqnarray}
v^{\rm PV}_{1,0}=&-&{\rm i} \frac{g_\pi h_\pi}{ 4\pi \sqrt{2} } \frac{m_\pi^2}{m} Y^\prime(m_\pi r)
({\bbox \sigma}_1 + {\bbox \sigma}_2) \cdot \hat{\bf r} \nonumber \\
&-&\frac{g_\omega h^1_\omega - g_\rho h^1_\rho}{4\pi} \frac{m_\rho}{m}
({\bbox \sigma}_1 + {\bbox \sigma}_2)\cdot \left[ {\bf p} \, , \, Y(m_\rho r)\right]_{+}  \ ,
\label{eq:DDH2}
\end{eqnarray}
and $v^{\rm PV}_{0,1}$=$v^{\rm PV\, \dagger}_{1,0}$.  In the equations above
the relative position and momentum are defined
as ${\bf r}={\bf r}_1-{\bf r}_2$ and ${\bf p}=({\bf p}_1-{\bf p}_2)/2$,
respectively, $\left[ \dots \, , \, \dots \right]_{+}$ denotes
the anticommutator, and $m$, $m_\pi$, $m_\rho$, and $m_\omega$ are the proton, pion,
$\rho$-meson, $\omega$-meson masses, respectively.  The Yukawa function
$Y(x_\alpha)$, suitably modified by the inclusion of
monopole form factors, is given by
 
\begin{equation}
Y(x_\alpha)= \frac{1}{x_\alpha}\Bigg\{ {\rm e}^{-x_\alpha}
-{\rm e}^{-(\Lambda_\alpha/m_\alpha) x_\alpha} \left[ 1+
\frac{1}{2}\frac{\Lambda_\alpha}{m_\alpha}
\left( 1-\frac{m^2_\alpha}{\Lambda^2_\alpha} \right)
x_\alpha \right] \Bigg\} \>\>,
\end{equation}
where $x_\alpha \equiv m_\alpha r$.  Note that
$Y^\prime(x)$ denotes ${\rm d}Y(x)/{\rm d}x$, and that
the terms proportional to $Y^\prime(x)$
in Eqs.~(\ref{eq:DDH1}) and~(\ref{eq:DDH2}) are usually written
in the form of a commutator, since
 
\begin{equation}
{\rm i} \left[ {\bf p} \, , \, Y(m_\alpha r) \right]_{-}
=m_\alpha Y^\prime (m_\alpha r)\, \hat{\bf r} \ .
\end{equation}
Finally, the values for the strong-interaction $\pi$-meson
pseudoscalar coupling constant $g_\pi$, and $\rho$- and $\omega$-meson vector
and tensor coupling constants $g_\alpha$ and $\kappa_\alpha$,
as well as for the cutoff parameters $\Lambda_\alpha$, are
taken from the BONN model~\cite{Machleidt01}, and
are listed in Table~\ref{tb:ddh}.  The weak-interaction
vector-meson coupling constants $h^{np}_\rho$ and $h^{np}_\omega$
correspond to the following combinations of DDH parameters
 
\begin{eqnarray}
h^{np}_\rho &=& (4T-3) h^0_\rho -\frac{2\, T}{\sqrt{6}} h^2_\rho \ , \\
h^{np}_\omega &=& h^0_\omega \ ,
\end{eqnarray}
where $T$=0,1.  The values for these and for $h_\pi$, $h^1_\rho$,
and $h^1_\omega$ are listed in Table~\ref{tb:ddh}.  Note that we have
taken the linear combination of $\rho$- and $\omega$-meson weak
coupling constants corresponding to $\vec p$$p$ elastic
scattering from the earlier analysis~\cite{Carlson02} of these
experiments.  The remaining couplings are the ``best value'' estimates,
suggested in Ref.~\cite{Desplanques80}.

In order to study the sensitivity of the calculated PV observables to
variations in the weak coupling constants, we also consider in the
present work a $v^{\rm PV}$ corresponding to the original ``best value''
estimates for these from Ref.~\cite{Desplanques80}, see Table~\ref{tb:ddhb}.

\section{Electromagnetic and Neutral Weak Current Operators}
\label{sec:cnt}

The electromagnetic current and charge operators, respectively ${\bf j}$ and $\rho$,
are expanded into a sum of one- and two-body terms 

\begin{equation}
{\bf j}({\bf q})=\sum_i {\bf j}_i({\bf q}) + \sum_{i<j} {\bf j}_{ij}({\bf q}) \ ,
\end{equation}
and similarly for $\rho({\bf q})$.  The one-body terms have the standard expressions
obtained from a non-relativistic reduction of the covariant single-nucleon
current~\cite{Viviani96}.  The two-body charge operators are those derived
in Ref.~\cite{Schiavilla90}; they only enter in the calculation of the
asymmetry in the deuteron electro-disintegration at quasi-elastic
kinematics, and will not be discussed further here. 

The two-body currents have terms associated with the parity-conserving (PC)
and parity-violating (PV) components of the interaction, respectively
${\bf j}_{ij}^{\rm PC}$ and ${\bf j}_{ij}^{\rm PV}$.  The operators
${\bf j}^{\rm PC}_{ij}$ were derived explicitly in Ref.~\cite{Schiavilla89},
and a complete listing of those relative to the Argonne $v_{18}$ (AV18)
interaction~\cite{Wiringa95} has been given most recently in Ref.~\cite{Viviani96}.
Only the two-body currents associated with $\pi$- and $\rho$-exchange
are retained in the case of the Bonn (BONN)~\cite{Machleidt01} and Nijmegen-I
(NIJM-I)~\cite{Stoks94} interactions.

In addition to these, the purely transverse two-body currents associated
with the excitation of $\Delta$-isobars and the $\rho \pi \gamma$ and
$\omega \pi \gamma$ mechanisms are included in all calculations.  Again
explicit expressions for these operators can be found in Ref.~\cite{Viviani96}.
Note, however, that the $\Delta$-isobar degrees of freedom are treated in
perturbation theory rather than with the transition-correlation-operator
method~\cite{Schiavilla92}, and that only effects due to single
$\Delta$ excitation are considered, according to Eqs.~(2.15) and~(3.4)
in Ref.~\cite{Viviani96}.

Before moving on to a discussion of the PV currents, we briefly review, for later
reference, the question of conservation of the electromagnetic current for the
case of the AV18.  As pointed out in Ref.~\cite{Viviani96}, the currents from its
$v_6$ part (specifically, its isospin-dependent central, spin-spin, and tensor
components) are strictly conserved.  In a one-boson-exchange model, which the
AV18 is not, these interaction components arise from $\pi$- and $\rho$-exchange.

The currents from the AV18 momentum-dependent ($p$-dependent) components---the
spin-orbit, ${\bf L}^2$, and quadratic-spin-orbit terms---are also
included.  In Ref.~\cite{Carlson90} and later papers, the currents from
the spin-orbit term were derived by generalizing the procedure used to obtain
the $v_6$ currents.  It was assumed that the isospin-independent (isospin-dependent)
central and spin-orbit interactions were due to $\sigma$ and $\omega$ exchanges  
($\rho$ exchange), and the associated two-body currents were constructed by
considering corresponding $\overline{N}$$N$-pair diagrams involving these meson
exchanges.  The currents from the ${\bf L}^2$ and quadratic-spin-orbit interactions
were obtained, instead, by minimal substitution~\cite{Viviani96,Schiavilla89}.

The currents from the $p$-dependent interactions are strictly not
conserved, as one can easily surmise by considering their commutator
with the charge density operator.  For example, in the case of the
isospin-dependent ${\bf L}^2$ and $({\bf L} \cdot {\bf S})^2$ interactions, this commutator
requires the presence of currents with the isospin
structure $(\bbox{\tau}_i \times \bbox{\tau}_j)_z$,
which cannot be generated by minimal substitution~\cite{Schiavilla89}.

We will return to this issue in Sec.~\ref{sec:r_np}.  Here, we only
want to emphasize that the currents from the $p$-dependent terms in the AV18
are short-ranged.  Their contributions to isovector observables, such as, for example, the
magnetic form factors of the trinucleons~\cite{Marcucci98}, are found to be numerically
much smaller than those due to the leading $v_6$ currents.  These currents
also lead to small, although non-negligible, corrections to isoscalar observables,
such as the deuteron magnetic moment and $B(q)$ structure function~\cite{Schiavilla91}.
However, in the case of the PV asymmetry in the $\vec n$$p$ radiative
capture at thermal energies under consideration in the present study,
they will turn out to play an important role (see Sec.~\ref{sec:r_np}). 
 
\subsection{Parity-violating currents}
\label{sec:cntpv}

\begin{figure}[bth]
\let\picnaturalsize=N
\def\picsize{3in}
\def\picfilenamea{figs/feyn.eps}
\epsfbox{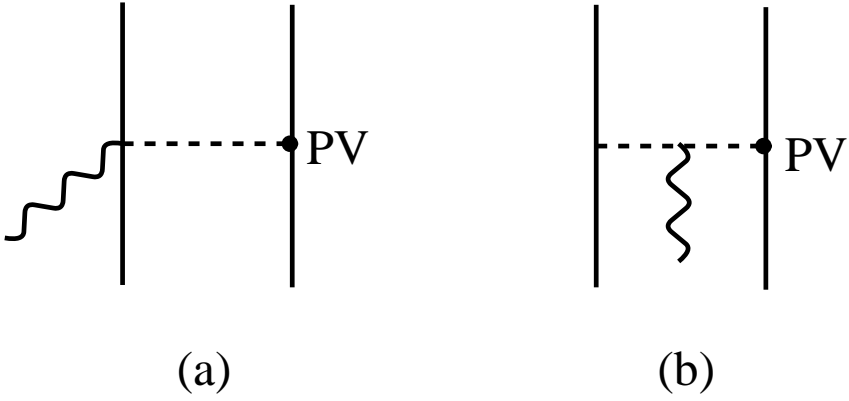}
\vspace{1cm}
\caption{Feynman diagram representation of the two-body currents associated
with pion exchange: solid lines, nucleons; dashed lines, pions; wavy 
lines, photons.  Note that one interaction vertex is parity-conserving, while
the other is parity-violating (PV).}
\label{fig:feyn}
\end{figure}

The DDH PV interaction~\cite{Desplanques80} is parameterized in terms of
$\pi$-, $\rho$-, and $\omega$-meson exchanges.  The meson-nucleon phenomenological
Lagrangian densities have been given most recently in Ref.~\cite{Haxton02}.  We adopt
here the notation and conventions of that work, except that we use pseudo-vector
coupling for the $\pi$$N$$N$ interaction Lagrangian, i.e.

\begin{equation}
L_{\pi NN}^{\rm PC} = -\frac{f_\pi}{m_\pi} \overline{N} \gamma_5 \gamma_\mu
\bbox{\tau} N \cdot \partial^\mu \bbox{\pi} \ ,
\end{equation}
with $f_\pi/m_\pi$=$g_\pi/(2 m)$.  The resulting $\gamma$$\pi$$N$$N$
coupling is given by

\begin{equation}
L_{\gamma \pi NN}^{\rm PC} =-e\frac{f_\pi}{m_\pi} \overline{N} \gamma_5 
\gamma_\mu (\bbox{\tau} \times \bbox{\pi})_z N A^\mu  \ ,
\end{equation}
and the $\gamma$$\pi$$N$$N$ current is then obtained from the Feynman amplitude
in Fig.~\ref{fig:feyn}, panel (a).  The complete PV $\pi$-exchange
current is derived from a non-relativistic reduction of both amplitudes
in Fig.~\ref{fig:feyn}, and to leading order reads:

\begin{eqnarray}
{\bf j}^{\rm PV}_{\pi,ij} ({\bf k}_i,{\bf k}_j)=
&-&\frac{f_\pi\, h_\pi}{\sqrt{2}\, m_\pi}
\left( \bbox{\tau}_i \cdot \bbox{\tau}_j - \tau_{z,i} \tau_{z,j} \right)
\Bigg[ v_\pi(k_j) \bbox{\sigma}_i + v_\pi(k_i) \bbox{\sigma}_j \nonumber \\
&-&\frac{ {\bf k}_i - {\bf k}_j }{ k_i^2 - k_j^2 }
\left[ v_\pi(k_j) - v_\pi(k_i) \right]
\left( \bbox{\sigma}_i \cdot {\bf k}_i - \bbox{\sigma}_j \cdot {\bf k}_j \right) \Bigg] \ ,
\end{eqnarray}
where $v_\pi(k)$ is defined as

\begin{equation}
v_\pi(k)=\left(\frac{ \Lambda_\pi^2-m_\pi^2 }{ k^2+\Lambda_\pi^2 }\right)^2
 \frac{1}{ k^2+m_\pi^2 } \ ,
\label{eq:vpl}
\end{equation}
${\bf k}_i$=${\bf p}_i^\prime - {\bf p}_i$ is the fractional momentum
delivered to nucleon $i$ (with these definitions
${\bf q}$=${\bf k}_i+{\bf k}_j$), and $\Lambda_\pi$ is a short-range cutoff.
In the limit of point-like couplings, the current
above is identical to that listed in Eqs.~(A7a) and~(A9a) of Ref.~\cite{Haxton02}.
$N$$\overline{N}$-pair terms arising from the photon coupling to the
nucleon line containing a PV $\pi$$N$$N$ vertex do not contribute to leading order.
Lastly, we note that ${\bf j}^{\rm PV}_\pi$ satisfies current conservation with the
PV $\pi$-exchange interaction, given in momentum space by

\begin{equation}
v^{\rm PV}_\pi({\bf k}_i,{\bf k}_j)= -{\rm i}\frac{f_\pi\, h_\pi}{\sqrt{2}\, m_\pi}
\left( \bbox{\tau}_i \times \bbox{\tau}_j \right)_z v_\pi(k_j)\, \bbox{\sigma}_i \cdot {\bf k}_j
+ i \rightleftharpoons j \ ,
\end{equation}
since

\begin{eqnarray}
\left[ v^{\rm PC}_{\pi}({\bf k}_i-{\bf q},{\bf k}_j) \, , \, P_i \right]&+&
\left[ v^{\rm PC}_{\pi}({\bf k}_i,{\bf k}_j-{\bf q}) \, , \, P_j \right] \nonumber \\
&=& -\frac{f_\pi\, h_\pi}{\sqrt{2}\, m_\pi}
\left( \bbox{\tau}_i \cdot \bbox{\tau}_j - \tau_{z,i} \tau_{z,j} \right)
\left[ v_\pi(k_i) \left( \bbox{\sigma}_i+\bbox{\sigma}_j \right) \cdot {\bf k}_i
      + i \rightleftharpoons j \right] \ , 
\end{eqnarray}
which is easily seen to be the same as $({\bf k}_i+{\bf k}_j) \cdot {\bf j}^{\rm PV}_\pi
({\bf k}_i,{\bf k}_j)$. Here $P_i$ denotes the isospin projection operator

\begin{equation}
P_i\equiv \frac{1+\tau_{z,i}}{2} \ .
\end{equation}

In the present work, the PV currents induced by $\rho$- and $\omega$-meson
exchanges have been neglected, since, due their short-range character, the
associated contributions are expected to be tiny.  We note, however,
that in Eq.~(A5) of Ref.~\cite{Haxton02}
a $\gamma$$\rho$$N$$N$ contact term, originating from gauging the $\rho$$N$$N$
tensor coupling, has been ignored.  It is given by

\begin{equation}
L_{\gamma\rho NN} = e \frac{g_\rho\, \kappa_\rho}{2\, m} \overline{N}
\sigma_{\mu \nu}(\bbox{\tau}\times \bbox{\rho}^\mu)_z N\, A^\nu \ ,
\end{equation}
and leads, in leading order, to an additional term in Eq.~(A7c)
of Ref.~\cite{Haxton02} of the form

\begin{equation}
-\frac{g_\rho \, \kappa_\rho}{2\, m} \left( h^0_\rho -\frac{h^2_\rho}{2\sqrt{6}}\right)
\left( \bbox{\tau}_i \times \bbox{\tau}_j \right)_z v_\rho(k_j)
\Bigg[ \bbox{\sigma}_i \times \bbox{\sigma}_j + \frac{ \bbox{\sigma}_j \cdot {\bf k}_j }
{m_\rho^2} \bbox{\sigma}_i \times {\bf k}_j \Bigg] + i \rightleftharpoons j \ ,
\end{equation}
with $v_\rho(k)$ defined similarly as in Eq.~(\ref{eq:vpl}).  The term above, when combined
with that having the same structure in the second line of Eq.~(A7c),
generates a contribution proportional to $1+\kappa_\rho$ which, in view of the large
value of the $\rho$$N$$N$ tensor coupling constant ($\kappa_\rho$=6.6), is expected
to be dominant in the PV $\rho$-exchange current. 

Finally, there is a PV one-body current originating from the nucleon's anapole
moment.  It can be derived, for example, by considering pion-loop diagrams
where one of the vertices involves a PV $\pi$$N$$N$ coupling; it has the 
structure, to leading order, given by

\begin{equation}
{\bf j}_i^{\rm PV}({\bf q})=-\frac{q_\mu^2}{2\, m^2}
\left[ a^S(q_\mu^2)+a^V(q_\mu^2) \tau_{z,i} \right] \bbox{\sigma}_i
{\rm e}^{ {\rm i} {\bf q} \cdot {\bf r}_i } \ ,
\end{equation}
where $q_\mu^2$ is the four-momentum transfer, $q_\mu^2$=$\omega^2-q^2$, and
the isoscalar and isovector anapole form factors are normalized as

\begin{equation}
a^{S,V}(0) = \frac{g_\pi h_\pi } { 4 \sqrt{2} \pi^2} \alpha^{S,V} \ ,
\end{equation}
with $\alpha^S$=1.6 and $\alpha^V$=0.4 from a calculation of pion-loop
contributions~\cite{Haxton89}.  More recent estimates of the nucleon
anapole form factors predict~\cite{Musolf91,Riska00,Maekawa00}
somewhat different values for $\alpha^{S,V}$.  A complete treatment would require
estimates of short-distance contributions~\cite{Zhu00} and electroweak radiative
corrections.  Thus, in view of the uncertainties in the quantitative estimate
of these effects, we will continue to use the values above in the present study
(see also Sec.~\ref{sec:r_dee}).  Note that ${\bf j}^{\rm PV}_i$ vanishes for
real-photon transitions.
\subsection{Neutral weak currents}
\label{sec:cntw}

In the Standard Model the vector part of the neutral
weak current, $j^{0,\sigma}$, is related to the isoscalar ($S$) and isovector
($V$) components of the electromagnetic current, denoted
respectively as $j^{\gamma,\sigma}_S$ and $j^{\gamma,\sigma}_V$, via

\begin{equation}
j^{0,\sigma}=-2 \,{\rm sin}^2\theta_W \, j^{\gamma,\sigma}_S
+(1-2\, {\rm sin}^2\theta_W)\, j^{\gamma,\sigma}_V \ ,
\end{equation}
where $\theta_W$ is the Weinberg angle,
and therefore the associated one- and two-body weak charge and
current operators are easily obtained from those given
in Sec~\ref{sec:cnt}.  The axial charge and current operators too
have one- and two-body terms.  Only the axial current

\begin{equation}
{\bf j}^5 ({\bf q})= \sum_i {\bf j}^5_i({\bf q})
             +\sum_{i<j} {\bf j}^5_{ij}({\bf q}) \label{eq2a} \>\>
\end{equation}
is needed in the present work.
The one- and two-body operators are essentially those
listed in Ref.~\cite{Marcucci00}, except for obvious changes 
in the isospin structure having to do with the fact that we are
dealing here with neutral rather than charge-raising/lowering weak currents, and for
the inclusion of nucleon and $N\Delta$ axial form factors---the parameterization
adopted for these is given in Ref.~\cite{Marcucci02}.
Note that in Ref.~\cite{Diaconescu01} the relativistic corrections in ${\bf j}^5_i$ 
and two-body axial currents were neglected, in line with the expectation,
confirmed in the present study, that the associated contributions were small.

Finally, the neutral weak currents given above are at tree-level, electro-weak
radiative corrections as well as strange-quark contributions to the vector and
axial-vector currents~\cite{Musolf94} have been ignored.  These effects
have been taken into account in recent calculations of the longitudinal asymmetry in
$d(\vec e,e^\prime)pn$ at quasi-elastic kinematics~\cite{Schiavilla03b}, however,
they will not be discussed further here.
\section{Formalism}
\label{sec:app}

In this section we discuss the $n$$p$ scattering- and bound-state
problems in the presence of a potential $v$ given by

\begin{equation}
v= v^{\rm PC}+v^{\rm PV} \>\>,
\end{equation}
where $v^{\rm PC}$ and $v^{\rm PV}$ denote the parity-conserving (PC)
and parity-violating (PV) components induced by the strong (including
electromagnetic) and weak interactions, respectively.  The formalism
and notation are similar to those developed in Ref.~\cite{Carlson02}.
 
\subsection{Partial-wave expansions of scattering state, $T$- and $S$-matrices}
\label{sec:pwest}

The Lippmann-Schwinger equation for the $N$$N$ scattering
state $\mid{\bf p},SM_S,T\rangle^{(\pm)}$, where ${\bf p}$
is the relative momentum, $S$, $M_S$, $T$, and $M_T$=0 specify
the pair spin, spin-projection, isospin, and isospin-projection
states (note that the label $M_T$=0 is unnecessary, since $v$
is diagonal in $M_T$), can be written as~\cite{Goldberger64}

\begin{equation}
\mid{\bf p},SM_S,T\rangle^{(\pm)}=
\mid {\bf p},SM_S,T\rangle_0 +
\frac{1}{E-H_0\pm {\rm i}\epsilon}\, v\,
\mid{\bf p},SM_S,T\rangle^{(\pm)} \>\>,
\label{eq:LS}
\end{equation}
where $H_0$ is the free Hamiltonian, and
$\mid \dots\rangle_0$ are the eigenstates of $H_0$,
namely plane waves, 

\begin{eqnarray}
\phi_{ {\bf p}, SM_S,T}({\bf r})&=&
\langle {\bf r}\mid {\bf p},SM_S,T\rangle_0 \nonumber \\
&=&\frac{1}{\sqrt{2}} \left[ {\rm e}^{{\rm i} {\bf p} \cdot {\bf r} }
-(-)^{S+T} {\rm e}^{-{\rm i} {\bf p} \cdot {\bf r} } \right]
\chi^S_{M_S} \eta^T_{M_T=0} \nonumber \\
&=&4\pi\sqrt{2} \sum_{JM_JL} {\rm i}^L\, \epsilon_{LST} \,\,
 j_L(pr) [Z_{LSM_S}^{JM_J}(\hat{\bf p})]^* \,
{\cal Y}_{LSJ}^{M_J}(\hat{\bf r})\eta^T_0 \>\>.
\label{eq:pwe}
\end{eqnarray}
Here $j_L(pr)$ denotes the regular spherical Bessel function,
and the following definitions have been introduced:

\begin{equation}
Z_{LSM_S}^{JM_J}(\hat{\bf p})\equiv
\sum_{M_L} \langle LM_L,SM_S\mid JM_J\rangle
\, Y_{LM_L}(\hat {\bf p}) \>\>,
\end{equation}
\begin{equation}
\epsilon_{LST} \equiv \frac{1}{2} \Big[1 -(-1)^{L+S+T} \Big] \>\>.
\end{equation}
The factor $\epsilon_{LST}$ ensures that the plane waves are properly
antisymmetrized.

The $T$-matrix corresponding to the potential $v$
is defined as~\cite{Goldberger64}

\begin{equation}
T({\bf p}^\prime,S^\prime M_S^\prime,T^\prime;{\bf p},SM_S,T)=\,
_0\langle {\bf p}^\prime,S^\prime M_S^\prime,T^\prime
\mid v \mid {\bf p},SM_S,T\rangle^{(+)} \>\>.
\label{eq:tmas}
\end{equation}
Insertion of the plane-wave states
$\mid {\bf p},SM_S,T\rangle_0$ into the
right-hand-side of the Lippmann-Schwinger equation leads to

\begin{eqnarray}
\mid{\bf p},SM_S,T\rangle^{(+)}&=&\mid{\bf p},SM_S,T\rangle_0 \nonumber \\
&+&\sum_{S^\prime M_S^\prime T^\prime}\int\frac{{\rm d}{\bf p}^\prime}{(2\pi)^3}
\frac{1}{2} \mid{\bf p}^\prime,S^\prime M_S^\prime,T^\prime\rangle_0
\, \frac{T({\bf p}^\prime,S^\prime M_S^\prime,T^\prime;{\bf p},SM_S,T)}
{E-p^{\prime 2}/(2\mu) +{\rm i}\epsilon} \>\>,
\label{eq:LSa}
\end{eqnarray}
from which the partial wave expansion of the scattering state
is easily obtained by first noting that
the potential, and hence the $T$-matrix, can be expanded as

\begin{eqnarray}
_0 \langle {\bf p}^\prime,S^\prime M_S^\prime,T^\prime
\mid v \mid {\bf p},SM_S,T\rangle_0 &=& 2(4\pi)^2\! \sum_{JM_J}
\sum_{L L^\prime}
\epsilon_{L^\prime S^\prime T^\prime}\,\, \epsilon_{LST} \,
Z_{L^\prime S^\prime M_S^\prime}^{JM_J}(\hat{\bf p}^\prime) \nonumber \\
&&[Z_{LSM_S}^{JM_J}(\hat{\bf p})]^*\,
v^J_{L^\prime S^\prime T^\prime,LST} (p^\prime;p) \>\>,
\label{eq:tmae}
\end{eqnarray}
with
\begin{equation}
v^J_{L^\prime S^\prime T^\prime,LST}(p^\prime;p)
={\rm i}^{L-L^\prime}\int {\rm d}{\bf r} \,
j_{L^\prime}(p^\prime r) {\cal Y}_{L^\prime S^\prime J}^{M_J \dagger} \,
\eta^{T^\prime \dagger}_0 \, v({\bf r}) \, \eta^T_0 \, {\cal Y}_{LSJ}^{M_J}\, j_L(p r) \>\>.
\label{eq:vpep}
\end{equation}
After insertion of the corresponding expansion for the $T$-matrix
into Eq.~(\ref{eq:LSa}) and a number of standard manipulations,
the scattering-state wave function can be written as

\begin{eqnarray}
\psi^{(+)}_{ {\bf p}, SM_S,T}({\bf r})&=&
4\pi\sqrt{2} \sum_{JM_J}\, \sum_{L L^\prime S^\prime T^\prime}
{\rm i}^{L^\prime}\, \epsilon_{L^\prime S^\prime T^\prime}
\, \epsilon_{LST} \, [Z_{LSM_S}^{JM_J}(\hat{\bf p})]^* \nonumber \\
&&\frac{w^J_{L^\prime S^\prime T^\prime,LST}(r;p)}{r}
{\cal Y}_{L^\prime S^\prime J}^{M_J}(\hat{\bf r})\, \eta^{T^\prime}_0 \>\>,
\label{eq:psipw}
\end{eqnarray}
with

\begin{equation}
\frac{w^J_{\alpha^\prime,\alpha}(r;p)}{r}=
\delta_{\alpha^\prime,\alpha}\, j_{L^\prime}(pr)
+\frac{2}{\pi}\int_0^\infty {\rm d}p^\prime p^{\prime\, 2}
j_{L^\prime}(p^\prime r) \frac{1}
{E-p^{\prime 2}/(2\mu)+{\rm i}\epsilon}
T^J_{\alpha^\prime ,\alpha}(p^\prime;p)  \ ,
\label{eq:psip}
\end{equation}
where the label $\alpha$ ($\alpha^\prime$) stands for the set of quantum numbers
$LST$ ($L^\prime S^\prime T^\prime$).  The (complex) radial wave function $w(r)$
behaves in the asymptotic region $r \rightarrow \infty$ as

\begin{equation}
\frac{w^J_{\alpha^\prime,\alpha}(r;p)}{r}\simeq \frac{1}{2} \Big[
\delta_{\alpha^\prime,\alpha} h^{(2)}_{L^\prime}(pr)
+h^{(1)}_{L^\prime}(pr) S^J_{\alpha^\prime , \alpha}(p) \Big] \>\>,
\label{eq:asy}
\end{equation}
where the on-shell ($p^\prime=p$) $S$-matrix has been introduced,

\begin{equation}
S^J_{\alpha^\prime , \alpha}(p)=\delta_{\alpha^\prime,\alpha}
-4{\rm i}\, \mu p \, T^J_{\alpha^\prime , \alpha} (p;p) \>\>,
\label{eq:sma}
\end{equation}
and the functions $h^{(1,2)}(pr)$ are defined in terms of the regular
and irregular ($n_L$) spherical Bessel functions as

\begin{equation}
h^{(1,2)}_L(pr)=j_L(pr) \pm {\rm i}\, n_L(pr) \>\>.
\end{equation}

\subsection{Schr\"odinger equation, phase-shifts, mixing angles, and the scattering amplitude}
\label{sec:phase}

The coupled-channel Schr\"odinger equations for the radial wave
functions $w(r)$ read:

\begin{equation}
\Bigg[ -\frac{ {\rm d}^2}{{\rm d}r^2} + \frac{L^\prime (L^\prime +1)}{r^2}
-p^2 \Bigg] w^J_{\alpha^\prime,\alpha}(r;p)
+\sum_{\beta} r \, v^J_{\alpha^\prime,\beta}(r)\, \frac{1}{r}
w^J_{\beta,\alpha}(r;p) = 0  \>\>,
\label{eq:cschr}
\end{equation}
with

\begin{equation}
v^J_{\alpha^\prime,\alpha}(r)={\rm i}^{L-L^\prime}\, 2\mu \,
\int{\rm d}\Omega\, {\cal Y}_{\alpha^\prime J}^{M_J \dagger}
\, \eta^{T^\prime\dagger}_0 \, v({\bf r})\, \eta^T_0 \, {\cal Y}_{\alpha J}^{M_J} \>\>,
\label{eq:vpe}
\end{equation}
where, because of time reversal invariance, the matrix
$v^J_{\alpha^\prime,\alpha}$ can be shown to be
real and symmetric (this is the reason for
the somewhat unconventional phase factor
in Eq.~(\ref{eq:vpe}); in order to maintain symmetry for
both the $v^{\rm PC}$- and $v^{\rm PV}$-matrices, and
hence the $S$-matrix, the states used here differ by a factor ${\rm i}^L$ from
those usually used in nucleon-nucleon scattering analyses).
The asymptotic behavior of the $w(r)$'s is given 
in Eq.~(\ref{eq:asy}), while explicit expressions for the
radial functions $v^{J,{\rm PV}}_{\alpha^\prime,\alpha}(r)$ can
be found in Ref.~\cite{Carlson02}---those associated with
$v^{\rm PC}$ are well known.

There are two coupled channels for $J$=0, and
four coupled channels for $J\geq 1$.  The situation
is summarized in Table~\ref{tb:chan}.
Again because of the invariance under time-inversion
transformations of $v^{\rm PC}+v^{\rm PV}$,
the $S$-matrix is symmetric (apart from also
being unitary), and can therefore be written
as~\cite{Goldberger64}

\begin{equation}
S^J = U^{\rm T} \, S^J_{\rm D} \, U \>\>,
\label{eq:sdiag}
\end{equation}
where $U$ is a real orthogonal matrix, and $S^J_D$ is a diagonal
matrix of the form

\begin{equation}
S^J_{\rm D;\alpha^\prime,\alpha} = \delta_{\alpha^\prime,\alpha}
{\rm e}^{2 {\rm i} \delta^J_\alpha} \>\>.
\end{equation}
Here $\delta^J_\alpha$ is the (real) phase-shift in
channel $\alpha$, which is function of the energy $E$ with $p=\sqrt{2\mu\, E}$.
The mixing matrix $U$ can be written as

\begin{eqnarray}
U &=& U^{(12)} \quad \qquad \qquad  J=0 \>\>, \\
  &=& \prod_{1 \leq i < j \leq 4} U^{(ij)}
\qquad J \geq 1 \>\>,
\label{eq:uma}
\end{eqnarray}
where $U^{(ij)}$ is the $2 \times 2$ or $4 \times 4$
orthogonal matrix, that includes the coupling between
channels $i$ and $j$ only, for example

\noindent
\centerline{
$
  U^{(13)}= \left[ \begin{array}{cccc}
           {\rm cos}\,\epsilon^J_{13} & 0 & {\rm sin}\,\epsilon^J_{13} & 0 \\
           0                          & 1 & 0                          & 0 \\
          -{\rm sin}\,\epsilon^J_{13} & 0 & {\rm cos}\,\epsilon^J_{13} & 0 \\
           0                          & 0 & 0                          & 1 
           \end{array} \right]
\simeq 1 +\epsilon^J_{13} \left[ \begin{array}{cccc}
           0 & 0 & 1 & 0 \\
           0 & 0 & 0 & 0 \\
          -1 & 0 & 0 & 0 \\
           0 & 0 & 0 & 0 
\end{array} \right ] \>\>.
$ }
\vskip 0.7cm

\noindent
Note that no coupling is allowed between channels 3 and 4
in the notation of Table~\ref{tb:chan}, and hence $U^{(34)}=1$.  Thus, for $J$=0
there are two phase-shifts and a mixing
angle, while for $J\geq 1$ there are
four phase-shifts and five mixing angles.  Of course, since
$| v^{\rm PV} | \ll | v^{\rm PC} |$, the mixing angles $\epsilon^J_{ij}$
induced by $v^{\rm PV}$ are $\ll 1$, a fact already exploited in the
last expression above for $U$.  Given the channel ordering in
Table~\ref{tb:chan}, Table~\ref{tb:mixing} specifies
which of the channel mixings are induced by $v^{\rm PC}$
and which by $v^{\rm PV}$.

The reality of the potential matrix elements
$v^J_{\alpha^\prime,\alpha}(r)$ makes
it possible to construct real solutions of the
Schr\"odinger equation~(\ref{eq:cschr}).
The problem is reduced to determining the
relation between these solutions and the complex $w(r)$'s functions.
Using Eq.~(\ref{eq:sdiag}) and $U^{\rm T} U=1$,
the $w(r)$'s can be expressed in the asymptotic region as

\begin{eqnarray}
\frac{w^J_{\alpha^\prime,\alpha}}{r} &\simeq&
\sum_\beta (U^{\rm T})_{\alpha^\prime \beta} \, {\rm e}^{ {\rm i}\delta^J_\beta}
\frac{ h^{(2)}_{\alpha^\prime} {\rm e}^{-{\rm i}\delta^J_\beta}
+  h^{(1)}_{\alpha^\prime} {\rm e}^{ {\rm i}\delta^J_\beta} }{2}
U_{\beta \alpha} \nonumber \\
&=& \sum_\beta (U^{\rm T})_{\alpha^\prime \beta} \,
{\rm e}^{ {\rm i}\delta^J_\beta} \,
\frac{ {\rm sin}[pr-L^\prime \, \pi/2 +\delta^J_\beta] }{pr}
\, U_{\beta \alpha} \>\>.
\end{eqnarray}
The expression above is real apart from
the ${\rm exp}({\rm i}\delta^J_\beta)$.  To eliminate
this factor, the following linear combinations of the
$w(r)$'s are introduced

\begin{eqnarray}
\frac{u^J_{\alpha^\prime,\alpha}}{r} &\equiv& 
{\rm e}^{-{\rm i}\delta^J_\alpha}\sum_\beta
\frac{w^J_{\alpha^\prime,\beta}}{r}(U^{\rm T})_{\beta \alpha} \nonumber \\
&\simeq&(U^{\rm T})_{\alpha^\prime \alpha} \left[
{\rm cos}\, \delta^J_\alpha \, j_{L^\prime}(pr)
     -{\rm sin}\, \delta^J_\alpha \, n_{L^\prime}(pr)\right] \>\>,
\label{eq:uasy}
\end{eqnarray}
and the $u(r)$'s are then the sought real solutions of Eq.~(\ref{eq:cschr}).

The asymptotic behavior of the $u(r)$'s can now be read off
from Eq.~(\ref{eq:uasy}) once the $U$-matrices above have been
constructed.  The latter can be written, up
to linear terms in the \lq\lq small\rq\rq mixing angles induced
by $v^{\rm PV}$, as

\noindent
\centerline{
$
  U= \left[ \begin{array}{cc}
            1             & \epsilon^0_{12} \\
           -\epsilon^0_{12} & 1
           \end{array} \right] \quad J=0 \>\>,
$}
\vskip 0.7cm
\centerline{
$
  U= \left[ \begin{array}{cccc}
    {\rm cos}\epsilon^J_{12}
  & {\rm sin}\epsilon^J_{12}
  &  \epsilon^J_{13}{\rm cos}\epsilon^J_{12}+\epsilon^J_{23}{\rm sin}\epsilon^J_{12} 
  &  \epsilon^J_{14}{\rm cos}\epsilon^J_{12}+\epsilon^J_{24}{\rm sin}\epsilon^J_{12} \\
   -{\rm sin}\epsilon^J_{12}
  & {\rm cos}\epsilon^J_{12}
  & -\epsilon^J_{13}{\rm sin}\epsilon^J_{12}+\epsilon^J_{23}{\rm cos}\epsilon^J_{12} 
  & -\epsilon^J_{14}{\rm sin}\epsilon^J_{12}+\epsilon^J_{24}{\rm cos}\epsilon^J_{12} \\
  -\epsilon^J_{13} & -\epsilon^J_{23} & 1  & 0 \\
  -\epsilon^J_{14} & -\epsilon^J_{24} & 0  & 1 \\
\end{array} \right ]  \quad J \geq 1 \>\>.
$}
\vskip 0.7cm

Inverting the first line of Eq.~(\ref{eq:uasy}),

\begin{equation}
\frac{w^J_{\alpha^\prime,\alpha}}{r} = \sum_\beta
{\rm e}^{ {\rm i}\delta^J_\beta}
\frac{u^J_{\alpha^\prime,\beta}}{r}\,U_{\beta \alpha} \>\>,
\end{equation}
and inserting the resulting expressions into Eq.~(\ref{eq:cschr})
leads to the Schr\"odinger equations satisfied by
the (real) functions $u(r)$.  They are identical to
those of Eq.~(\ref{eq:cschr}), but for the $w(r)$'s being replaced
by the $u(r)$'s.  These equations are then solved by standard numerical
techniques.  Note that: i) $v^J_{\alpha,\alpha} =
v^{J,\, {\rm PC}}_{\alpha,\alpha}$, since the diagonal
matrix elements of $v^{\rm PV}$ vanish because of parity selection rules;
ii) terms of the type $r \,v^{J,\, {\rm PV}}_{\alpha^\prime,\beta}(r)
u^J_{\beta,\alpha}(r)/r$ involving the product of a PV 
potential matrix element with a $v^{\rm PV}$-induced wave function are neglected.

Finally, the physical amplitude for $n$$p$ elastic scattering from an initial
state with spin projections $m_n$, $m_p$ to a final state
with spin projections $m^\prime_n$, $m^\prime_p$ is given by
 
\begin{eqnarray}
\langle m_n^\prime m_p^\prime \mid M \mid m_n m_p\rangle&=&\frac{1}{2}
\sum_{S^\prime M_S^\prime T^\prime,SM_S T} (-)^{T+T^\prime}
\langle \frac{1}{2} m_n^\prime ,
        \frac{1}{2} m_p^\prime \mid S^\prime M_S^\prime\rangle
\langle \frac{1}{2} m_n ,\frac{1}{2} m_p \mid S M_S\rangle \nonumber \\
&&  M_{S^\prime M_S^\prime T^\prime, SM_S T}(E,\theta) \>\>,
\end{eqnarray}
where the amplitude $M$ is related to the
$T$-matrix defined in Eq.~(\ref{eq:tmas}) via
 
\begin{equation}
M_{S^\prime M_S^\prime T^\prime, SM_S T}(E,\theta) =-\frac{\mu}{2\pi}
T({\bf p}^\prime,S^\prime M_S^\prime,T^\prime;p\hat{\bf z},SM_S,T) \>\>,
\end{equation}
and the factor $(-)^{T+T^\prime}/2$ comes from the Clebsch-Gordan coefficients
combining the neutron and proton states to total initial (final)
isospin $T$ ($T^\prime$).  Note that the direction of the initial momentum
${\bf p}$ has been taken to define the spin quantization axis
(the $z$-axis), $\theta$ is the angle between
$\hat {\bf p}$ and $\hat{\bf p}^\prime$, the direction
of the final momentum, and the energy $E=p^2/(2\mu)$
($= p^{\prime 2}/(2\mu)$).  Using the expansion of
the $T$-matrix, Eq.~(\ref{eq:tmae}) with $v^J_{L^\prime S^\prime T^\prime,LST}$
replaced by $T^J_{L^\prime S^\prime T^\prime,LST}$, and the relation
between the $S$- and $T$-matrices, Eq.~(\ref{eq:sma}), the amplitude induced
by $v^{\rm PC}+v^{\rm PV}$ can be expressed as
 
\begin{eqnarray}
M_{S^\prime M_S^\prime T^\prime, SM_S T}(E,\theta)&=& \sqrt{4\pi} \sum_{JLL^\prime}
\sqrt{2L+1}\, \epsilon_{\alpha^\prime}\, \epsilon_\alpha \,
\langle L^\prime (M_S-M^\prime_S), S^\prime M^\prime_S\mid J M_S\rangle \nonumber \\
&&\langle L 0, SM_S\mid J M_S\rangle Y_{L^\prime (M_S-M_S^\prime)}(\theta)\,
\frac{S^J_{\alpha^\prime,\alpha}(p) - \delta_{\alpha^\prime,\alpha}}
{{\rm i} p} \>\>,
\label{eq:am}
\end{eqnarray}
where again $\alpha (\alpha^\prime) = LST (L^\prime S^\prime T^\prime)$.

\subsection{Momentum-space formulation} 
\label{sec:p-space}

In order to consider the PC momentum-space Bonn (BONN)~\cite{Machleidt01}
and Nijmegen (NIJM-I)~\cite{Stoks94} potentials, it is useful to formulate the $n$$p$ scattering
problem in $p$-space.  One way to accomplish this is to solve for
the $K$-matrix~\cite{Goldberger64} 

\begin{eqnarray}
K^J_{\alpha^\prime,\alpha}(p^\prime;p)&=&
v^J_{\alpha^\prime,\alpha}(p^\prime;p)\nonumber \\
&+&\frac{4\mu}{\pi} \int_0^\infty {\rm d}kk^2 \sum_\beta
v^J_{\alpha^\prime,\beta}(p^\prime;k)
\frac{\cal P}{p^2-k^2}K^J_{\beta,\alpha}(k;p) \>\>,
\label{eq:kma}
\end{eqnarray}
where ${\cal P}$ denotes a principal-value integration,
and the $p$-space matrix elements of the potential
are defined in Eq.~(\ref{eq:vpep}).  The integral
equations~(\ref{eq:kma}) are discretized, and the resulting
systems of linear equations are solved by direct numerical
inversion.  The principal-value integration is eliminated by
a standard subtraction technique~\cite{Gloeckle83}.  Once the
$K$-matrices in the various channels have been determined, the
corresponding (on-shell) $S$-matrices are obtained from

\begin{equation}
S^J(p)=\left[ 1+2{\rm i} \, \mu p\, K^J(p;p) \right]^{-1}
\left[ 1-2{\rm i} \, \mu p\, K^J(p;p) \right] \ ,
\label{eq:skma}
\end{equation}
and from these the amplitudes $M_{S^\prime M_S^\prime T^\prime, SM_ST}(E,\theta)$,
Eq.~(\ref{eq:am}), are constructed.

Some of the studies of PV effects in the $n$$p$ system of interest
here, specifically those relative to the $\vec n$$p$ radiative capture, $d(\vec\gamma,n)p$ 
photo-disintegration at threshold and $d(\vec e,e^\prime)$$n$$p$ electro-disintegration
in quasi-elastic kinematics are more conveniently carried out in $r$-space, and therefore require
$r$-space wave functions.  To this end, one first re-writes Eq.~(\ref{eq:psip})
in a compact notation as

\begin{eqnarray}
\frac{w^J(r;p)}{r}= j(pr)
&-&2{\rm i}\, \mu p\, j(pr)\, T^J(p;p) \nonumber \\
&+&\frac{4\mu}{\pi}\int_0^\infty {\rm d}k k^2
j(kr) \frac{\cal P} {p^2-k^2}
T^J(k;p) \ ,
\label{eq:psir}
\end{eqnarray}
where the matrices $[w^J(r;p)]_{\alpha^\prime,\alpha}\equiv w^J_{\alpha^\prime,\alpha}(r;p)$
and $[j(pr)]_{\alpha^\prime,\alpha}\equiv \delta_{\alpha^\prime,\alpha} j_{L^\prime}(pr)$
have been introduced for ease of presentation.  Then, by making use of the following
relation between the off-shell $T$- and $K$-matrices

\begin{equation}
T^J(p^\prime;p)=K^J(p^\prime;p)-2{\rm i}\, \mu p\, K^J(p^\prime;p)\,T^J(p;p) \ , 
\end{equation}
which on-shell leads to

\begin{equation}
T^J(p;p)=\Big[ 1+ 2{\rm i}\, \mu p\, K^J(p;p)\Big]^{-1} K^J(p;p) \ ,
\end{equation}
one can simply express the $w^J(r;p)$ matrix of solutions in terms of the previously
determined $K$-matrix as

\begin{eqnarray}
\frac{w^J(r;p)}{r}&=&\Bigg[ j(pr)
+\frac{4\mu}{\pi}\int_0^\infty {\rm d}k\, k^2\,
j(kr) \frac{\cal P} {p^2-k^2}
K^J(k;p)\Bigg]\nonumber \\
&\times& \Bigg[ 1+ 2{\rm i}\, \mu p\, K^J(p;p)\Bigg]^{-1} \ .
\label{eq:psirk}
\end{eqnarray}
The Bessel transforms above are carried out numerically by Gaussian integration 
over a uniform $p$-grid extending up to momenta $\simeq$ 125 fm$^{-1}$.  The computer
programs have been successfully tested by comparing, for the PC
Argonne $v_{18}$ (AV18)~\cite{Wiringa95} and PV DDH~\cite{Desplanques80}
potentials, $r$-space wave functions
as obtained from Eq.~(\ref{eq:psirk}) and by direct solution of
the Schr\"odinger equations, Eq.~(\ref{eq:cschr}).

\subsection{The deuteron wave function} 
\label{sec:deut}

The deuteron state has $J$=1 and its normalized wave function is
written in $r$-space as

\begin{equation}
\psi_{d,m_d}({\bf r})=\sum_{LST} {\rm i}^L \, \epsilon_{LST} \, u_{LST}(r) \, 
{\cal Y}_{LS,J=1}^{m_d}(\hat{\bf r})\, \eta^T_0 \ .
\label{eq:dw}
\end{equation}
It has PC components with $L$$S$$T$=010 and 210, the standard $^3$S$_1$
and $^3$D$_1$ waves (however, note again the unconventional phase factor ${\rm i}^L$,
which makes the sign of the D-wave opposite to that of the S-wave), and PV
components with $L$$S$$T$=100 and 111, the $^1$P$_1$ and $^3$P$_1$ waves
(which are real functions because of the phase choice above).
The radial functions are determined by solving the Schr\"odinger
equation~(\ref{eq:cschr}) in $J$=1 channel with the boundary conditions
$u_{LST}(r)\propto r^L$ in the limit $r \rightarrow 0$ and   

\begin{eqnarray}
u_{010}(r) &\propto& \frac{{\rm e}^{-\kappa r}}{r} \ , \\
u_{210}(r) &\propto& \frac{{\rm e}^{-\kappa r}}{r}
\Bigg[ 1 +\frac{3}{\kappa r}+\frac{3}{(\kappa r)^2}\Bigg]\ , \\
u_{100}(r) &\propto& \frac{{\rm e}^{-\kappa r}}{r} \Bigg( 1 +\frac{1}{\kappa r}\Bigg)\ , 
\end{eqnarray}
in the asymptotic region.  The asymptotic behavior of $u_{111}(r)$ is
identical to that of $u_{100}(r)$ above, and the constant $\kappa$
denotes the combination $\sqrt{2\mu |E_d|}$, where $|E_d|$ is the deuteron
binding energy (2.225 MeV).

In $p$-space the deuteron wave function is obtained from solutions of the
homogeneous integral equations

\begin{equation}
\overline{u}_{LST}(p)=\frac{1}{E_d-p^2/(2\mu)} \frac{2}{\pi}
\int_0^\infty {\rm d}k\, k^2 \sum_{L^\prime S^\prime T^\prime}
v^{J=1}_{LST,L^\prime S^\prime T^\prime}(p;k)\, \overline{u}_{L^\prime S^\prime T^\prime}(k) \ ,
\end{equation}
and from these 

\begin{equation}
u_{LST}(r)=\frac{2}{\pi} \int_0^\infty {\rm d}p\, p^2\, j_L(pr)\, \overline{u}_{LST}(p) \ .
\end{equation}

\begin{figure}[bth]
\let\picnaturalsize=N
\def\picsize{5in}
\def\picfilenamea{figs/deut.eps}
\epsfbox{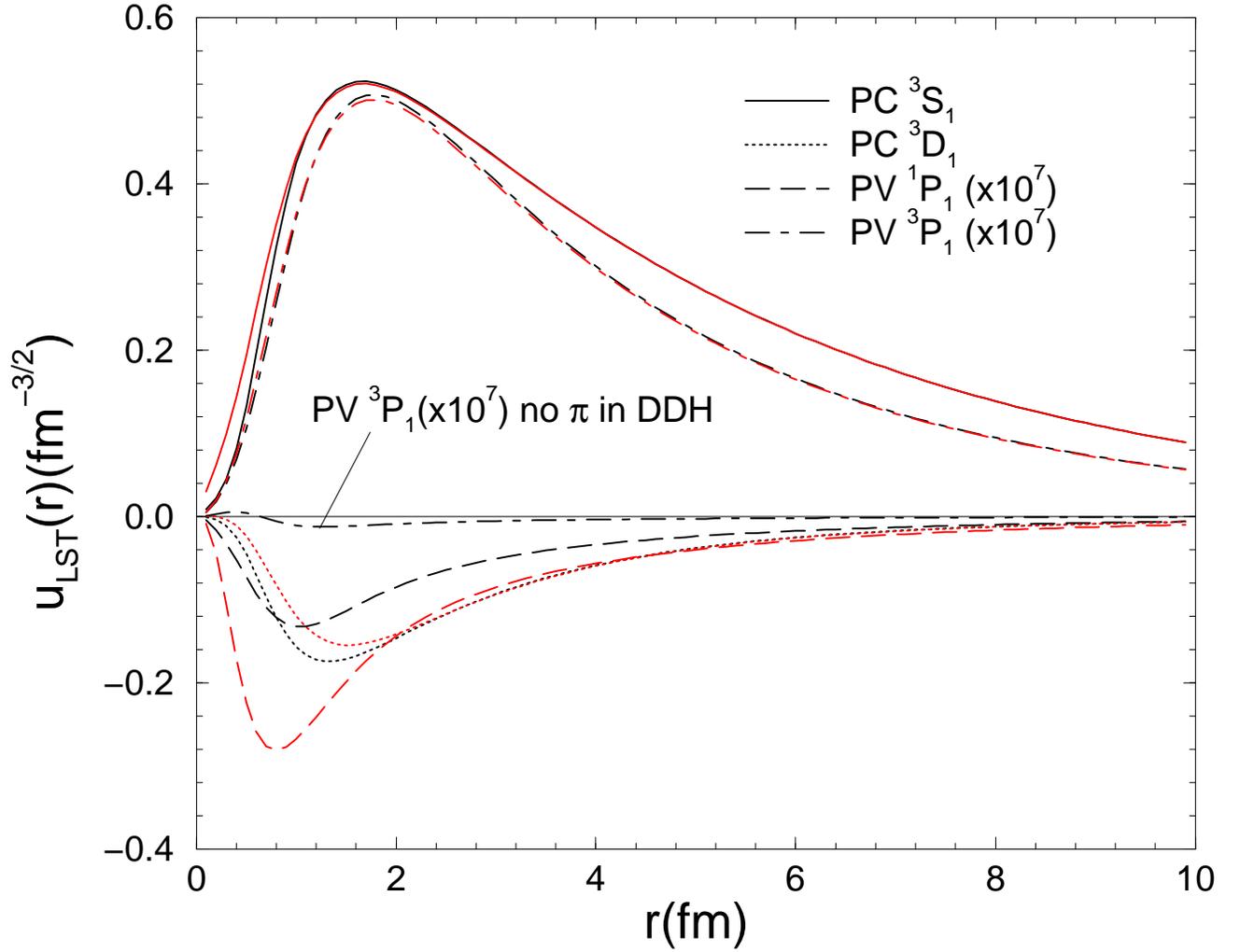}
\caption{The deuteron PC $^3$S$_1$ and $^3$D$_1$ and PV $^1$P$_1$ and $^3$P$_1$
radial wave functions obtained with the (PC) AV18 (black) or BONN (red) and (PV)
DDH potentials.  Also shown is the $^3$P$_1$ wave obtained with the AV18 and a
truncated DDH potential (labeled \lq\lq no $\pi$ in DDH\rq\rq), including only
$\rho$- and $\omega$-meson exchange contributions.  For the phase convention,
see text.}
\label{fig:deut}
\end{figure}

Figure~\ref{fig:deut} displays the functions $u_{LST}(r)$ obtained with
the PC AV18~\cite{Wiringa95} (BONN~\cite{Machleidt01}) and
PV DDH~\cite{Desplanques80} potentials (the values for the coupling constants
and cutoff parameters in the DDH potential are those listed in Table~\ref{tb:ddh}).
The PC $^3$S$_1$ and $^3$D$_1$ components are not very sensitive to the input
PC potential.  For example, most of the difference between the AV18
and BONN $^3$D$_1$ waves is due to non-localities present in the
one-pion-exchange (OPE) part of the BONN potential.  In fact, it
has been known for over two decades~\cite{Friar77}, and recently re-emphasized
by Forest~\cite{Forest00}, that the local and non-local OPE interactions
are related to each other by a unitary transformation.  Therefore the differences
between local and non-local OPE cannot be of any consequence for the
prediction of observables, such as binding energies and electromagnetic form
factors, provided, of course, that three-body interactions and/or two-body
currents generated by the unitary transformation are also included---see Ref.~\cite{Schiavilla01} for
a recent demonstration of this fact within the context of a calculation
of the deuteron structure function $A(q)$ and tensor observable $T_{20}(q)$
based on the local AV18 and non-local BONN potentials and associated (unitarily
consistent) electromagnetic currents.  This
point was also stressed in Ref.~\cite{Carlson02}.

The PV $^3$P$_1$ component is, in magnitude, much larger than the $^1$P$_1$.
This is easily understood, since the long-range $\pi$-exchange term in the
DDH potential is non-vanishing only for transitions in which $|T-T^\prime\,|$=1,
and therefore does not contribute in the $^1$P$_1$ channel.  In this channel,
however, the DDH $\rho$- and $\omega$-exchange terms play a role.  Note that,
because of the short-range character of the associated dynamics, the AV18
and BONN $^1$P$_1$ waves show considerably more model-dependence than the corresponding
$^3$P$_1$ waves. 

Finally, in Fig.~\ref{fig:deut} the PV $^3$P$_1$ wave
obtained with the AV18 and a truncated DDH potential,
retaining only the short-range $\rho$- and $\omega$-exchanges,
is also shown.  The comparison between the $^3$P$_1$
waves corresponding to the full and truncated DDH potentials
demonstrates that this channel is indeed dominated by the $\pi$-exchange
term in the DDH.
\section{Parity-Violating Observables}
\label{sec:obs}

In this section we give explicit expressions for
parity-violating (PV) observables in the $n$$p$ system, including
the longitudinal asymmetry and spin rotation
in $\vec n$$p$ elastic scattering, the photon
asymmetry in $\vec n$$p$ radiative capture, and the asymmetries
in deuteron photo-disintegration $d(\vec \gamma,n)p$ in the threshold region and
electro-disintegration $d(\vec e,e^\prime)np$ in quasi-elastic kinematics.

\subsection{Longitudinal asymmetry and spin rotation in $\vec n$$p$ elastic scattering}
\label{sec:nn}
 
The differential cross section for scattering of a neutron with initial
polarization $m_n$ is given by
 
\begin{equation}
\sigma_{m_n}(E,\theta)=\frac{1}{2} \sum_{m_p} \sum_{m_n^\prime m_p^\prime}
\mid \langle m_n^\prime m_p^\prime \mid M \mid m_n m_p\rangle\mid^2 \>\>,
\end{equation}
and the longitudinal asymmetry is defined as
 
\begin{equation}
A(E,\theta)=\frac{\sigma_{+}(E,\theta) - \sigma_{-}(E,\theta)}
                 {\sigma_{+}(E,\theta) + \sigma_{-}(E,\theta)} \>\>,
\label{eq:athe}
\end{equation}
where $\pm$ denote the initial polarizations $\pm 1/2$.  The
total asymmetry $A(E)$, integrated over the solid angle, then reads 
 
\begin{equation}
A(E)= \frac{\int {\rm d}\Omega\, \sigma(E,\theta) \, A(E,\theta) }
{\int{\rm d}\Omega \, \sigma(E,\theta)} \>\>,
\label{eq:asye}
\end{equation}
where $\sigma= \left( \sigma_{+}+\sigma_{-} \right)/2$
is the spin-averaged differential cross section.  The optical theorem
allows $A(E)$ to be simply expressed as
 
\begin{eqnarray}
A(E)\!&=&\!{\rm Im}\sum_{T\, T^\prime}\Bigg[
\left[ (-)^{T+T^\prime} -1 \right] M_{11T^\prime,11T}(E,0)
+\frac{(-)^{T+T^\prime}}{2} \sum_{S\, S^\prime} \left[1 - (-)^{S+S^\prime} \right]
M_{S^\prime 0 T^\prime,S0T}(E,0) \Bigg] \nonumber \\
&\times& \frac{1} {{\rm Im} \sum_{SM_ST} M_{SM_ST,SM_ST}(E,0) } \>\>,
\label{eq:asy_t}
\end{eqnarray}
where in the equation above use has been made of the symmetry property
 
\begin{equation}
M_{S^\prime,M_S^\prime, T^\prime; S,M_S,T}(E,\theta)
=(-)^{T+T^\prime} (-)^{M_S-M_S^\prime}
M_{S^\prime,-M_S^\prime,T^\prime;S,-M_S,T}(E,\theta) \>\>.
\end{equation}
It is clear that the numerator of $A(E)$ would vanish
in the absence of PV interactions, since
$v^{\rm PC}$, in contrast to $v^{\rm PV}$,
cannot change the total spin $S$ or isospin $T$ of the $n$$p$ pair.
In particular, the long-range part of $v^{\rm PV}$ due
to pion exchange can only contribute to the first term in the numerator
of Eq.~(\ref{eq:asy_t}), since it is diagonal in $S$, but non-vanishing
for transitions $\mid T-T^\prime \mid$=1.
 
The transmission of a low-energy neutron beam through
matter is described in terms of an index of refraction.  A
heuristic argument, outlined in Ref.~\cite{Fermi50}, and the
more rigorous---although less transparent---derivation presented
in Ref.~\cite{Goldberger64} show that a neutron with spin
projection $\mid m_n\rangle$, after traversing a slab of width
$d$ of matter, is described by an asymptotic wave function given by
 
\begin{equation}
{\rm e}^{{\rm i} p\, (z-d)} \,
{\rm e}^{{\rm i} p\, d\, n_{m_n}} \mid m_n\rangle \>\>,
\end{equation}
where ${\bf p}$=${\bf p}_n/2$ is the initial relative momentum (assumed along
the $z$-axis), and the index of refraction $n_{m_n}$ is related to the density $\rho$
of scattering centers in matter and the forward scattering amplitude.  For
the specific case under consideration here---neutron scattering from hydrogen---this
relation reads:
 
\begin{equation}
n_{m_n}-1 = \frac{2 \pi \rho}{p^2}\, \frac{1}{2} \sum_{m_p}
\langle m_n m_p \mid M \mid m_n m_p\rangle \Big\vert_{\theta=0} \>\>.
\end{equation}
Thus a neutron, initially polarized in the $x$-direction,
 
\begin{equation}
{\rm e}^{{\rm i} p z} \left( \mid +\rangle + \mid -\rangle \right)/\sqrt{2} \>\>,
\end{equation}
having traversed a slab of matter, is described in the asymptotic
region by a wave function given by
 
\begin{equation}
{\rm e}^{{\rm i} p\, (z-d) }
{\rm e}^{{\rm i} p\, d\, (n_+ + n_-)/2 }
\left[ {\rm e}^{ {\rm i} p\,d\, (n_+ - n_-)/2 } \mid +\rangle +
       {\rm e}^{-{\rm i} p\,d\, (n_+ - n_-)/2 } \mid -\rangle \right]/\sqrt{2} \ .
\end{equation}
In the absence of $v^{\rm PV}$, the difference $n_+ - n_-$ vanishes, since
it is proportional to the sum over $T,T^\prime$ in the numerator of
Eq.~(\ref{eq:asy_t}), while
 
\begin{equation}
\frac{{\rm Im} (n_+ + n_-)}{2} =  \frac{\rho}{2\, p} \sigma \>\>,
\end{equation}
where $\sigma$ is the spin-averaged cross section introduced above, and
hence there will be an attenuation in the beam flux proportional
to exp($-\rho \, d \, \sigma$).  The real part of $(n_+ + n_-)/2$ instead
generates an un-observable phase factor.
 
However, if PV interactions are present, then the real
part of the (now non-vanishing) difference $n_+ - n_-$ leads to a
rotation of the neutron polarization by an angle $\phi$ explicitly
given by~\cite{Michel64,Stodolsky74,Avishai84}
\begin{equation}
\phi = -\frac{2 \pi\, \rho\, d}{p}\, \frac{1}{2} \sum_{m_p}
{\rm Re} \Bigg[ \langle +, m_p \mid M \mid +, m_p\rangle
               -\langle -, m_p \mid M \mid -, m_p\rangle \Bigg]_{\theta=0} \>\>.
\label{eq:nphi}
\end{equation}

\subsection{Photon asymmetry in $\vec n$$p$ radiative capture}
\label{sec:npr}

In the center-of-mass (CM) frame, the radiative transition amplitude
between an initial continuum state with neutron and proton in spin-projection
states $m_n$ and $m_p$, respectively, and in relative momentum ${\bf p}$,
and a final deuteron state in spin-projection state $m_d$, recoiling with momentum
$-{\bf q}$, is given by 

\begin{equation}
j^{(+)}_{\lambda m_d, m_n m_p}(p \hat{\bf z},{\bf q}) =
\langle -{\bf q};m_d \mid \hat{\bbox \epsilon}_\lambda^*({\bf q}) \cdot
{\bf j}^\dagger({\bf q})\mid p \hat{\bf z},m_n m_p\rangle^{(+)} \ ,
\end{equation}
where ${\bf q}$ is the momentum of the emitted photon and
$\hat{\bbox \epsilon}_\lambda({\bf q})$, $\lambda=\pm 1$,
are the spherical components of its polarization vector,
and ${\bf j}({\bf q})$ is the nuclear electromagnetic current
operator.  Note that $\hat{\bf p}$ has been taken along the $z$-axis,
the spin-quantization axis.

The initial $n$$p$ continuum state, satisfying outgoing-wave boundary
conditions, is related to that constructed in Sec.~\ref{sec:pwest} via

\begin{eqnarray}
\mid\! p\hat{\bf z},m_n m_p\rangle^{(+)}\!\!&=&\!\frac{1}{\sqrt{2}}\sum_{ST} (-)^{T+1}
\langle \frac{1}{2} m_n ,\frac{1}{2} m_p \mid\! S M_S\rangle 
\mid p\hat{\bf z},SM_S,T\rangle^{(+)} \nonumber \\ 
&=&\!\sqrt{4 \pi} \!\sum_{J \alpha}\! \overline{\epsilon}_\alpha \, \sqrt{2L+1}
\langle \frac{1}{2} m_n ,\frac{1}{2} m_p \mid\! SM_S\rangle \langle L0 ,SM_S \mid\! JM_S\rangle
|JM_S, \alpha \rangle^{(+)} \ ,
\label{eq:psia}
\end{eqnarray}
where in the first line the factor $(-)^{T+1}/\sqrt{2}$ is from a Clebsch-Gordan coefficient
combining the neutron and proton states to total isospin $T$, and in the second line
$\overline{\epsilon}_\alpha\equiv -(-)^T\, \epsilon_\alpha$ and 
the states $|JM_J, \alpha \rangle^{(+)}$ have wave functions 

\begin{equation}
\langle{\bf r} \mid JM_J,\alpha\rangle^{(+)} = \sum_{\alpha^\prime}
{\rm i}^{L^\prime}\, \epsilon_{\alpha^\prime}
\frac{w^J_{\alpha^\prime ,\alpha}(r;p)}{r}
{\cal Y}_{L^\prime S^\prime J}^{M_J}(\hat{\bf r})\, \eta^{T^\prime}_0 \ ,
\label{eq:psiap}
\end{equation}
with $\alpha$=$L$$S$$T$ and similarly for $\alpha^\prime$.  The quantum
numbers $\alpha$ and $\alpha^\prime$ characterize the incoming and outgoing
waves, respectively.

The CM differential cross-section for capture of a neutron with spin projection 
$m_n$ is then written as

\begin{equation}
\sigma^\gamma_{m_n}(\theta)=\frac{\sigma_0}{2} \sum_{m_p,\lambda,m_d}
\mid j^{(+)}_{\lambda m_d, m_n m_p}(p\hat{\bf z},{\bf q})\mid^2 \ ,
\end{equation}
where $\theta$ is the angle between $\hat {\bf z}$ and $\hat {\bf q}$ and

\begin{equation}
\sigma_0=\frac{\alpha}{2\pi\, v} \frac{q}{1+q/m_d} \ .
\end{equation}
Here $\alpha$ is the fine-structure constant,
$m_d$ is the deuteron mass, $v$ is the relative velocity, $v=p/\mu$,
and the photon energy $q$ is given by 

\begin{equation}
q=m_d \Bigg[ -1 + \sqrt{ 1 + \frac{2}{m_d} \left( |E_d|+\frac{p^2}{2\mu}\right) } \Bigg]
  \simeq |E_d|+\frac{p^2}{2\mu} \ .
\end{equation}

The photon asymmetry $A^\gamma(\theta)$ is defined as in Eq.~(\ref{eq:athe})
with $\sigma_{m_n}(\theta)$ replaced by $\sigma^\gamma_{m_n}(\theta)$.  By
expanding the matrix elements of the current operator in terms of reduced matrix
elements (RMEs) of electric ($E_l$) and magnetic ($M_l$) multipole operators
as~\cite{Marcucci00}

\begin{eqnarray}
\langle -{\bf q};m_d \mid \hat{\bbox \epsilon}_\lambda^*({\bf q}) \cdot
{\bf j}^\dagger({\bf q})\mid JM_J, \alpha \rangle^{(+)} &=&-\sqrt{2\pi} \sum_{ll_z}
(-{\rm i})^l \sqrt{\frac{2l+1}{3}} \langle JM_J,ll_z\mid1m_d \rangle \,
d^l_{l_z,-\lambda}(-\theta) \nonumber \\
&&\Big[E_l(J,\alpha)+\lambda M_l(J,\alpha)\Big]  \ ,
\label{eq:mult}
\end{eqnarray}
with
\begin{equation}
X_l(J,\alpha) \equiv \langle d,J\!=\!1\mid\mid X_l\mid\mid J,\alpha\rangle^{(+)} 
\end{equation}
and $X_l$=$E_l$ or $M_l$, one finds, by retaining the $^1$S$_0$
and $^3$S$_1$ channels in the sum over $\alpha$
in Eq.~(\ref{eq:psia}), the only relevant incoming waves in the energy regime
of interest here (fractions of eV),

\begin{equation}
A^\gamma(\theta) = a^\gamma \, {\rm cos}\, \theta \ ,
\end{equation}
where $a^\gamma$ is given by

\begin{equation}
a^\gamma=\frac{ -\sqrt{2}\, {\rm Re}\, \Big[ M_1^*(^1{\rm S}_0) E_1(^3{\rm S}_1) +
                                             E_1^*(^1{\rm S}_0) M_1(^3{\rm S}_1)\Big]
+ {\rm Re}\,\Big[E_1^*(^3{\rm S}_1) M_1(^3{\rm S}_1)\Big] }
{|M_1(^1{\rm S}_0)|^2 + |E_1(^1{\rm S}_0)|^2 + |M_1(^3{\rm S}_1)|^2 +  |E_1(^3{\rm S}_1)|^2} \ ,
\end{equation}
and in Eq.~(\ref{eq:mult}) the $d^l_{l_z,l_z^\prime}$
are standard rotation matrices~\cite{Edmonds57}.  The $E_l(J,\alpha)$ and
$M_l(J,\alpha)$ RMEs should carry a superscript $(+)$; it has been dropped for ease of
presentantion.

Several comments are now in order.  Firstly, the photon asymmetry
has the expected dependence on ${\rm cos}\, \theta$, since
$A^\gamma(\theta) \propto {\bbox \sigma}_n \cdot \hat{\bf q}$.
Note that the contributions of higher order multipole operators
with $l$=2 have been ignored in the equation above.

Secondly, because of the definition of the states in Eq.~(\ref{eq:psiap}),
a generic RME $X_l(J\alpha)$ is expressed as 

\begin{equation}
X_l(J\alpha)=\sum_{\alpha^\prime} X_l(J\alpha^\prime, \alpha) \ ,
\end{equation}
namely as a sum over the contributions of outgoing channels
$\alpha^\prime$ corresponding to an incoming channel $\alpha$; for example,

\begin{equation}
M_1(^1{\rm S}_0)=M_1(^1{\rm S}_0,^1{\rm S}_0)+M_1(^3{\rm P}_0,^1{\rm S}_0) \ .
\label{eq:m1def}
\end{equation}

Thirdly, in the absence of parity-violating interactions, the only surviving RMEs
are the $M_1(^1{\rm S}_0,^1{\rm S}_0)$, $M_1(^3{\rm S}_1,^3{\rm S}_1)$, and
$M_1(^3{\rm D}_1,^3{\rm S}_1)$, and therefore the parameter $a^\gamma$ vanishes.
Furthermore, the $M_1(^3{\rm S}_1,^3{\rm S}_1)$ RME also vanishes due to orthogonality of the
initial and final states (in the limit in which isoscalar two-body currents
are neglected), while the $M_1(^3{\rm D}_1,^3{\rm S}_1)$ RME is suppressed
in the energy regime of interest here.  Thus, the standard result for the
spin-averaged radiative capture cross-section, integrated over the solid
angle, follows:

\begin{equation}
\sigma^\gamma= (4 \pi)^2\, \sigma_0\, |M_1(^1{\rm S}_0,^1{\rm S}_0)|^2 \ .
\end{equation}

Lastly, when PV interactions are present, the analysis is more delicate, since
then, in addition to ad-mixing small opposite-parity components into the wave
functions corresponding to $v^{\rm PC}$, these interactions also induce two-body
terms in the electromagnetic current operator, as discussed in Sec.~\ref{sec:cnt}.
Thus, ${\bf j}$=${\bf j}^{\rm PC} +{\bf j}^{\rm PV}$, where ${\bf j}^{\rm PC}$
includes the convection and spin-magnetization currents of single nucleons as
well as the two-body currents associated with $v^{\rm PC}$, while ${\bf j}^{\rm PV}$
includes those terms generated by $v^{\rm PV}$.  The multipole operators can then
be written as $X_{ll_z}$=$X_{ll_z}^{\rm PC} +X_{ll_z}^{\rm PV}$, and those constructed
from ${\bf j}^{\rm PV}$ have un-natural parities, namely $(-)^l$ for $M^{\rm PV}_{ll_z}$ 
and $(-)^{l+1}$ for $E^{\rm PV}_{ll_z}$.  Therefore, for example, the RMEs
$M^{\rm PC}_1(^1{\rm S}_0,^1{\rm S}_0)$ and $M^{\rm PV}_1(^1{\rm S}_0,^1{\rm S}_0)$
connect the PC $^1$S$_0$ state to, respectively, the PC and PV components of the deuteron.
A straight-forward analysis then shows that, up to linear terms in effects
induced by $v^{\rm PV}$ either in the wave functions or currents, the parameter
$a^\gamma$ is given by

\begin{equation}
a^\gamma=\frac{ -\sqrt{2}\, {\rm Re}\, \Bigg[ M_1^*(^1{\rm S}_0,^1{\rm S}_0)
\Big[ E_1(^3{\rm S}_1,^3{\rm S}_1) + E_1(^3{\rm D}_1,^3{\rm S}_1) +
      E_1(^3{\rm P}_1,^3{\rm S}_1) \Big] \Bigg] }
{|M_1(^1{\rm S}_0,^1{\rm S}_0)|^2 } \ ,
\end{equation}
where again terms containing the RMEs
$M_1(^3{\rm S}_1,^3{\rm S}_1)$ and $M_1(^3{\rm D}_1,^3{\rm S}_1)$
have been neglected.  In the expression above, the RME $E_1(^1{\rm P}_1,^3{\rm S}_1)$
has also been neglected, since transitions induced by the isoscalar electric dipole
operator are strongly suppressed~\cite{Viviani00,Nollett01}.  Thus, the only relevant
transitions are those connecting the $^3$P$_1$ PV $n$$p$ state to the PC deuteron
component and the $^3$S$_1$ and $^3$D$_1$ PC $n$$p$ states to the $^3$P$_1$ PV
deuteron component.

\subsection{Helicity-dependent asymmetry in $d(\vec \gamma,n)p$ photo-disintegration}
\label{sec:gdnp}

The relevant matrix element in the photo-disintegration of a deuteron initially at
rest in the laboratory is

\begin{equation}
j^{(-)}_{m_n m_p,\lambda m_d}({\bf p},{\bf q}) =
^{(-)}\!\! \langle {\bf q};{\bf p},m_n m_p \mid \hat{\bbox \epsilon}_\lambda({\bf q}) \cdot
{\bf j}({\bf q})\mid m_d\rangle 
\label{eq:jmin}
\end{equation}
in the notation of Sec.~\ref{sec:npr} above.  Here $\mid {\bf q};{\bf p},m_n m_p\rangle^{(-)}$
represents an $n$$p$ scattering state with total momentum ${\bf q}$ and relative
momentum ${\bf p}$, satisfying incoming wave boundary conditions.  Its (internal)
wave function has the same partial wave expansion given in Eq.~(\ref{eq:psipw}),
except for the replacement $w^J_{\alpha^\prime,\alpha}(r)
\rightarrow [w^J_{\alpha^\prime,\alpha}(r)]^*$.

The cross section for absorption of a photon of helicity $\lambda$, summed over the
final states and averaged over the initial spin projections of the deuteron, reads:

\begin{eqnarray}
\sigma^\gamma_\lambda &=& \overline{\sum_i} \sum_f  2\pi\, \delta\left( q+E_d-E_f \right)
\frac{2\pi\, \alpha}{q} \mid j^{(-)}_{m_n m_p,\lambda m_d}({\bf p},{\bf q}) \mid^2 \nonumber \\
&=& \frac{8\pi^2\, \alpha}{3} \mu\, p\, \sum_{J\alpha}\sum_{l\geq 1} \epsilon_\alpha
\mid \lambda \, M_l^{(-)}(J,\alpha) + E_l^{(-)}(J,\alpha) \mid^2  \ ,
\end{eqnarray}
where $E_f$ is the energy of the final state,

\begin{equation}
E_f=\frac{q^2}{2\, (m_n+m_p)}+\frac{p^2}{2\, \mu} \ ,
\end{equation}
and

\begin{equation}
\overline{\sum_i} \sum_f \equiv \frac{1}{3} \sum_{m_d}\sum_{m_n m_p}
\int \frac{{\rm d}{\bf p}}{(2\pi)^3} \frac{1}{2} \ .
\end{equation}
Note that the factor 1/2 above is introduced to avoid double-counting the final states, and
that the dependence upon the boundary condition of the continuum
wave functions, the superscript $(-)$, has been reinserted in the
RMEs of the electric and magnetic multipoles, namely

\begin{equation}
X^{(-)}_l(J,\alpha) \equiv\, ^{(-)}\!\langle J,\alpha \mid\mid X_l\mid\mid d,J\!=\!1\rangle
\end{equation}
and $X_l$=$E_l$ or $M_l$.

The resulting PV asymmetry, defined as
$P^\gamma=(\sigma^\gamma_+ - \sigma^\gamma_-)/(\sigma^\gamma_+ + \sigma^\gamma_-)$,
is expressed as

\begin{equation}
P^\gamma = \frac{\sum_{J\alpha} \sum_{l \geq 1} \epsilon_\alpha \, 
\left[\, M_l^{(-)}(J,\alpha) \, E^{(-)\, *}_l(J,\alpha) +{\rm c.c.}  \, \right]}
{\sum_{J\alpha} \sum_{l \geq 1} \epsilon_\alpha \,
\left[\,  \mid\! M_l^{(-)}(J,\alpha)\! \mid^2 +\mid \! E_l^{(-)}(J,\alpha) \! \mid^2 \, \right]}
\label{eq:pg}
\end{equation}
and therefore vanishes unless (i) the initial and/or final states
do not have definite parity (as is the case here because of the presence
of PV $N$$N$ interactions) and/or (ii) the electric and magnetic
multipole operators have unnatural parities $(-)^{l+1}$ and
$(-)^l$, respectively, because of two-body PV electromagnetic currents
associated with PV $N$$N$ interactions~\cite{Schiavilla03a}.

It is easily shown that in the inverse process $p(n,\vec{\gamma})d$ the
expression for the photon circular polarization parameter is identical
to that given above, but for the RMEs $E_l^{(-)}$ and $M_l^{(-)}$ 
being replaced by the corresponding $E_l^{(+)}$ and $M_l^{(+)}$, defined
in the previous section.  Indeed, by making use of the transformation
properties of the states and electric and magnetic multipole
operators under time reversal ${\cal T}$, 

\begin{eqnarray}
{\cal T} \mid d,m_d\rangle &=& (-)^{m_d-1}  \mid d,-m_d\rangle \ , \\
{\cal T} \mid J,M_J;\alpha\rangle^{(+)} &=&(-)^{M_J-J} \mid J,-M_J;\alpha\rangle^{(-)} \ , \\
{\cal T} X_{l,l_z} {\cal T}^\dagger &=& (-)^{l_z} X_{l,-l_z}  \ ,
\end{eqnarray}
one finds the following relation for the RMEs:

\begin{equation}
E_l^{(+)}(J,\alpha) = (-)^{J+l} E_l^{(-)}(J,\alpha) \ ,
\end{equation}
and similarly for the $M_l$'s.  Hence, the circular polarizations measured in
the direct and inverse processes are the same.

\subsection{Longitudinal asymmetry in $d(\vec{e},e^\prime)np$ electro-disintegration}
\label{sec:deep}

The longitudinal asymmetry in the inclusive scattering of polarized
electrons off a nuclear target results from the interference of
amplitudes associated with photon and $Z^0$ exchanges as well as
from the presence of parity-violating components in the nucleon-nucleon
interaction.  For completeness, we summarize below the relevant formulae.
The initial and final electron (nucleus)
four-momenta are labeled by $k^\mu$ and $k^{\prime \mu}$ ($P^\mu$
and $P^{\prime \mu}$), respectively, while the four-momentum transfer
$q^\mu$ is defined as $q^\mu \equiv k^\mu-k^{\prime \mu}
\equiv (\omega,{\bf q})$.  The amplitudes for the $\gamma$-
and $Z$-exchange processes are then given by~\cite{Musolf92}

\begin{eqnarray}
M &=& -\frac{4\pi \alpha}{q_\mu^2}(M^\gamma + M^Z) \ , \\
M^\gamma &=& \overline{u}^{\, \prime} \gamma^{\sigma} u\,
            j^\gamma_{\sigma,fi} \ , \\
M^Z &=&\frac{1}{4 \pi\sqrt{2}} \frac{G_\mu q_\mu^2}{\alpha}
 \overline{u}^{\, \prime} \gamma^{\sigma}
( g_V^{(e)}+g_A^{(e)} \gamma_5) u\, j^Z_{\sigma,fi} \ ,
\end{eqnarray}
where $G_\mu$ is the Fermi constant for muon decay,
$g_V^{(e)}=-1+4\, {\rm sin}^2\theta_W$ and $g_A^{(e)}=1$ are
the Standard Model values for the neutral-current couplings
to the electron given in terms of the Weinberg angle $\theta_W$,
$u$ and $u^\prime$ are the initial and final electron spinors,
and $j^{\gamma,\sigma}_{fi}$ and $j^{Z,\sigma}_{fi}$ denote matrix
elements of the electromagnetic and weak neutral currents, i.e.
 
\begin{equation}
j^{\gamma,\sigma}_{fi} \equiv \langle f\vert j^{\gamma,\sigma}(0) \vert
i\rangle
\equiv ( \rho^\gamma_{fi}({\bf q}), {\bf j}^\gamma_{fi}({\bf q})) \ ,
\end{equation}
and similarly for $j^{Z,\sigma}_{fi}$.  Here $\vert i\rangle$ and
$\vert f\rangle$ represent the initial deuteron state and final
$n$$p$ scattering state with incoming-wave boundary conditions (the $(-)$
solution), respectively.  Note that in the amplitude $M^Z$ the $q_\mu^2$
dependence of the $Z^0$ propagator has been ignored, since $|q_\mu^2| \ll m_Z^2$.

The parity-violating asymmetry $A$ is given by the
ratio of the difference over the sum of the inclusive cross sections
$d\sigma_h/d\Omega d\omega$ for incident electrons with helicities $h=\pm 1$.
It depends on the three-momentum and energy transfers, $q$ and $\omega$,
and scattering angle, $\theta_e$, of the electron and conveniently expressed

\begin{equation}
A = A_{\gamma \gamma} + A_{\gamma Z} \ .
\end{equation}
Standard manipulations then lead to the following expression
for the asymmetry in the extreme relativistic limit for the
electron~\cite{Diaconescu01,Musolf92}
 
\begin{eqnarray}
A_{\gamma \gamma} &=& \frac{ v_{T^\prime} R_{T^\prime}^{\gamma,\gamma} }
      { v_L R_L^{\gamma,\gamma} + v_T R_T^{\gamma,\gamma} } \ , 
\label{eq:agg} \\
A_{\gamma Z} &=& \frac{1}{2\sqrt{2}\, \pi}\frac{G_\mu Q^2}{\alpha}
\frac{ g_A^{(e)} v_L R_L^{\gamma,0} +
       g_A^{(e)} v_T R_T^{\gamma,0} +
       g_V^{(e)} v_{T^\prime} R_{T^\prime}^{\gamma,5} }
      { v_L R_L^{\gamma,\gamma} +
        v_T R_T^{\gamma,\gamma} } \ ,
\label{eq:agz}
\end{eqnarray}
where the $v$'s are defined in terms of electron kinematical variables,
 
\begin{eqnarray}
v_L &=& \frac{q_\mu^4}{q^4} \ , \\
v_T &=& {\rm tan}^2 (\theta_e/2) +\frac{|q_\mu^2|}{2\, q^2} \ , \\
v_{T^\prime} &=& {\rm tan}(\theta_e/2)
 \sqrt{ {\rm tan}^2(\theta_e/2) +\frac{|q_\mu^2|}{q^2} } \ .
\end{eqnarray}
The $R$'s are the nuclear electro-weak response functions, which
depend on $q$ and $\omega$, to be defined below.  To this end,
it is first convenient to separate the weak current
$j^{Z,\sigma}$ into its vector $j^{0,\sigma}$ and axial-vector $j^{5,\sigma}$
components, and to write correspondingly

\begin{equation}
j^{Z,\sigma}_{fi}=j^{0,\sigma}_{fi} + j^{5,\sigma}_{fi}
                 \equiv ( \rho^0_{fi}({\bf q}) , {\bf j}^0_{fi}({\bf q}) )
                       +( \rho^5_{fi}({\bf q}) , {\bf j}^5_{fi}({\bf q}) ) \ .
\end{equation}
The response functions can then be expressed as
 
\begin{eqnarray}
R_L^{\gamma,{\rm a}}(q,\omega) &=& \overline{\sum_i} \sum_f \delta(\omega+E_d-E_f)
\, {\rm Re} \left [ \rho^\gamma_{fi}({\bf q})
               \rho^{{\rm a} *}_{fi}({\bf q}) \right ] \ ,\label{eq:rl} \\
R_T^{\gamma,{\rm a}}(q,\omega) &=& \overline{\sum_i} \sum_f \delta(\omega+E_d-E_f)
\, {\rm Re} \left [ {\bf j}^\gamma_{T,fi}({\bf q}) \cdot
                    {\bf j}^{{\rm a} *}_{T,fi}({\bf q}) \right] \ ,\label{eq:rt} \\
R_{T^\prime}^{\gamma,{\rm b}}(q,\omega)&=&\overline{\sum_i} \sum_f \delta(\omega+E_d-E_f)
\, {\rm Im}  \left[ {\bf j}^\gamma_{fi}({\bf q})
               \times {\bf j}^{{\rm b} *}_{fi}({\bf q}) \right]_z \ , \label{eq:rtpp} 
\end{eqnarray}
where $E_d$ is the ground-state energy of the deuteron (assumed at rest in the laboratory),
$E_f$ is the energy of the final scattering state, and in Eqs.~(\ref{eq:rl}) and (\ref{eq:rt})
[Eq.~(\ref{eq:rtpp})] the superscript ${\rm a}$ (${\rm b}$) is either $\gamma$ or
$0$ ($\gamma$ or 5).  Note that there is a sum over the final states and
an average over the initial spin projections of the deuteron.  In the expressions
above for the $R$'s, it has been assumed that the three-momentum transfer ${\bf q}$ is along
the $z$-axis, which defines the spin quantization axis for the nuclear states.

The asymmetry induced by hadronic weak interactions, $A_{\gamma \gamma}$, is
easily seen to be proportional to the interference of electric and magnetic
multipole contributions as in Eq.~(\ref{eq:pg}), indeed

\begin{equation}
\frac{(2 \pi)^2 \alpha}{q} \, R_{T^\prime}^{\gamma,\gamma}
= \sigma^\gamma_+ -\sigma^\gamma_- \ ,
\end{equation}
namely $R_{T^\prime}^{\gamma,\gamma}$ is related, of course for $\omega$=$q$,
to the difference of helicity-dependent photo-disintegration cross
sections.  Similar considerations to those in Sec.~\ref{sec:gdnp} allow one to
conclude that this response would vanish in the absence of PV $N$$N$ interactions
(note, however, that in the present case there is, in addition to two-body PV
currents, also a one-body PV term originating from radiative electro-weak corrections,
the anapole current~\cite{Haxton89}). 

\section{Calculation}
\label{sec:cal}

In this section we briefly review the techniques used to calculate
the PV observables in the $n$$p$ system---these are similar to those
discussed most recently in Ref.~\cite{Diaconescu01}.

The deuteron wave function in Eq.~(\ref{eq:dw}) is written, for each
spatial configuration ${\bf r}$, as a vector in the spin-isospin
space of the two nucleons,

\begin{equation}
\psi_{d,m_d}({\bf r})=\sum_{n=1}^8 \psi_{d,m_d}^{(n)}({\bf r})\, \mid\! n\rangle \ ,
\end{equation}
where $\mid\! n\rangle = (p\!\uparrow)_1\, (n\!\uparrow)_2, (n\!\uparrow)_1\, (p\!\uparrow)_2, \dots,
(n\!\downarrow)_1\, (p\!\downarrow)_2$ and $\psi_{d,m_d}^{(n)}$ are the components
of $\psi_{d,m_d}$ in this basis.  The scattering wave function in Eq.~(\ref{eq:psipw})
is first approximated by retaining PC and PV interaction effects in all channels up
to a certain pre-selected $J_{\rm max}$ and by using spherical Bessel function for channels
with $J > J_{\rm max}$, and is then expanded, for any given ${\bf r}$, in the same basis
$\{\mid \! n\rangle\}$ defined above.  The radial functions $w_{\alpha^\prime,\alpha}$ are obtained
with the methods discussed in Sec.~\ref{sec:p-space}.

Matrix elements of the electromagnetic (and neutral-weak) current operators are
written schematically as

\begin{equation}
\langle\psi_f\mid O \mid \psi_i\rangle = \int{\rm d}{\bf r}\sum_{m,n}
\psi_f^{(m)\,*}({\bf r}) O_{m,n}({\bf r}) \psi_i^{(n)}({\bf r}) \ .
\end{equation}
The spin-isospin algebra is performed exactly with techniques similar to
those developed in Ref.~\cite{Schiavilla89}, while the ${\bf r}$-space
integrations are carried out efficiently by Gaussian quadratures.  Note 
that no multipole expansion of the transition operators is required.

Extensive and independent tests of the computer programs have been
completed successfully.

\section{Results and Discussion}
\label{sec:res}

In this section we present results for the longitudinal asymmetry and spin rotation
in $\vec n$$p$ elastic scattering, the photon asymmetry in the $\vec n$$p$ radiative
capture at thermal energies, and the asymmetries in the threshold photo-disintegration
and quasi-elastic electro-disintegration of the deuteron.  To provide an estimate
for the model dependence of these results, we consider several different high-quality
interactions fit to strong-interaction data, the Argonne $v_{18}$ (AV18)~\cite{Wiringa95}, Bonn
(BONN)~\cite{Machleidt01} and Nijmegen-I (NIJM-I)~\cite{Stoks94} interactions.  We adopt
the standard DDH~\cite{Desplanques80} one-boson exchange model of the parity-violating
(PV) interaction, and solve the Schr\"odinger equation for the scattering
state and deuteron bound state with the methods discussed in Sec.~\ref{sec:app}.
The values for the meson-nucleon coupling constants
and cutoff parameters in the DDH model are those
listed in Table~\ref{tb:ddh}.  Note that we have taken the linear combination of $\rho$-
and $\omega$-meson weak coupling constants corresponding to $\vec p$$p$ elastic scattering from
an earlier analysis~\cite{Carlson02} of these experiments.  The remaining couplings are 
the DDH ``best value'' estimates.  As in the earlier analysis of $\vec{p}p$
scattering, the cutoff values in the meson-exchange interaction are taken from the BONN
potential.  

It is also useful to introduce here some of
the notation adopted in the following subsections to denote variations
on the DDH model defined above.  The PV interaction denoted as DDHb
corresponds to a DDH model with $\pi$, $\rho$ and $\omega$ weak coupling
constants as specified by the ``best value'' set of Ref.~\cite{Desplanques80},
while the PV interaction denoted as DDH$\pi$ includes only the $\pi$-exchange
term in the DDH model with the ``best value'' for the weak $\pi$$N$$N$
coupling constant.  The remaining pion and vector-meson strong interaction
coupling constants and short-range cutoff parameters are as given in
Table~\ref{tb:ddhb} for the DDHb model and are the same as in Table~\ref{tb:ddh}
for the DDH$\pi$ model.

Finally, while the short-range contributions to the PV interaction should not be viewed as
resulting solely from the exchange of single mesons, the six parameters of the DDH model
are still useful in characterizing all the low-energy PV mixings.  For example, two-pion
exchange could play a role~\cite{Pirner73}, however we assume that its effects can be included,
at least at low energy, through the present combination of pion- and short-range terms.

\subsection{Longitudinal asymmetry and neutron spin rotation}
\label{sec:r_a}

In this section we present results for the longitudinal asymmetry and
neutron spin rotation in $\vec n$$p$ elastic scattering.

\begin{figure}[bth]
\let\picnaturalsize=N
\def\picsize{5in}
\def\picfilenamea{figs/epsj0.eps}
\epsfbox{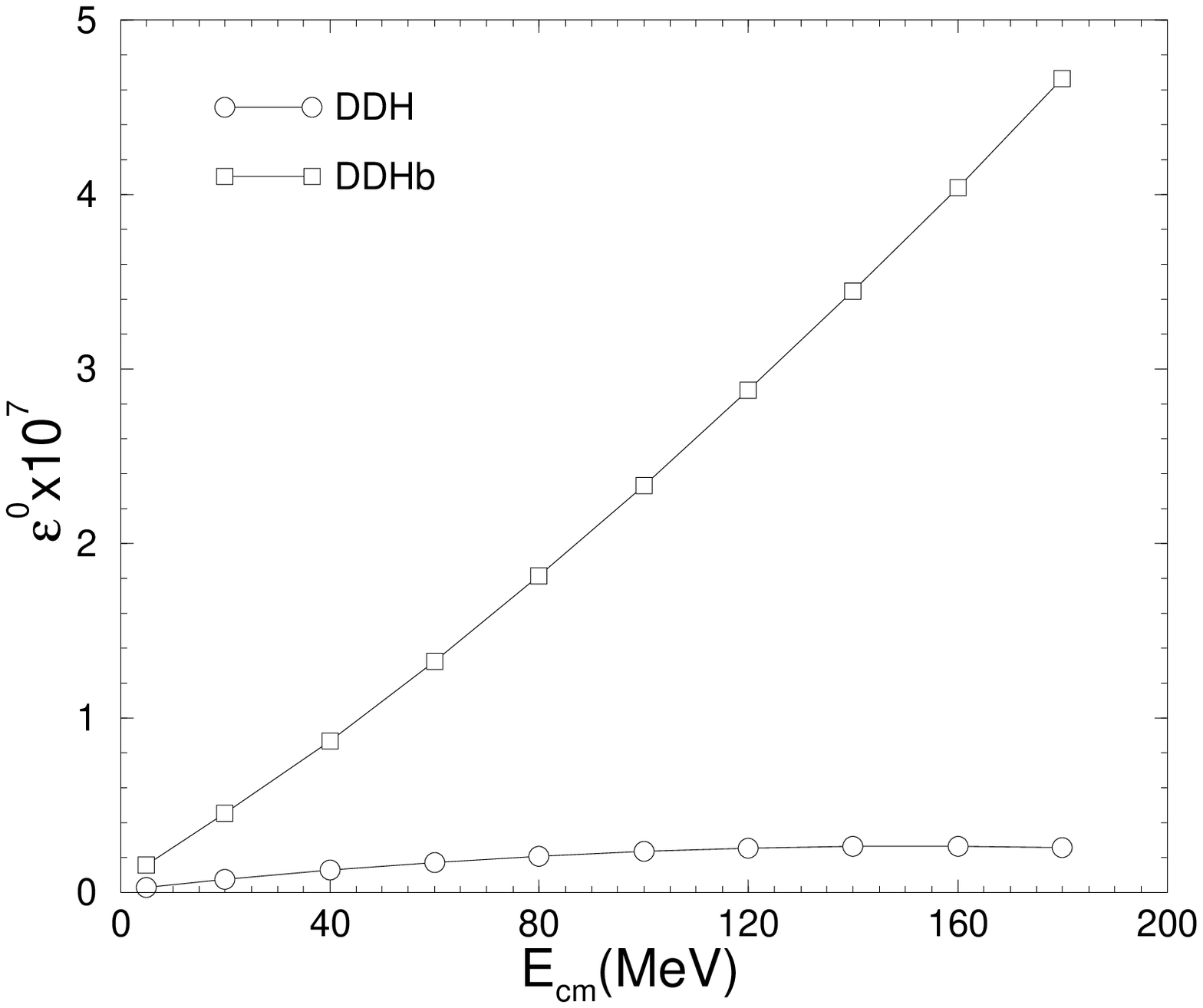}
\caption{The $^1$S$_0$-$^3$P$_0$ mixing parameter obtained with the AV18 model
in combination with either of two variations of the DDH model, labeled DDH and
DDH ``best values'', see text.}
\label{fig:e0}
\end{figure}

\begin{figure}[bth]
\let\picnaturalsize=N
\def\picsize{5in}
\def\picfilenamea{figs/epsj1b.eps}
\epsfbox{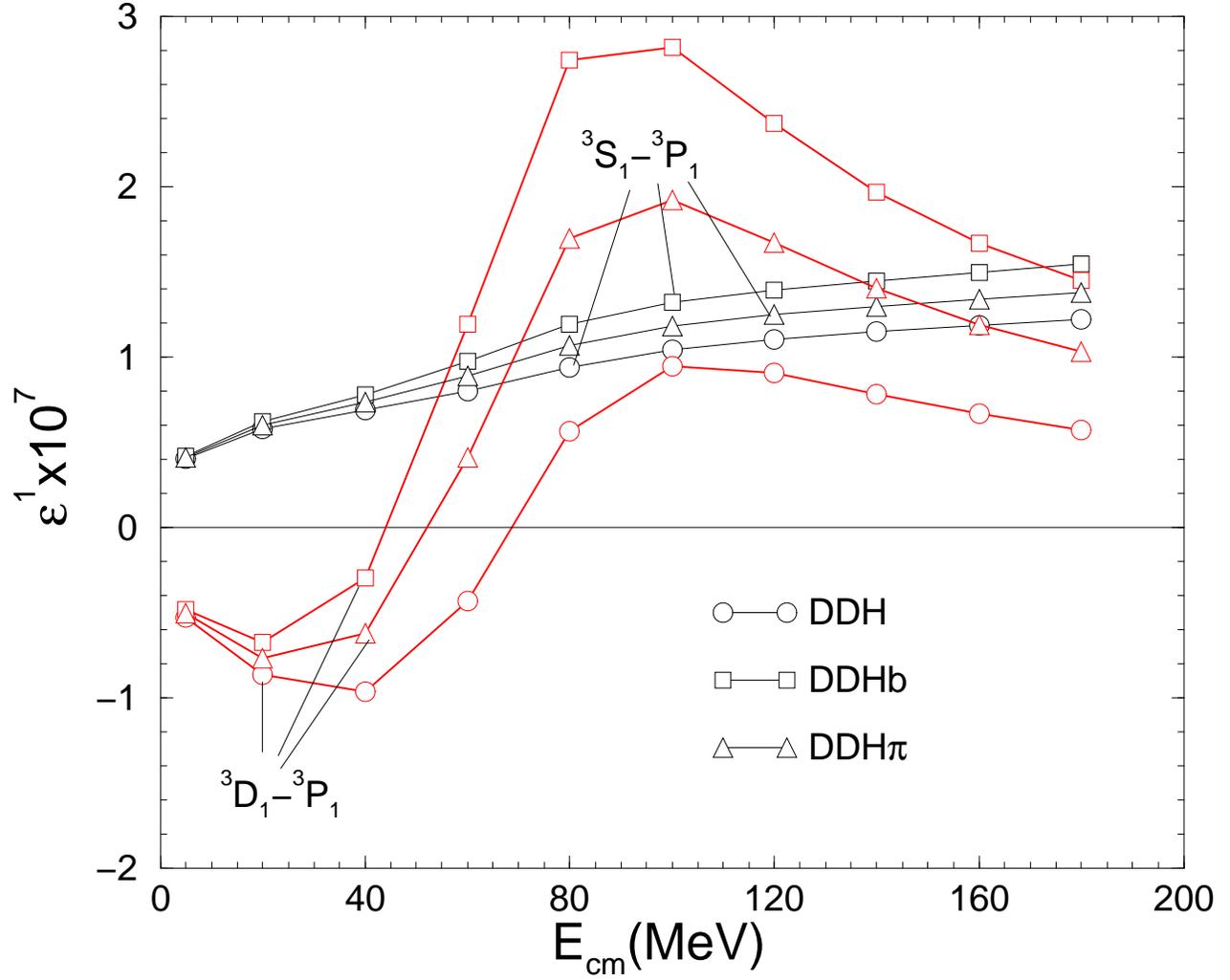}
\caption{The $^3$S$_1$-$^3$P$_1$ and $^3$D$_1$-$^3$P$_1$ mixing parameters
obtained with the AV18 model in combination with either of three variations
of the DDH model, labeled DDH, DDH ``best values'', and DDH $\pi$-only,
see text.}
\label{fig:e1b}
\end{figure}

\begin{figure}[bth]
\let\picnaturalsize=N
\def\picsize{5in}
\def\picfilenamea{figs/epsj1a.eps}
\epsfbox{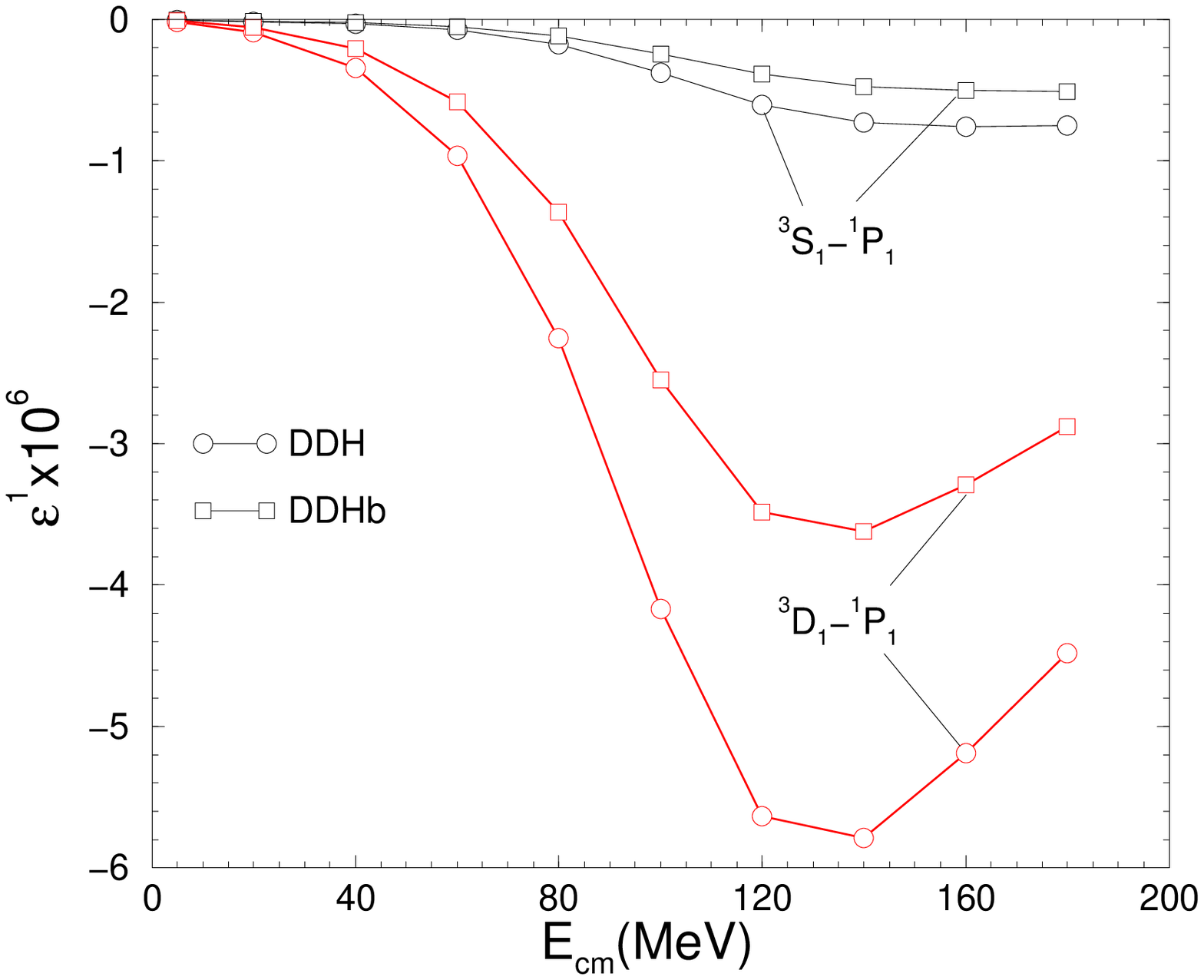}
\caption{The $^3$S$_1$-$^1$P$_1$ and $^3$D$_1$-$^1$P$_1$
mixing parameters obtained with the AV18 model
in combination with either of two variations of the DDH model, labeled DDH and
DDH ``best values'', see text.}
\label{fig:e1a}
\end{figure}

\begin{figure}[bth]
\let\picnaturalsize=N
\def\picsize{5in}
\def\picfilenamea{figs/epsj2b.eps}
\epsfbox{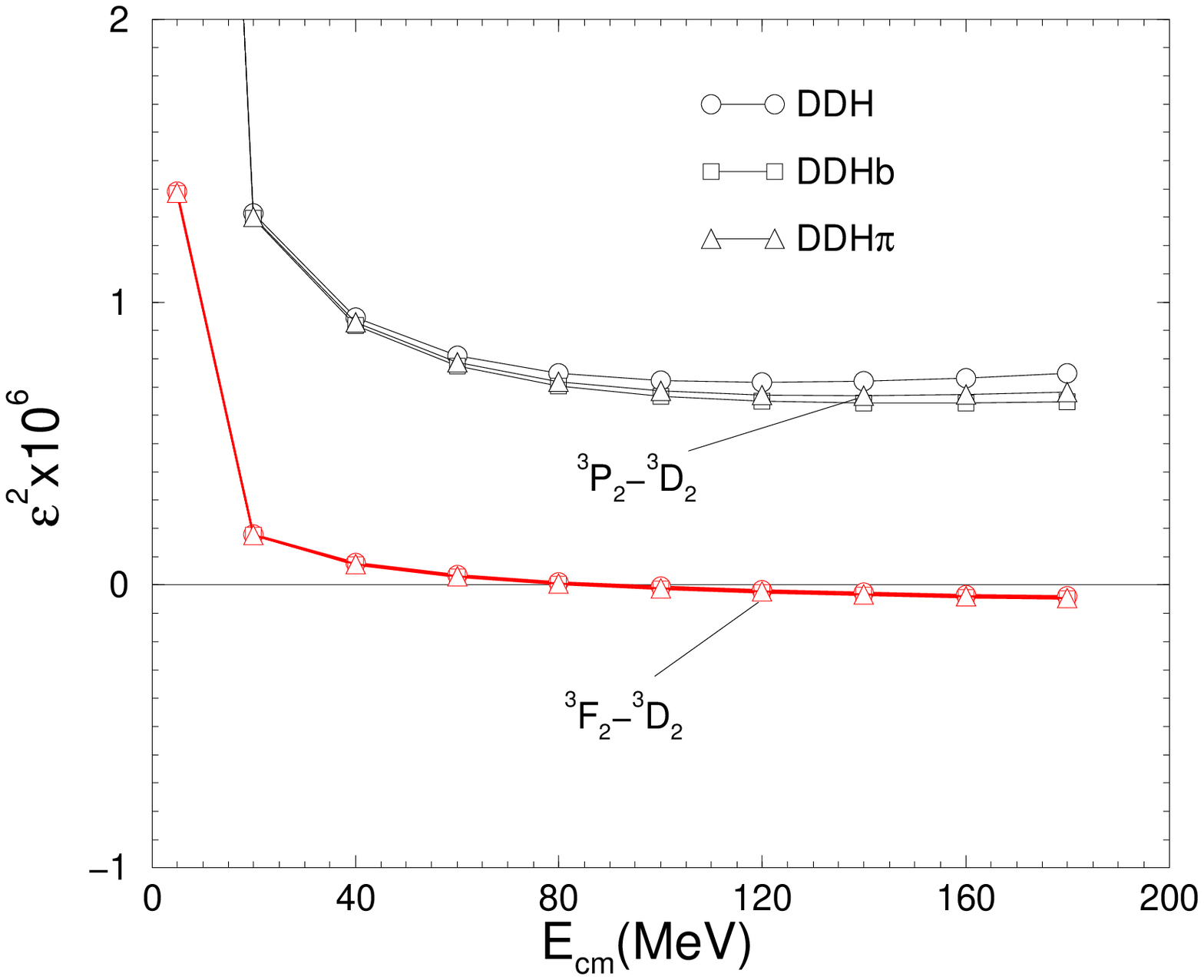}
\caption{Same as in Fig.~\protect\ref{fig:e1b}, but for the mixing parameters
$^3$P$_2$-$^3$D$_2$ and $^3$F$_2$-$^3$D$_2$.}
\label{fig:e2b}
\end{figure}

\begin{figure}[bth]
\let\picnaturalsize=N
\def\picsize{5in}
\def\picfilenamea{figs/epsj2a.eps}
\epsfbox{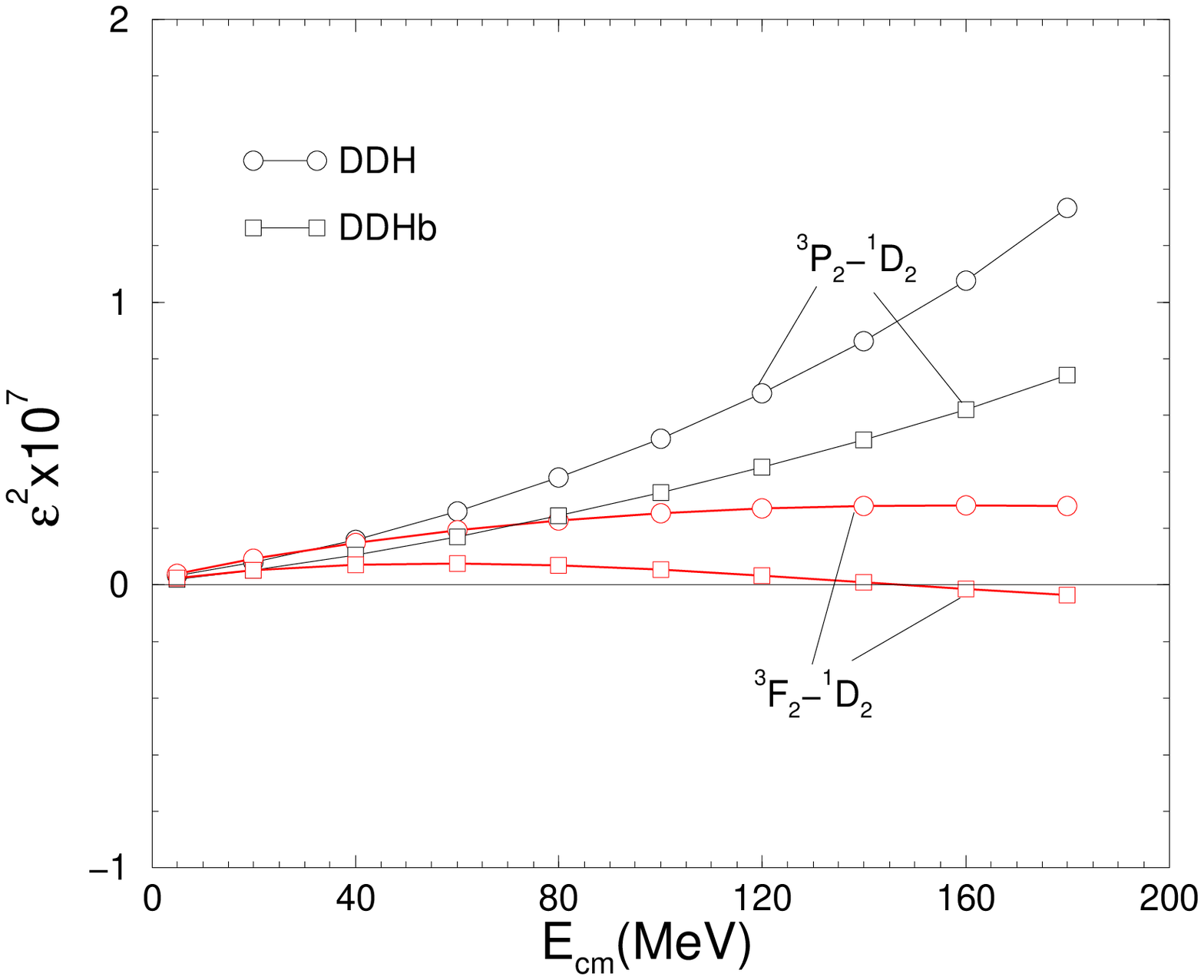}
\caption{Same as in Fig.~\protect\ref{fig:e1a}, but for the mixing parameters
$^3$P$_2$-$^1$D$_2$ and $^3$F$_2$-$^1$D$_2$.}
\label{fig:e2a}
\end{figure}

Figures~\ref{fig:e0}--\ref{fig:e2a} show the mixing parameters $\epsilon^J_{lm}$
induced by AV18 model in combination with the DDH, DDHb, and
DDH$\pi$ interactions.  Only those $\epsilon^J_{lm}$ induced by the PV
interaction are displayed, namely for $J$=0 $\epsilon^0_{12}$ and for $J \geq 1$
$\epsilon^J_{lm}$ with $l$=1,2 and $m$=3,4 in the notation of Tables~\ref{tb:chan}
and~\ref{tb:mixing}.  

The definitions adopted for the phase-shifts and mixing
parameters are those introduced in Sec.~\ref{sec:phase}.  Up to linear terms
in $v^{\rm PV}$, the $\delta_\alpha^J$ and $\epsilon^{J\geq 1}_{12}$ values are
not affected by weak interactions, and are determined solely by the strong
interaction.  They are identical to those listed in Ref.~\cite{Wiringa95},
but for two differences.  Firstly, the Blatt-Biedenharn parameterization
is used here for the $S$-matrix~\cite{Blatt52} rather than the bar-phase
parameterization of it~\cite{Stapp57} employed in Ref.~\cite{Wiringa95}.
Secondly, because of the phase choice in the potential components
(see Eq.~(\ref{eq:vpe}) and comment below it), the mixing parameters
$\epsilon^{J\geq 1}_{12}$ have opposite sign relative to those listed in
Ref.~\cite{Wiringa95}.

The coupling between channels with the same pair isospin $T$ is
induced by the short-range part of the DDH interaction, associated
with vector-meson exchanges; its long-range component, due to pion
exchange, vanishes in this case.  As a result, the mixing parameters
in Figs.~\ref{fig:e0},~\ref{fig:e1a}, and~\ref{fig:e2a}, calculated
with the DDH and DDHb models, are rather different,
reflecting the large differences in the values for the some of the
strong and weak coupling constants and short-range cutoffs between these
two models, see Tables~\ref{tb:ddh} and~\ref{tb:ddhb}.

The mixing parameters between channels with $|\Delta T|$=1,
Figs.~\ref{fig:e1b} and~\ref{fig:e2b}, in which the pion-exchange
term is present, are still rather sensitive to the short-range
behavior of the PV interaction, as reflected again by the differences
in the DDH and DDHb predictions.  However, this sensitivity is
much reduced for the more peripheral waves, such as the $^3$P$_2$-$^3$D$_2$
and $^3$F$_2$-$^3$D$_2$ channels.

\begin{figure}[bth]
\let\picnaturalsize=N
\def\picsize{5in}
\def\picfilenamea{figs/epsj0_ab.eps}
\epsfbox{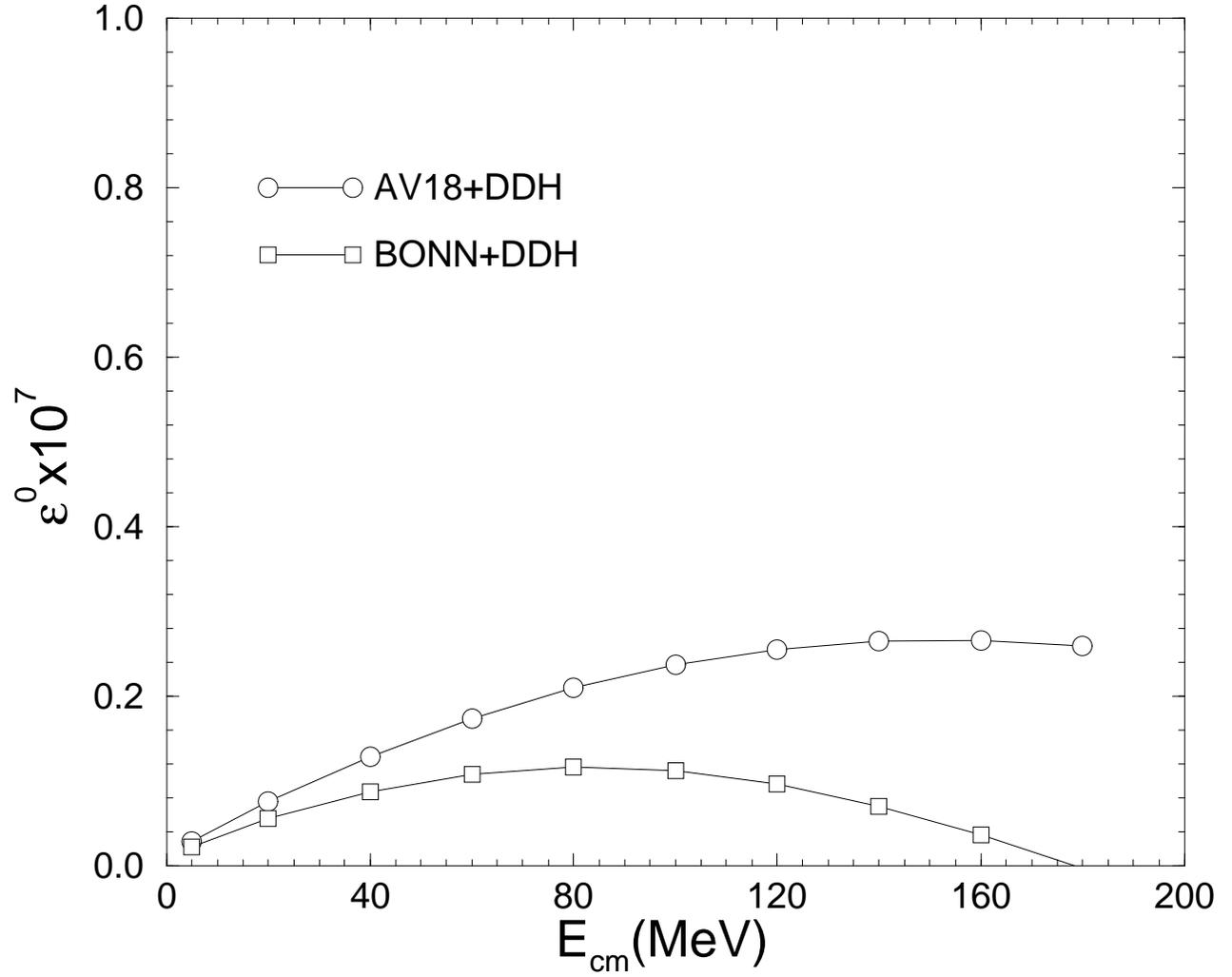}
\caption{The $^1$S$_0$-$^3$P$_0$ mixing parameter obtained with the DDH model 
in combination with either the AV18 or BONN model.}
\label{fig:e0ab}
\end{figure}

\begin{figure}[bth]
\let\picnaturalsize=N
\def\picsize{5in}
\def\picfilenamea{figs/epsj1b_ab.eps}
\epsfbox{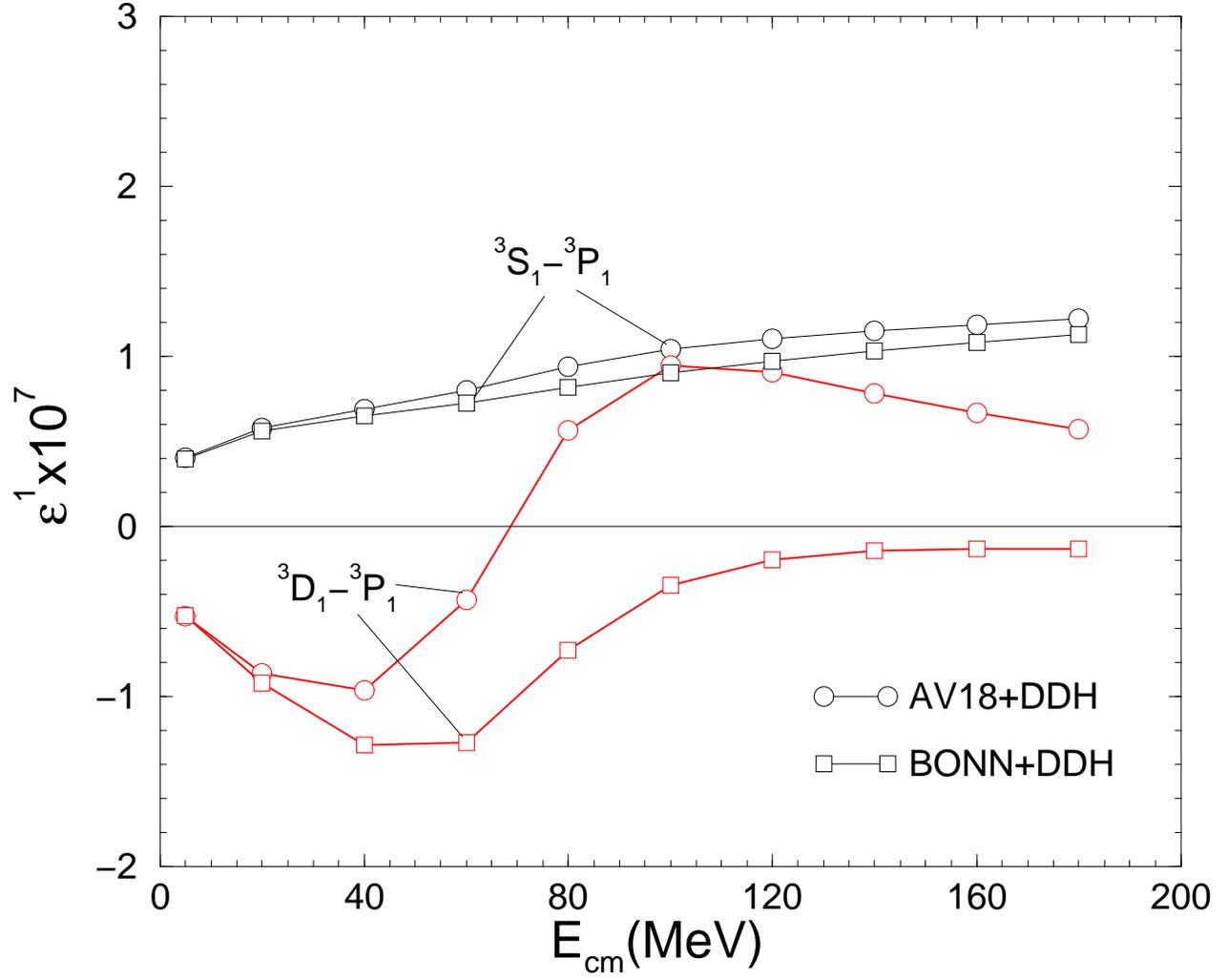}
\caption{The $^3$S$_1$-$^3$P$_1$ and $^3$D$_1$-$^3$P$_1$ mixing parameters
obtained with the DDH model in combination with either of the AV18 or BONN model.}
\label{fig:e1bab}
\end{figure}

\begin{figure}[bth]
\let\picnaturalsize=N
\def\picsize{5in}
\def\picfilenamea{figs/epsj1a_ab.eps}
\epsfbox{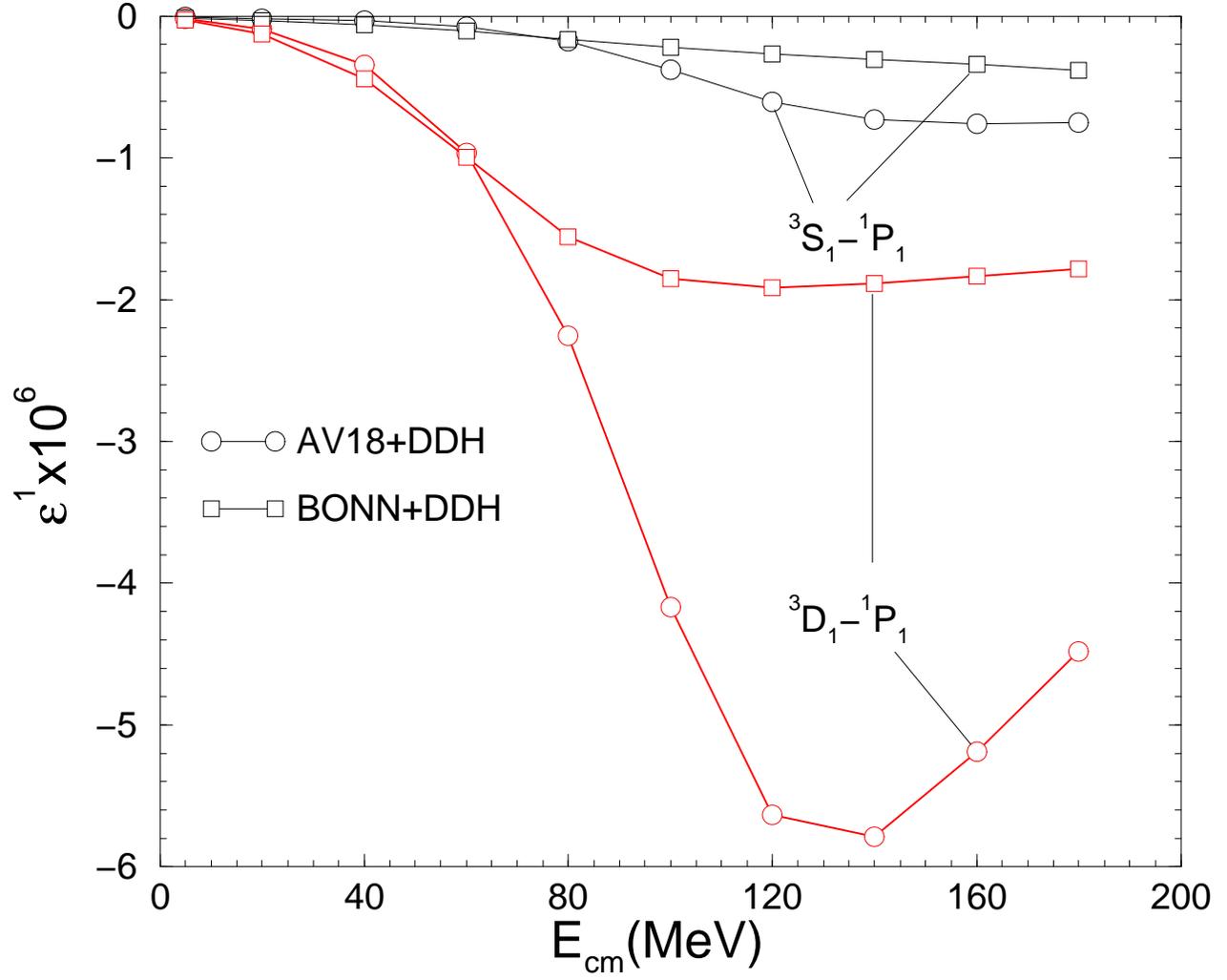}
\caption{Same as in Fig.~\protect\ref{fig:e1bab} but for the $^3$S$_1$-$^1$P$_1$
and $^3$D$_1$-$^1$P$_1$ mixing parameters.}
\label{fig:e1aab}
\end{figure}

Figures~\ref{fig:e0ab}--\ref{fig:e1aab} are meant to illustrate the
sentitivity of the mixing angles to the input strong-interaction
potential, which can be quite large, particularly in channels,
such as the $^3$D$_1$-$^1$P$_1$.

\begin{figure}[bth]
\let\picnaturalsize=N
\def\picsize{5in}
\def\picfilenamea{figs/asy_np.eps}
\epsfbox{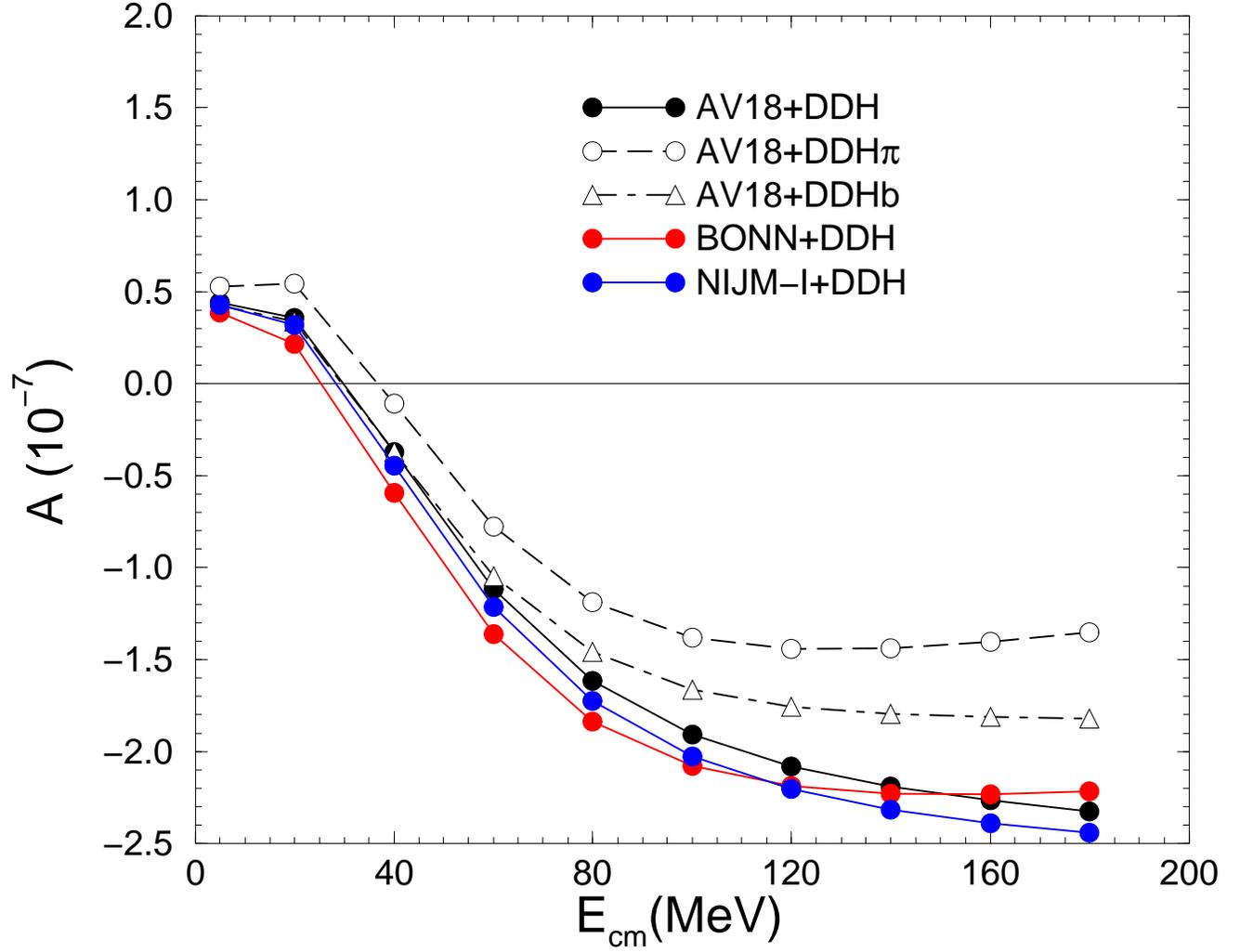}
\caption{The neutron asymmetry obtained with various combinations
of strong- and weak interaction potentials, as function
of the center-of-mass energy.}
\label{fig:anp}
\end{figure}

The total longitudinal asymmetry, defined in Eq.~(\ref{eq:asye}), is shown
in Fig.~\ref{fig:anp} for a number of combinations of strong- and
weak-interaction potentials.  The asymmetries were calculated
by retaining in the partial wave expansion for the amplitude,
Eq.~(\ref{eq:am}), all channels with $J$ up to $J_{\rm max}$=6.
There is very little sensitivity to the input strong-interaction
potential.  As also remarked in Ref.~\cite{Carlson02}, this reduced
sensitivity is undoutbly a consequence of the fact that present
potentials are fitted to extended $p$$p$ and $p$$n$ databases with
high accuracy.

\begin{figure}[bth]
\let\picnaturalsize=N
\def\picsize{5in}
\def\picfilenamea{figs/asy_np_j.eps}
\epsfbox{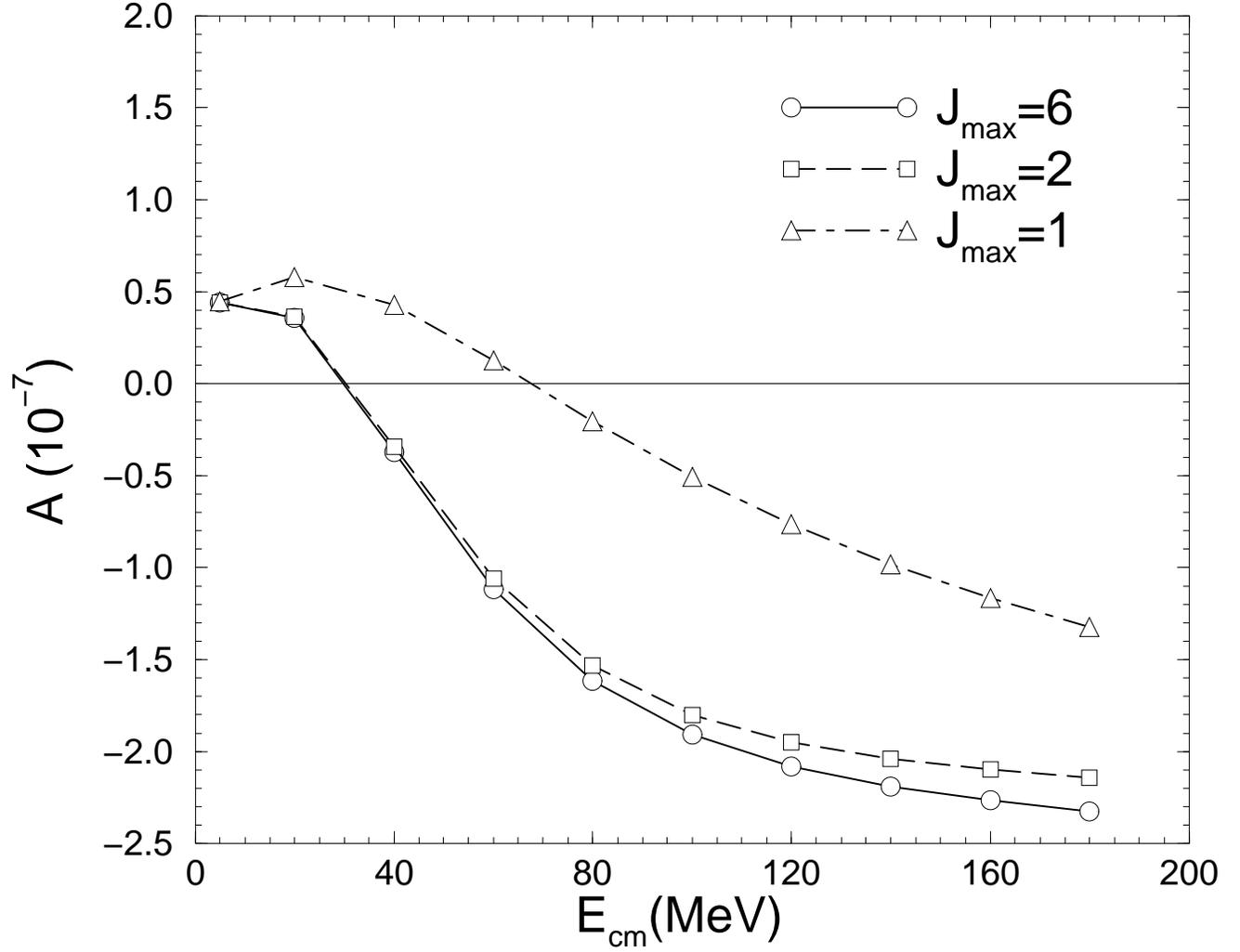}
\caption{Contributions to the neutron asymmetry obtained by including only
the $J$=0 and $J$=1 channels, and by adding the $J$=2 channels, and finally
all $J$ channels up to $J_{\rm max}$=6.  The AV18+DDH potential combination
is used, black solid line in Fig.~\protect\ref{fig:anp}.}
\label{fig:anpj}
\end{figure}

Figure~\ref{fig:anpj} shows that the total asymmetries obtained by
including only the $J$=0 and 1 channels ($^1$S$_0$-$^3$P$_0$,
$^3$S$_1$-$^3$P$_1$, $^3$D$_1$-$^3$P$_1$, $^3$S$_1$-$^1$P$_1$,
and $^3$D$_1$-$^1$P$_1$) and, in addition, the $J$=2 channels,
and finally all $J$ channels up to $J_{\rm max}$=6.  In the energy
range (0--200) MeV the asymmetry is dominated by the $J$=0--2
contributions.

\begin{figure}[bth]
\let\picnaturalsize=N
\def\picsize{5in}
\def\picfilenamea{figs/sgm_np.eps}
\epsfbox{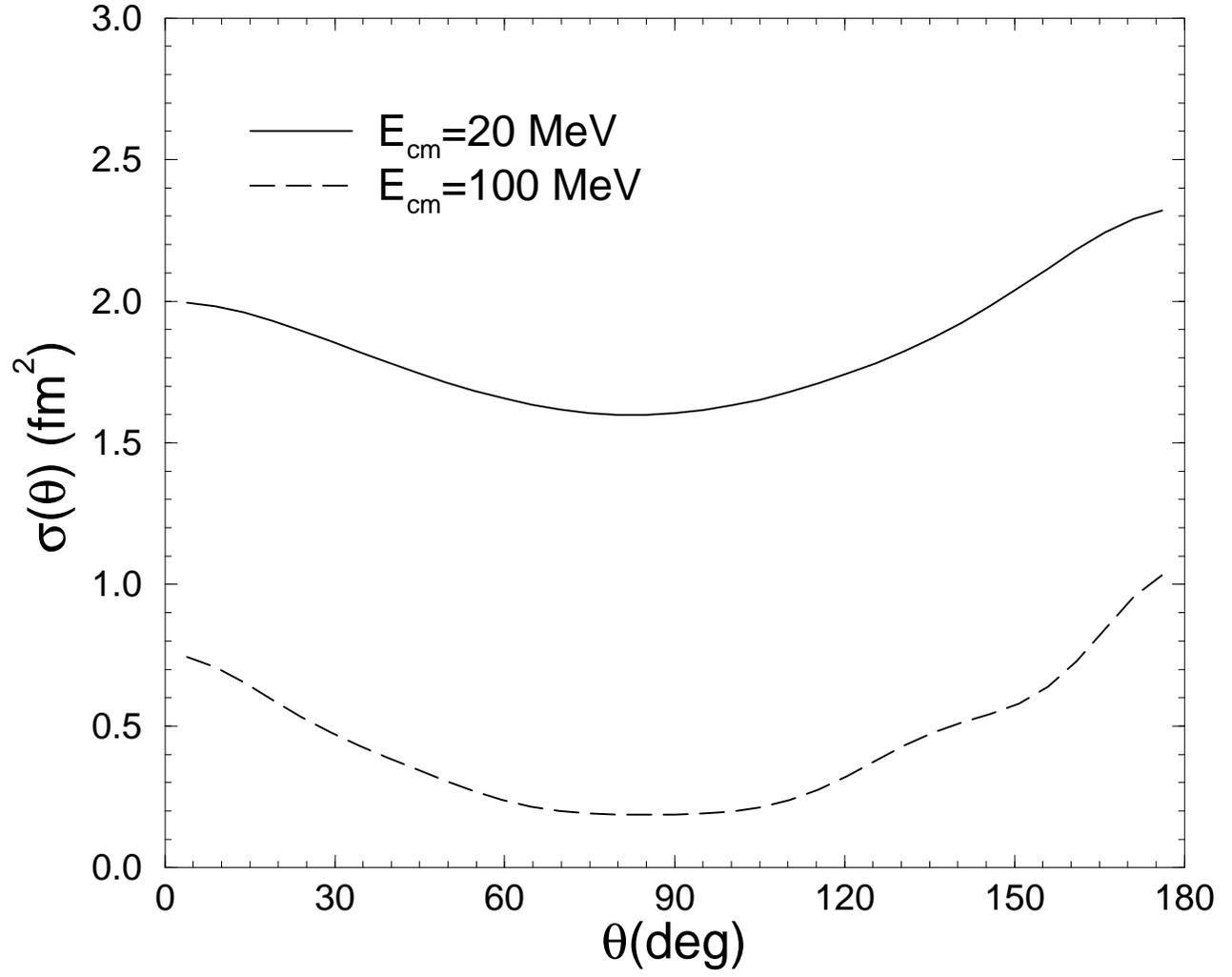}
\caption{Angular distributions for the spin-averaged $n$$p$ (strong-interaction)
cross section at center-of-mass energies of 20 MeV and 100 MeV, corresponding
to the AV18 potential.}
\label{fig:snpa}
\end{figure}

\begin{figure}[bth]
\let\picnaturalsize=N
\def\picsize{5in}
\def\picfilenamea{figs/asy_np_ang.eps}
\epsfbox{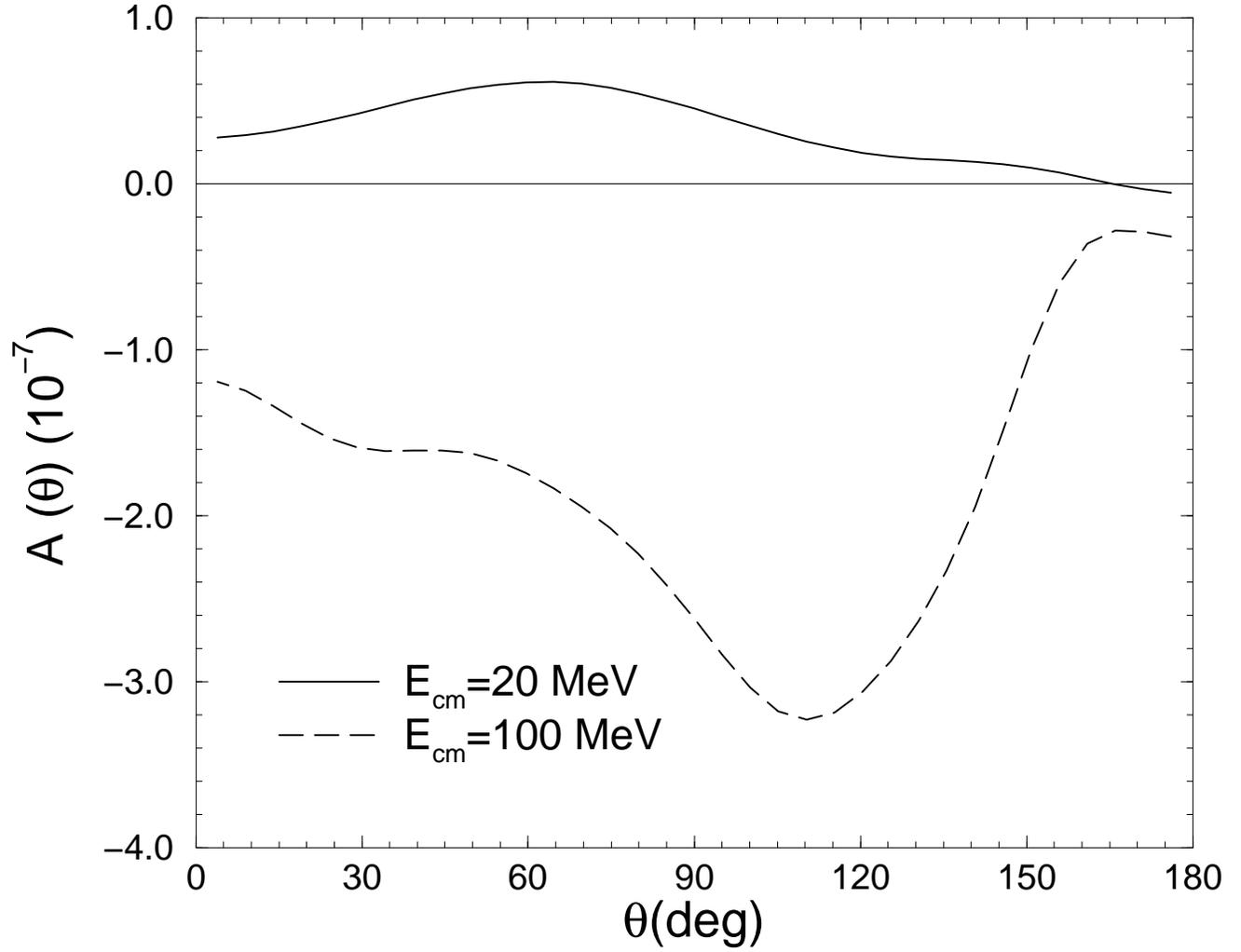}
\caption{Angular distributions for the neutron asymmetry at center-of-mass
energies of 20 MeV and 100 MeV.  The AV18+DDH potential combination
is used, black solid line in Fig.~\protect\ref{fig:anp}.}
\label{fig:anpa}
\end{figure}

For completeness, we present in Figs.~\ref{fig:snpa} and~\ref{fig:anpa}
results for the angular distributions of the (PC) spin-averaged differential
cross section and (PV) longitudinal asymmetry at center-of-mass energies
of 20 MeV and 100 MeV.  The asymmetry $A(E,\theta)$ is defined
in Eq.~(\ref{eq:athe}).

The predictions for the neutron spin rotation per unit length, ${\rm d}\phi/{\rm d}d$
with $\phi$ defined in Eq.~(\ref{eq:nphi}), are listed in Table~\ref{tb:nspin} in
the limit of vanishing incident neutron energy.  The density of liquid hydrogen
is taken as $\rho$=$0.4\times 10^{23}$ atoms-cm$^{-3}$.

While results corresponding to different input strong interactions are within
$\leq$ 10\% of each other, the calculated values show significant sensitivity
to the short-range bahavior of the PV interaction, columns labeled DDH and DDHb.
It is worth re-emphasizing that the longitudinal asymmetry in $\vec p$$p$
elastic scattering predicted by the DDHb model is at variance with that
observed experimentally~\cite{Carlson02}.  The short-range cutoff parameters
and combinations of $\rho$- and $\omega$-meson PV coupling constants in $TT_z$=11,
respectively $h_\rho^0+h_\rho^1+h^2_\rho/\sqrt{6}$ and $h^0_\omega+h^1_\omega$,
were constrained, in the DDH model, to reproduce this (measured) asymmetry~\cite{Carlson02}.
An additional difference between the DDH and DDHb models is in the values adopted for
the (PC) $\rho$-meson tensor coupling to the nucleon, 6.1 in the DDH (from the BONN
interaction) and 3.7 in the DDHb (consistent with estimates from vector-meson dominance).
Hence, the DDHb results are not realistic.  Comparison between the DDH and DDH$\pi$
predictions, however, indicates that the neutron spin
rotation is sensitive to the long-range part of $v^{\rm PV}$, and therefore
a measurement of this observable would be useful in constraining the PV $\pi$$N$$N$
coupling constant.

Finally, there is a sign difference between the present results and those reported
in Ref.~\cite{Avishai84}.  It is not due to the different strong interaction
potential used in that calculation.  Indeed, with the Paris potential~\cite{Lacombe80}
in combination with the DDHb model we obtain ${\rm d}\phi/{\rm d}d$=$+8.88\times 10^{-9}$
rad-cm$^{-1}$, the same magnitude but opposite sign than given in Ref.~\cite{Avishai84}.

In order to understand this discrepancy, we have carried out a calculation of the
neutron spin rotation, which ignores strong-interaction effects.  It is equivalent
to a first-order (in $v^{\rm PV}$) perturbative estimate of this observable, and
the corresponding results, listed in the last row of Table~\ref{tb:nspin} (row labeled
``plane waves''), demonstrate that strong-interaction distorsion effects are crucial,
in fact they are responsible for flipping the sign of $\phi$.  This is in contradiction
with the statement reported in the first paragraph after Eq.~(6) of Ref.~\cite{Avishai84}:
Avishai and Grange claim that the ``plane-wave'' prediction with the DDHb model
is $-6.0\times 10^{-9}$ rad-cm$^{-1}$, namely it has the same sign as in their full calculation.

The sign difference between the predictions obtained by either including or neglecting
strong-interaction distorsion effects can easily be understood.  For simplicity,
consider the DDH$\pi$ model, in which case the relevant matrix element contributing
to $\phi$ is $\langle ^3$P$_1\mid$$v^{\rm PV}({\rm DDH}\pi)$$\mid ^3$S$_1\rangle$,
connecting the continuum $T$=0 $^3$S$_1$ and $T$=1 $^3$P$_1$ channels.
The essential difference between the un-distorted and distorted $^3$S$_1$ wave functions
is the presence of a node in the latter, thus ensuring its orthogonality to the
deuteron $^3$S$_1$ component.  It is this node that causes the sign flip.
\subsection{Photon asymmetry in $\vec n$$p$ radiative capture at low energies}
\label{sec:r_np}

The PV asymmetry $a^\gamma$ in the $^1$H($\vec n$,$\gamma$)$^2$H reaction at thermal
neutron energies is calculated for the AV18, BONN and NIJM-I interactions.  
The asymmetry is expected to be constant for low-energy
neutrons up to energies well beyond the 1--15 meV averaged
in the experiment currently running at the LANSCE facility~\cite{Bowman04}.  Each strong
interaction model has associated two-body currents.  For the AV18 we consider the currents
from the momentum-independent terms---the $\pi$- and $\rho$-exchange currents from its
$v_6$ part---as well as from the momentum-dependent terms, as reviewed in
Sec.~\ref{sec:cnt}.  Further discussion of the AV18 currents is given below.
For the BONN and NIJM-I interactions, we retain only the $\pi$- and $\rho$-exchange currents
with cutoff parameters taken from the BONN model ($\Lambda_\pi$=1.72 GeV
and $\Lambda_\rho$=1.31 GeV), while we neglect contributions from other meson exchanges.
In all calculations, however, the currents associated with the $\Delta$
excitation and $\omega$$\pi$$\gamma$ transition have been included.
 
The total cross section $\sigma^\gamma$ is due to the well-known $M_1$
transition connecting the PC $^1$S$_0$ $n$$p$ state to the PC deuteron state.
The calculated values for each model are given in Table~\ref{tb:np_x},
both for one-body (impulse) currents alone and for the one- and two-body
currents.  In each case the largest two-body contribution, approximately
two-thirds of the total, comes from the currents associated with pion
exchange.  The total cross section is in good agreement with experimental
results, which are variously quoted as 334.2(0.5) mb~\cite{Cox65} or 332.6(0.7)
mb~\cite{Mughabghab81}.  It would be possible to adjust, for example, the transition
magnetic moment $\mu_{\gamma N\Delta}$ of the $\Delta$-excitation current to precisely
fit one of these values, here we simply choose a $\mu_{\gamma N\Delta}$ of 3 n.m.,
which is consistent with an analysis of $\gamma$-$N$ data at resonance.
 
As discussed Sec.~\ref{sec:npr}, the PV asymmetry arises from an interference between
the $M_1$ term above and the $E_1$ transition, connecting the $^3$P$_1$ PV $n$$p$ state
to the PC deuteron state and the $^3$S$_1$ PC $n$$p$ state to the $^3$P$_1$
PV deuteron state.  The $E_1$ transitions proceeding through the PV $^1$P$_1$
$n$$p$ or deuteron states are suppressed, because of an isospin selection rule
forbidding isoscalar electric-dipole transitions and also because of
spin-state orthogonality.  In principle, there is a relativistic correction
to the electric dipole operator, associated with the definition
of the center-of-energy~\cite{Nollett01}.  However, its contribution in transitions
proceeding through the $^1$P$_1$ channel vanishes too,
since the associated operator is diagonal in the pair spin.

The calculated asymmetries are listed in Table~\ref{tb:np_x}.
The results are consistent with earlier~\cite{Desplanques75,McKellar75} and
more recent~\cite{Desplanques01,Hyun01} estimates, and are in agreement with each other 
at the few-per-cent level, which is also the magnitude of the contributions from the
short-range terms.  In particular, they show that this observable is very sensitive
to the weak PV $\pi$$N$$N$ coupling constant, while it is essentially unaffected by
short-range contributions (in this context, see also Fig.~\ref{fig:deut}).
The $E_1$ transition has been calculated in the long-wavelength
approximation (LWA) with the Siegert form of the $E_1$ operator (see, for example,
Eq.~(4.5) of Ref.~\cite{Viviani00}), thus eliminating many of the model dependencies
and leaving only simple (long-range) matrix elements.  In the notation of Sec.~\ref{sec:npr},
the associated reduced matrix elements (RMEs) are explicitly given by

\begin{eqnarray}
E_1(^3{\rm S}_1,^3{\rm S}_1) + E_1(^3{\rm D}_1,^3{\rm S}_1)&=&
{\rm i} \frac{q}{2\,\sqrt{6 \pi}}\int_0^\infty {\rm d}r\, r^2\Bigg[
u(r;^3\!{\rm P}_1)\left[               w(r;^3\!{\rm S}_1)
                 -\frac{1}{\sqrt{2}} w(r;^3\!{\rm D}_1) \right] \nonumber \\
&-&w(r;^3\!{\rm P}_1) \left[                    u(r;^3\!{\rm S}_1)
                          -\frac{1}{\sqrt{2}} u(r;^3\!{\rm D}_1) \right] \Bigg] \ ,
\end{eqnarray}
where the $w$'s and $u$'s denote the $n$$p$ continuum and deuteron radial wave functions
defined, respectively, as in Secs.~\ref{sec:pwest} and~\ref{sec:deut} (only the
outgoing channel quantum numbers are displayed for the $w$'s).  Corrections beyond
the LWA terms in $E_1$ transitions have been found to be quite small.  For completeness, we also
give the well known expression for the $M_1$ RME, as calculated to leading order
in $q$ and in the limit in which only one-body currents are retained,

\begin{equation}
M_1(^1{\rm S}_0,^1{\rm S}_0)={\rm i} \frac{q}{2\, \sqrt{2\pi} \, m}(\mu_p -\mu_n)
\int_0^\infty {\rm d}r\, r \, u(r;^3\!{\rm S}_1)\, w(r;^1\!{\rm S}_0) \ ,
\end{equation}
where the combination $\mu_p -\mu_n$=4.706 n.m. is the nucleon isovector magnetic moment.

We have also calculated the $E_1$ contributions with the full current
density operator ${\bf j}({\bf x})$, namely by evaluating matrix elements of

\begin{equation}
E_{1\lambda} = \frac{1}{q} \int {\rm d}{\bf x}\, {\bf j}({\bf x}) \cdot
\nabla \times j_1(qx) {\bf Y}_{1\lambda}^{11}(\hat {\bf x}) \ ,
\label{eq:e1op}
\end{equation}
where ${\bf Y}_{1\lambda}^{11}$ are standard vector spherical harmonics.
To the extent that retardation corrections beyond the LWA of the $E_1$ operator
are negligible~\cite{Viviani00}, this should produce identical results
{\it provided} the current is exactly conserved.  In order to satisfy current
conservation, currents from both the strong (PC) and weak (PV) interactions
are required, as discussed in Sec.~\ref{sec:cnt}.  In the following we
keep only the $\pi$-exchange term in the DDH interaction (with their
``best guess'' for the weak $\pi$$N$$N$ coupling constant), and use the
AV18 strong-interaction model.

As reviewed in Sec.~\ref{sec:cnt}, the PC two-body currents constructed
from the $v_6$ part of the AV18 interaction (the $\pi$- and
$\rho$-exchange currents) exactly satisfy current
conservation with it.  The same holds true for the PV $\pi$-exchange
currents derived from the DDH interaction in Sec.~\ref{sec:cntpv}.
However, the PC two-body currents originating from the isospin- and
momentum-dependent terms of the AV18 are strictly not conserved (see
below).  The associated contributions, while generally
quite small, play here a crucial role because of the large cancellation
between the (PC) $v_6$ currents from the AV18 and the (PV) $\pi$-currents
from the DDH.  This point is illustrated in Table~\ref{tb:cntre}.
Note that the PC currents from $\Delta$-excitation and $\omega$$\pi$$\gamma$
transition are transverse and therefore do not affect the $E_1$
matrix element.  However, they slightly reduce the PV asymmetry, since their
contributions increase the $M_1$ matrix element by $\simeq 1$\%.
They are not listed in Table~\ref{tb:cntre}.
 
The asymmetry is given by the sum of the two columns in Table~\ref{tb:cntre},
namely $+0.17 \times 10^{-8}$ (last row).  This value should be
compared to $-5.02 \times 10^{-8}$, obtained with the Siegert
form of the $E_1$ operator for the same interactions (and currents
for the $M_1$ matrix element).  As already mentioned, we have explicitly
verified that retardation corrections in the $E_1$ operator are too small
to account for the difference.  Thus the latter is to be ascribed to the lack of
current conservation, originating from the isospin- and momentum-dependent
terms of the AV18.

To substantiate this claim, we have carried out
a calculation based on a $v_8$ reduction~\cite{Wiringa02}
of the AV18 (denoted as AV8), constrained to reproduce the binding energy of the deuteron
and the isoscalar combinations of the S- and P-wave phase shifts (note,
however, that we do include in the AV8 the electromagnetic terms from the AV18,
omitted in Ref.~\cite{Wiringa02}).  For the AV8 model, the $\pi$- and
$\rho$-exchange currents from the isospin-dependent central, spin-spin,
and tensor interaction components are constructed as for the AV18, and therefore
are exactly conserved.  However, the currents from the isospin-independent
interaction, $v^b(r_{ij}) {\bf L}\cdot {\bf S}$, are derived by minimal substitution,

\begin{equation}
{\bf p}_i \rightarrow {\bf p}_i - e\, P_i \, {\bf A}({\bf r}_i) \ ,
\end{equation}
where $e$ and ${\bf A}$ are the electric charge
and vector potential, respectively,
and $P_i$ is the proton projection operator.
The linear terms in ${\bf A}$ are written as
$-\int {\rm d}{\bf x}\, {\bf j}({\bf x}) \cdot {\bf A}({\bf x})$,
and the resulting spin-orbit current density---or, rather, its Fourier
transform---reads
 
\begin{equation}
{\bf j}^{\rm PC}_{b,ij}({\bf q}) = \frac{v^b(r_{ij})}{2}
\left( {\rm e}^{ {\rm i} {\bf q} \cdot {\bf r}_i}\, P_i -
       {\rm e}^{ {\rm i} {\bf q} \cdot {\bf r}_j}\, P_j \right) \,
{\bf S} \times {\bf r}_{ij} \ .
\end{equation}

In the case of the isospin-dependent terms, after symmetrizing
$v^{b\tau}(r) \left[ {\bf L}\cdot {\bf S}\, , \,
\bbox{\tau}_i \cdot \bbox{\tau}_j \right]_+/2$, one obtains
 
\begin{equation}
{\bf j}^{\rm PC}_{b\tau,ij}({\bf q}) = \frac{v^{b\tau}(r_{ij})}{2}
\left( {\rm e}^{ {\rm i} {\bf q} \cdot {\bf r}_i}\, Q_j -
       {\rm e}^{ {\rm i} {\bf q} \cdot {\bf r}_j}\, Q_i \right) \,
{\bf S} \times {\bf r}_{ij} \ ,
\end{equation}
where
 
\begin{equation}
Q_i \equiv \frac{\bbox{\tau}_i \cdot \bbox{\tau}_j + \tau_{z,i} }{2} \ .
\end{equation}
While minimal substitution ensures that the
current is indeed conserved for the isospin-independent interaction, i.e.
 
\begin{equation}
{\bf q} \cdot {\bf j}^{\rm PC}_{b,ij}({\bf q}) =
\left[ v^b(r_{ij}) {\bf L}\cdot {\bf S}\,  , \, \rho_i({\bf q})+\rho_j({\bf q}) \right] \ ,
\label{eq:j_con}
\end{equation}
this prescription does not lead to a conserved current for the isospin-dependent
one, since the commutator above generates an isovector term of the type

\begin{equation}
{\rm i} (\bbox{\tau}_i \times \bbox{\tau}_j)_z
\frac{v^{b\tau}(r_{ij})}{2} \left[ {\bf L}\cdot {\bf S}\,  , \,
 {\rm e}^{ {\rm i} {\bf q} \cdot {\bf r}_i} -
 {\rm e}^{ {\rm i} {\bf q} \cdot {\bf r}_j} \right]_+ \ .
\end{equation}
Physically, this corresponds to the fact that isospin-dependent interactions
are associated with the exchange of charged particles, which an electromagnetic
field can couple to.  One can enforce current conservation
by introducing an additional term~\cite{Marcucci03}, which in the case
of ${\bf j}^{\rm PC}_{b\tau,ij}$ is taken as
 
\begin{equation}
{\rm i} (\bbox{\tau}_i \times \bbox{\tau}_j)_z
\frac{v^{b\tau}(r_{ij})}{2} \left[ {\bf L} \cdot {\bf S}\,  , \,
{\bf r}_{ij} \frac{ {\rm e}^{ {\rm i} {\bf q} \cdot {\bf r}_i} -
 {\rm e}^{ {\rm i} {\bf q} \cdot {\bf r}_j} }{{\bf q} \cdot {\bf r}_{ij} } \right]_+ \ .
\end{equation}

The results obtained for the total cross section and PV asymmetry with the
AV8 and pion-only DDH interactions and associated (exactly conserved) currents
are listed in Table~\ref{tb:cnte1}.  A few comments are in order.  Firstly, the
$M_1$ cross section in impulse approximation is $\simeq$ 30\% smaller than predicted
with the AV18 interaction.  This is due to the fact that the $n$$p$ singlet 
scattering length obtained with the AV8 (truncated) model is --19.74 fm, and
so is about 15\% smaller in magnitude than its physical value, --23.75 fm, reproduced
by the AV18 within less than 0.1\%~\cite{Wiringa95}.

Secondly, the enhancement of the $M_1$ cross section in impulse approximation
due to (PC) two-body currents, 9.3\%, is essentially consistent with that
predicted with the AV18.

Lastly, the PV asymmetry obtained with the full currents is close
to that calculated with the Siegert form of the $E_1$ operator.
The remaning $\simeq 1$\% difference is due to numerical inaccuracies
as well as additional corrections from retardation terms and higher
order multipoles.  Both of these effects are included in the full-current
calculation.  Note the crucial role played by the spin-orbit currents
constructed above.

\subsection{Deuteron threshold disintegration with circularly polarized photons}
\label{sec:r_dg}

\begin{figure}[bth]
\let\picnaturalsize=N
\def\picsize{5in}
\def\picfilenamea{figs/dgpn/xdgpn_s.eps}
\epsfbox{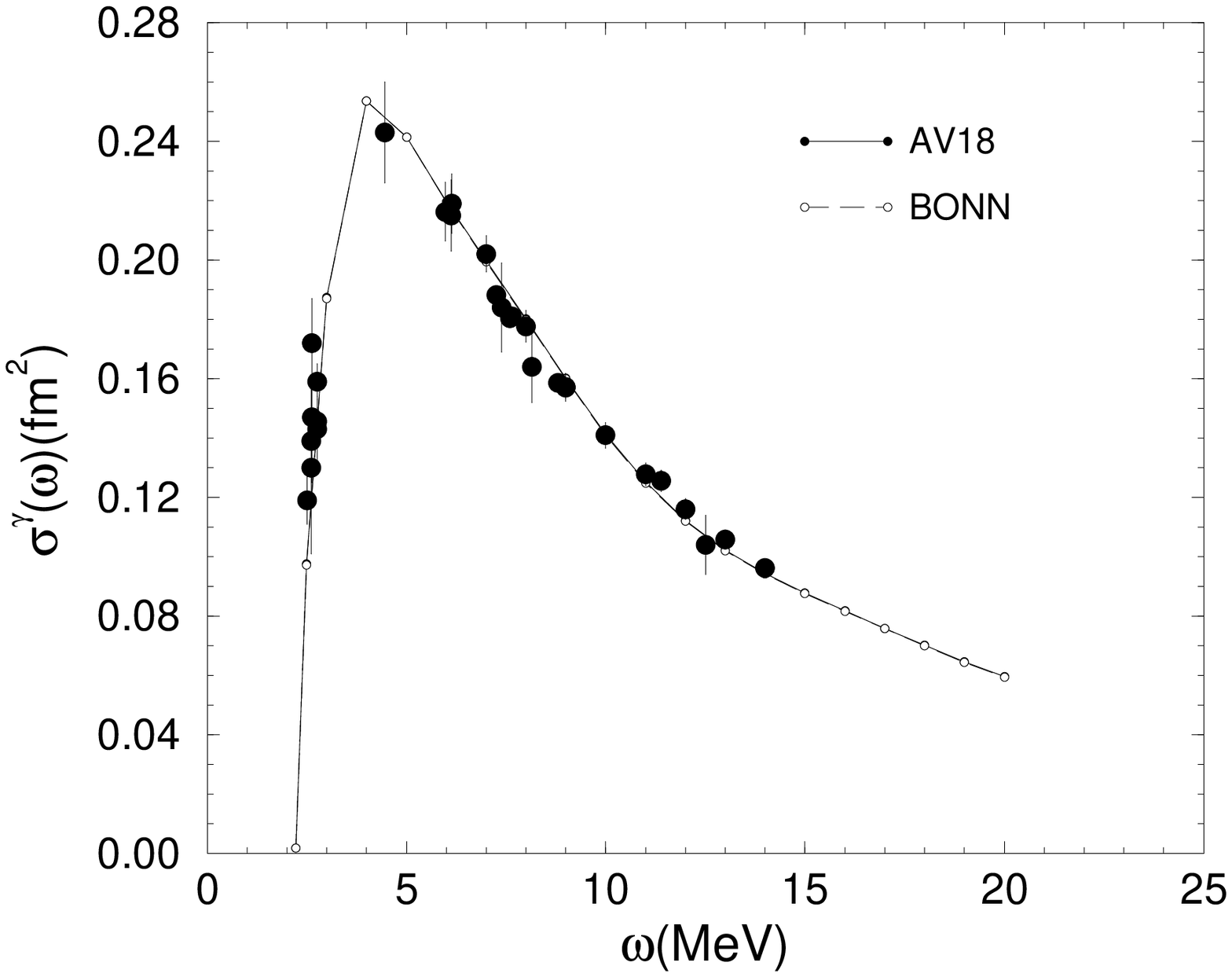}
\caption{The deuteron photo-disintegration cross sections, calculated
with the AV18 and BONN interactions, are compared to data.  Note that
the AV18 and BONN results are indistinguishable.}
\label{fig:xdg}
\end{figure}

The photo-disintegration cross sections calculated with the AV18 and
BONN models from threshold to 20 MeV photon energies are in excellent
agreement with data~\cite{Bishop50}--\cite{DeGraeve92}, see Fig.~\ref{fig:xdg}.
The model dependence between the AV18 and BONN results is negligible.
In the calculations the final $n$$p$ states include interaction effects
in all channels up to $J$=5 and spherical Bessel functions for $J > 5$,
as discussed in the next section.

In the energy regime of interest here, the (total) cross section is
dominated by the contributions of $E_1$ transitions connecting the
deuteron to the $n$$p$ triplet P-waves.  The Siegert form is used for
the $E_1$ operator.  Because of the way the calculations are carried out
(see Sec.~\ref{sec:cal}), it is conveniently implemented by making use
of the following identity for the current density operator ${\bf j}({\bf x})$,
or rather its Fourier transform ${\bf j}({\bf q})$, 

\begin{eqnarray}
{\bf j}({\bf q})&=&{\bf j}({\bf q})-{\bf j}({\bf q}\!=\!0)-\int{\rm d}{\bf x} \, {\bf x}
\, \nabla \cdot {\bf j}({\bf x}) \nonumber \\
&=&{\bf j}({\bf q})-{\bf j}({\bf q}\!=\!0)+{\rm i}\, \left[ H \, , \, \int{\rm d}{\bf x}
\, {\bf x} \, \rho({\bf x}) \right] \ ,
\label{eq:j_s}
\end{eqnarray}
where in the first line the volume integral of ${\bf j}({\bf x})$
has been re-expressed in terms of the divergence of the current,
ignoring vanishing surface contributions, and in the second line
use has been made of the continuity equation.  Here $\rho({\bf x})$
is the charge density operator.  In evaluating the matrix elements
in Eq.~(\ref{eq:jmin}) the commutator term reduces to

\begin{equation}
{\rm i} \int{\rm d}{\bf x} \, {\bf x}
\left[ H\, , \, \rho({\bf x}) \right] \rightarrow {\rm i} q \,
\int{\rm d}{\bf x} \, {\bf x} \, \rho({\bf x}) \simeq {\rm i} q \, \sum_i P_i \, {\bf r}_i \ ,
\end{equation}
where $P_i$ is the proton projection operator introduced earlier, and
relativisitc corrections to $\rho({\bf x})$, such as those associated
with spin-orbit and pion-exchange contributions~\cite{Viviani00,Nollett01},
have been neglected.

\begin{figure}[bth]
\let\picnaturalsize=N
\def\picsize{5in}
\def\picfilenamea{figs/dgpn/xdgpn_cmp.eps}
\epsfbox{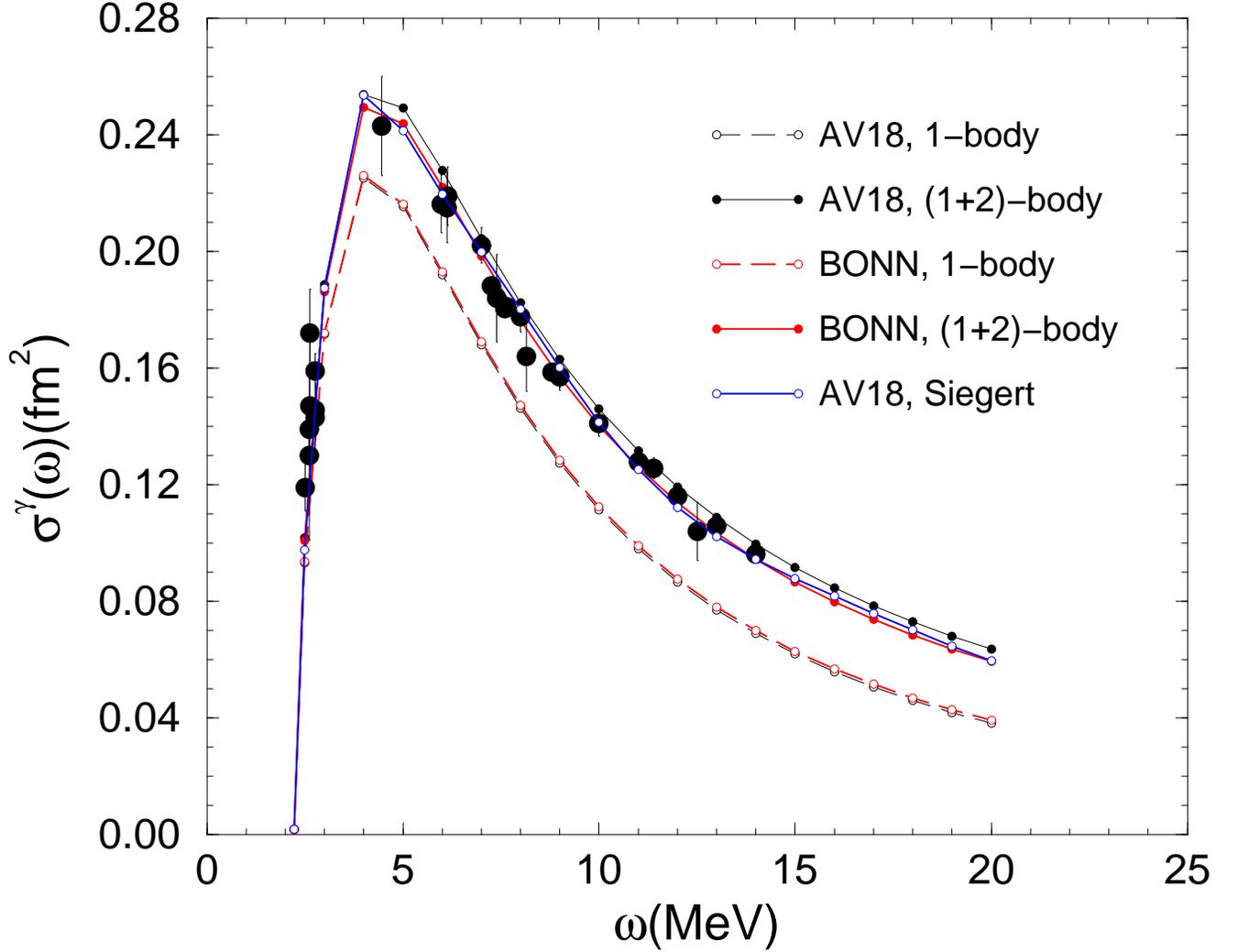}
\caption{The deuteron photo-disintegration cross sections, calculated
with the AV18 and BONN interactions, are compared to data.  Results
obtained by including only one-body terms and both one- and two-body terms
in the electromagnetic current are shown along with those calculated
by using the right-hand-side of Eq.~(\protect\ref{eq:j_s}).  The latter
are the same as in Fig.~\protect\ref{fig:xdg}.}
\label{fig:xdgc}
\end{figure}

We have also calculated the photo-disintegration cross section by using
the expression given in Eq.~(\ref{eq:e1op}) for the $E_1$ operator, or
equivalently by calculating matrix elements of the current ${\bf j}({\bf q})$
without resorting to the identity in Eq.~(\ref{eq:j_s}).  The results
obtained by including only the one-body terms and both the one- and
two-body terms in ${\bf j}({\bf q})$ are compared with those obtained in 
the Siegert-based calculation (as well as with data) in Fig.~\ref{fig:xdgc}.
The same conclusions as in the previous section remain valid here.  Had the
current been exactly conserved, then the Siegert-based and full ${\bf j}({\bf q})$
calculations would have produced identical results.  The small differences in
the case of the AV18 model, as an example, are to be ascribed to missing
isovector currents associated with its momentum-dependent interaction components
(see previous section).

\begin{figure}[bth]
\let\picnaturalsize=N
\def\picsize{5in}
\def\picfilenamea{figs/dgpn/asy_dgpn_s.eps}
\epsfbox{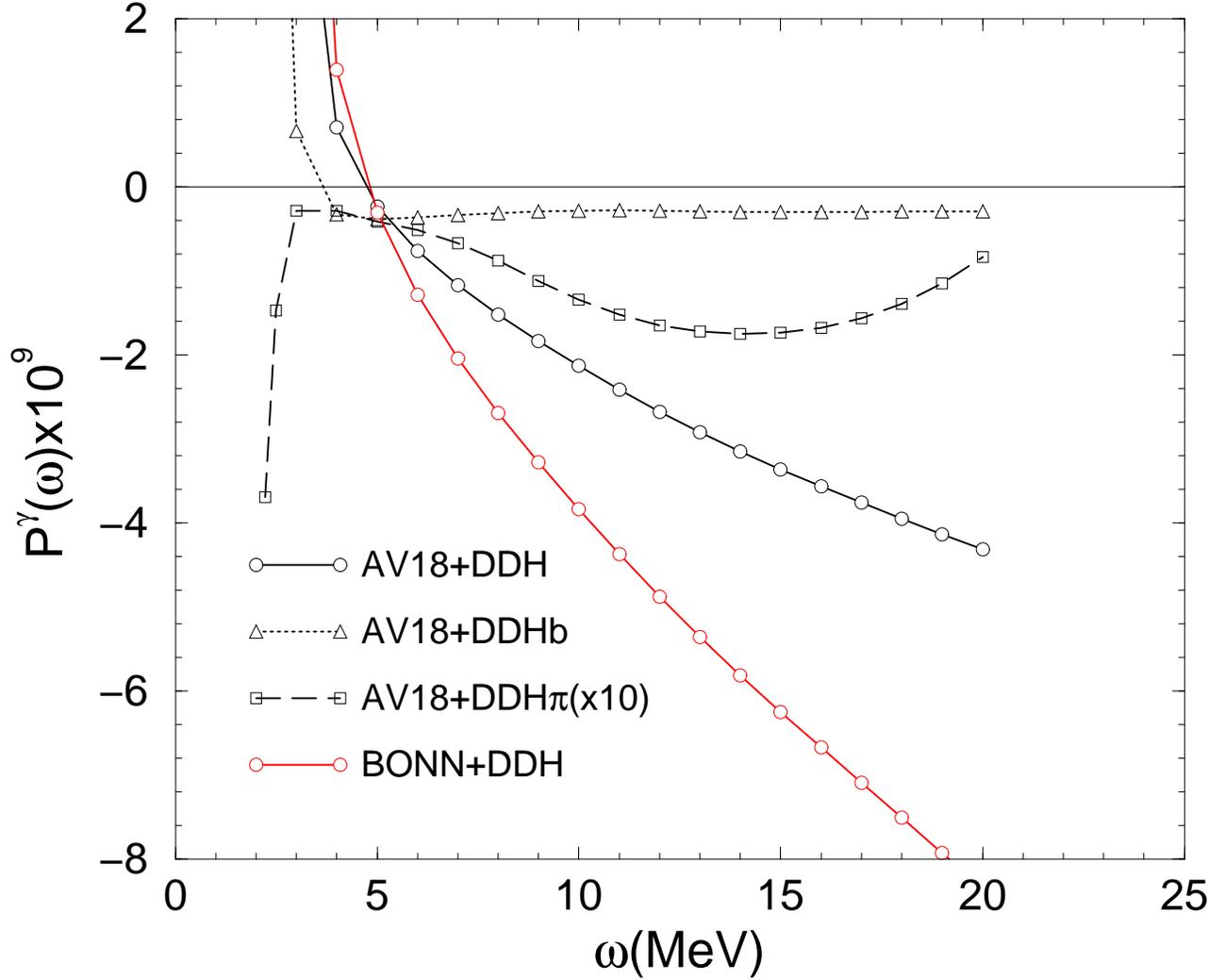}
\caption{The photon helicity-dependent asymmetries obtained with various combinations
of strong- and weak interaction potentials.  Note that the predictions corresponding
to the AV18+DDH$\pi$ potential combination are suppressed by roughly one order of
magnitude relative to those corresponding to the AV18+DDH and AV18+DDHb models.
All results are obtained by using the right-hand-side of Eq.~(\protect\ref{eq:j_s}).}
\label{fig:adg}
\end{figure}

The PV photon polarization parameter $P^\gamma$, obtained with various
combinations of PC and PV interactions, is displayed in Fig.~\ref{fig:adg},
while its value at a photon energy $\simeq$ 1.3 keV above breakup threshold
is listed in Table~\ref{tb:tadg}.  All results presented below
use the current operator in the form given on the right-hand-side of
Eq.~(\ref{eq:j_s}).  Note that, as discussed in Sec.~\ref{sec:gdnp}, the
parameters $P^\gamma$ for the direct $d(\vec{\gamma},n)p$ and inverse
$p(n,\vec{\gamma})d$ processes are the same.  In the threshold region, a
few keV above breakup, the expression for $P^\gamma$ reduces to 

\begin{equation}
P^\gamma = \frac{ 2 \, {\rm Re}\,\left[ M_1(^1{\rm S}_0) E_1^*(^1{\rm S}_0)\right] }
{|M_1(^1{\rm S}_0)|^2} \ ,
\end{equation}
where, in the notation of the previous section, the $M_1(^1{\rm S}_0)$ RME
is defined as in Eq.~(\ref{eq:m1def}) and similarly for $E_1(^1{\rm S}_0)$.
In this energy region, the only relevant channel in the final $n$$p$ state
has $J$=0, see discussion at the end of Sec.~\ref{sec:npr}.  Note that the
combination of RMEs occurring in $P^\gamma$ is different from that in
$a^\gamma$, the photon angular asymmetry parameter measured in $\vec{n}$$p$
radiative capture.  Indeed, in contrast to $a^\gamma$, the photon polarization
parameter is almost entirely determined by the short-range part of the DDH
interaction, mediated by vector-meson exchanges (and having isoscalar and
isotensor character~\cite{Danilov65}), see Table~\ref{tb:tadg} and
Fig.~\ref{fig:adg}.  This is easily understood, since in the $^1$S$_0$-$^3$P$_0$
channel the pion-exchange component of the DDH interaction vanishes.
Furthermore, the $E_1$ transition connecting the $^1$S$_0$ $n$$p$ continuum
state to the PV $^3$P$_1$ component of the deuteron, which is predominantly
induced by the pion-exchange interaction, is strongly suppressed, to leading
order, by spin-state othogonality.  Higher order corrections, associated with
retardation effects and relativistic contribution to the electric dipole operator,
were estimated in Ref.~\cite{Friar83} and were found to be of the order
of a few \% of the leading result arising from vector-meson exchanges.
Some of these corrections are retained in the present study.

The predictions in Table~\ref{tb:tadg} and in Fig.~\ref{fig:adg} display
great sensitivity both to the strengths of the PV vector-meson couplings
to the nucleon and to differences in the short-range behavior of the
strong-interaction potentials, thus reinforcing the conclusion that
these short-ranged meson couplings are not in themselves physical
observables, rather the parity-violating mixings are the physically
relevant parameters.

Note that the $\simeq$ 5\% decrease
in $P^\gamma$ values between the rows labelled ``impulse'' and ``full''
is due to the corresponding 5\% enhancement of the $M_1$ transition connecting
the PC $^1$S$_0$ and deuteron states, due to two-body terms in the
electromagnetic current included in the ``full'' calculation.

The results in Table~\ref{tb:tadg} are consistent both in sign
and order of magnitude with those of ealier
studies~\cite{Hadjimichael71,Lassey75,Craver76,Lee78}, remaining
numerical differences are to be ascribed to different strong-
and weak-interaction potentials adopted in these earlier works.  Indeed,
we have explicitely verified that by using the PC AV18 potential
and the Cabibbo model for the PV potential~\cite{Henley69} we obtain
$P^\gamma$ values close to those reported in Ref.~\cite{Craver76,Lee78}.
However, our results seem to be at variance with those of Ref.~\cite{Oka83}
at photon energies a few MeV above above the breakup threshold.  In particular,
Table~II in that paper suggests that at 10 and 20 MeV the dominant contribution
to $P^\gamma$ is from the PV pion-exchange interaction and that $P^\gamma$ has 
the values $-2.66\times 10^{-8}$ and $-4.54\times 10^{-8}$, respectively.  This
is in contrast to what reported in Fig.~\ref{fig:adg} of the present work,
curves labelled AV18+DDHb and AV18+DDH$\pi$.  There is a two-order of magnitude
difference between the values referred to above and those obtained here.
It is not obvious whether this is due to the use in Ref.~\cite{Oka83}
of the Hamada-Johnston~\cite{Hamada62} PC potential---a PV potential
``close'' to our model DDHb is adopted.

The results in Table~\ref{tb:tadg} are consistent with the latest experimental
determination, $P^\gamma$=$(1.8\pm 1.8) \times 10^{-7}$~\cite{Knyaz'kov84},
but about two orders of magnitude smaller than an earlier
measurement~\cite{Lobashov72}.

\begin{figure}[bth]
\let\picnaturalsize=N
\def\picsize{5in}
\def\picfilenamea{figs/dgpn/asy_dgpn_j_s.eps}
\epsfbox{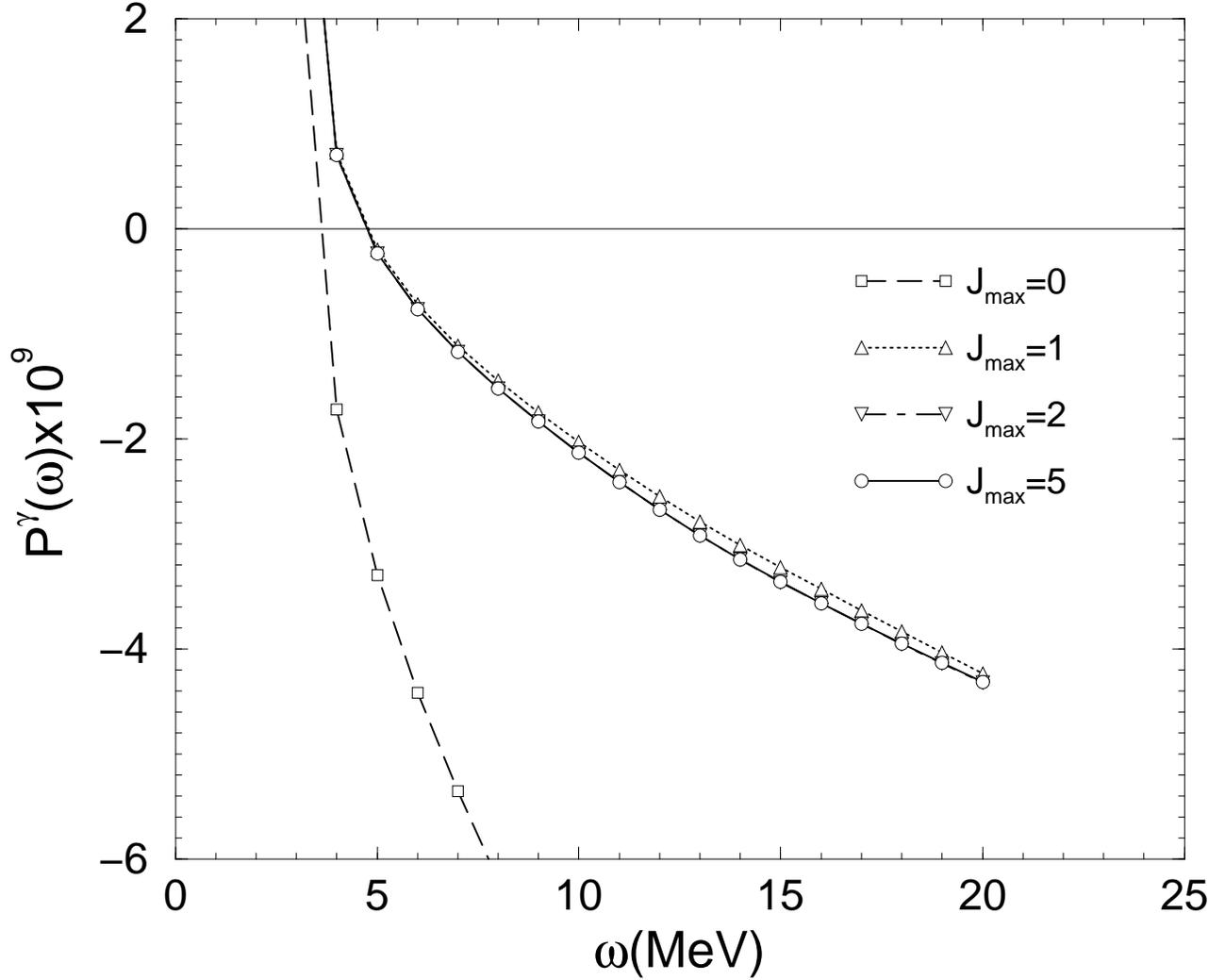}
\caption{Contributions to the photon helicity-dependent asymmetry obtained by
including PV admixtures in the wave functions of all channels up to $J_{\rm max}$,
with $J_{\rm max}$=0, 1, 2, and 5.  The AV18+DDH potential combination is used,
black solid line in Fig.~\protect\ref{fig:adg}.  Note that the curves labeled
$J=2$ and $J=5$ are indistinguishable.}
\label{fig:adgj}
\end{figure}

Figure~\ref{fig:adgj} shows the photon-polarization parameter
obtained by including PV admixtures in the $n$$p$ continuum wave functions
of all channels with $J \leq J_{\rm max}$ and $J_{\rm max}$=0, 1, 2 and 5.
In the energy range explored so far, $P_\gamma$ is essentially given by
the contributions of the $J$=0 and 1 channels.

\begin{figure}[bth]
\let\picnaturalsize=N
\def\picsize{5in}
\def\picfilenamea{figs/dgpn/asy_dgpn_mec_s.eps}
\epsfbox{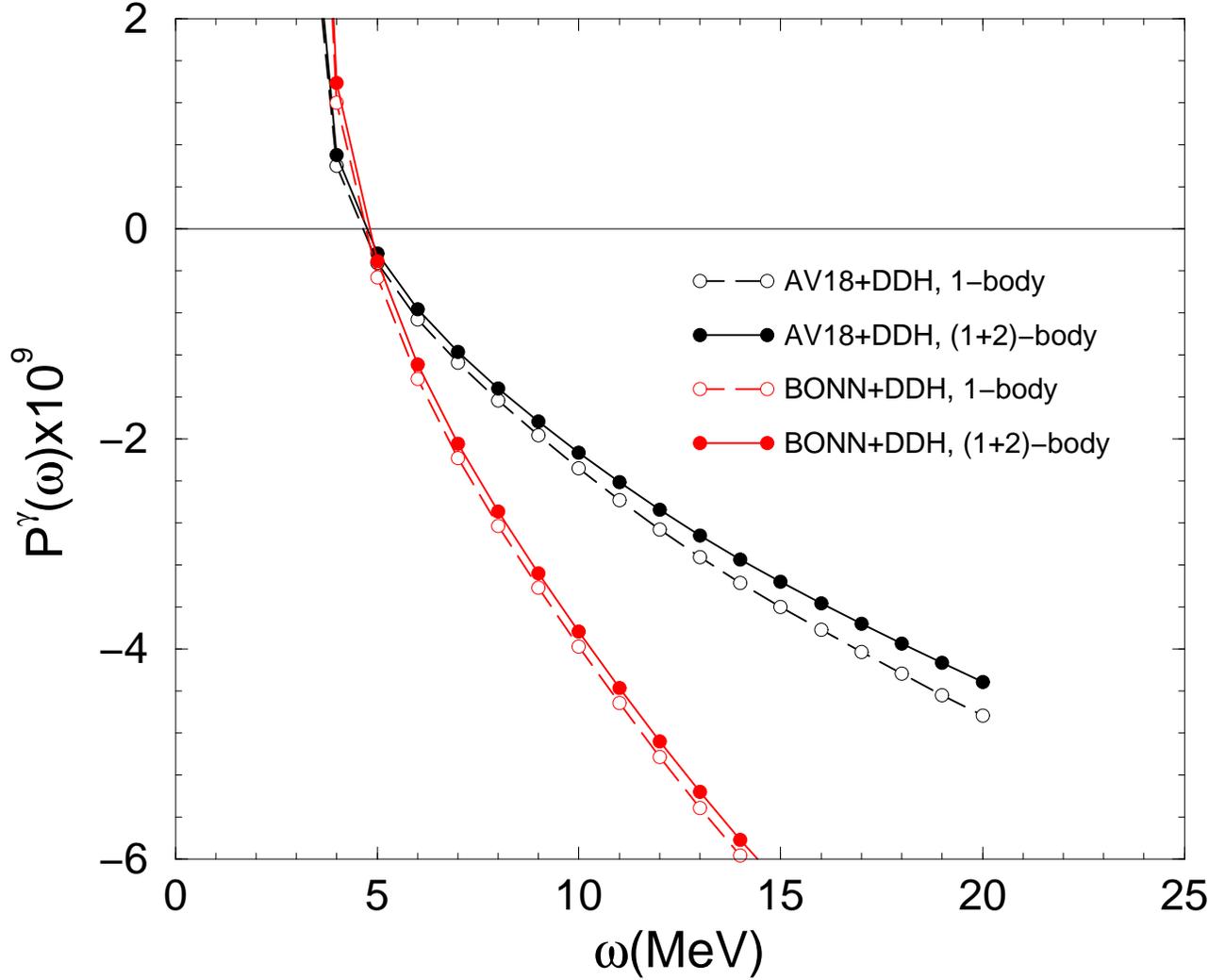}
\caption{The photon helicity-dependent asymmetries obtained with the AV18+DDH and
BONN+DDH potential combinations by including only one-body terms and both one-
and two-body terms in the right-hand-side of Eq.~(\protect\ref{eq:j_s}).}
\label{fig:adg2}
\end{figure}

Finally, Fig.~\ref{fig:adg2} illustrates the effects of two-body terms
in the electromagnetic current, written as in the right-hand-side of
Eq.~(\ref{eq:j_s}).  The associated contributions are of the order
of a few \% relative to those from one-body terms.

\subsection{Deuteron electro-disintegration at quasi-elastic kinematics}
\label{sec:r_dee}

In this section we present results for the asymmetries
$A_{\gamma \gamma}$ and $A_{\gamma Z}$ obtained by including
one- and two-body terms in the electromagnetic and
neutral-weak currents.  Note, however, that only the PV two-body
terms associated with $\pi$-exchange in the DDH interaction
are considered in the present calculations (in addition, of course, to the
PC terms discussed in Sec.~\ref{sec:cnt}).  The PV currents
from $\rho$- and $\omega$-exchange have been neglected, since they
are expected to play a minor role due to their short-range character.
One should also observe that at the higher momentum transfers
of interest here, 100--300 MeV/c, relevant for the SAMPLE
experiments~\cite{Hasty00,Beck01}, it is not possible to include
the contributions of electric multipole operators through the 
Siegert theorem, these must be calculated explicitly from
the full current.

The $A_{\gamma Z}$ contribution was recently studied in Ref.~\cite{Diaconescu01},
where it was shown that two-body terms in the nuclear electromagnetic and weak
neutral currents only produce (1--2)\% corrections to the asymmetry due to the
corresponding single-nucleon currents.  The present study---a short account
of which has been published in Ref.~\cite{Schiavilla03a}---investigates the
asymmetry originating from hadronic weak interactions.  It updates and sharpens earlier
predictions obtained in Refs.~\cite{Hwang80,Hwang81}---for example, these
calculations did not include the effects of two-body currents induced
by PV interactions.

The present calculation proceeds as discussed in Sec.~\ref{sec:cal}.
We have used the AV18 or BONN models (and associated currents) in
combination with the full DDH interaction (with coupling and cutoff
values as given in Table~\ref{tb:ddh}).  The final state, labeled
by the relative momentum ${\bf p}$, pair spin and $z$-projection $S M_S$,
and pair isospin $T$ ($M_T$=0), is expanded in partial waves; PC
and PV interaction effects are retained in all partial waves
with $J \leq 5$, while spherical Bessel functions are employed
for $J>5$.  In the quasi-elastic regime of interest here, it has
been found that interaction effects are negligible for $J>5$.
 
\begin{figure}[bth]
\let\picnaturalsize=N
\def\picsize{5in}
\def\picfilenamea{figs/xs_iii.eps}
\epsfbox{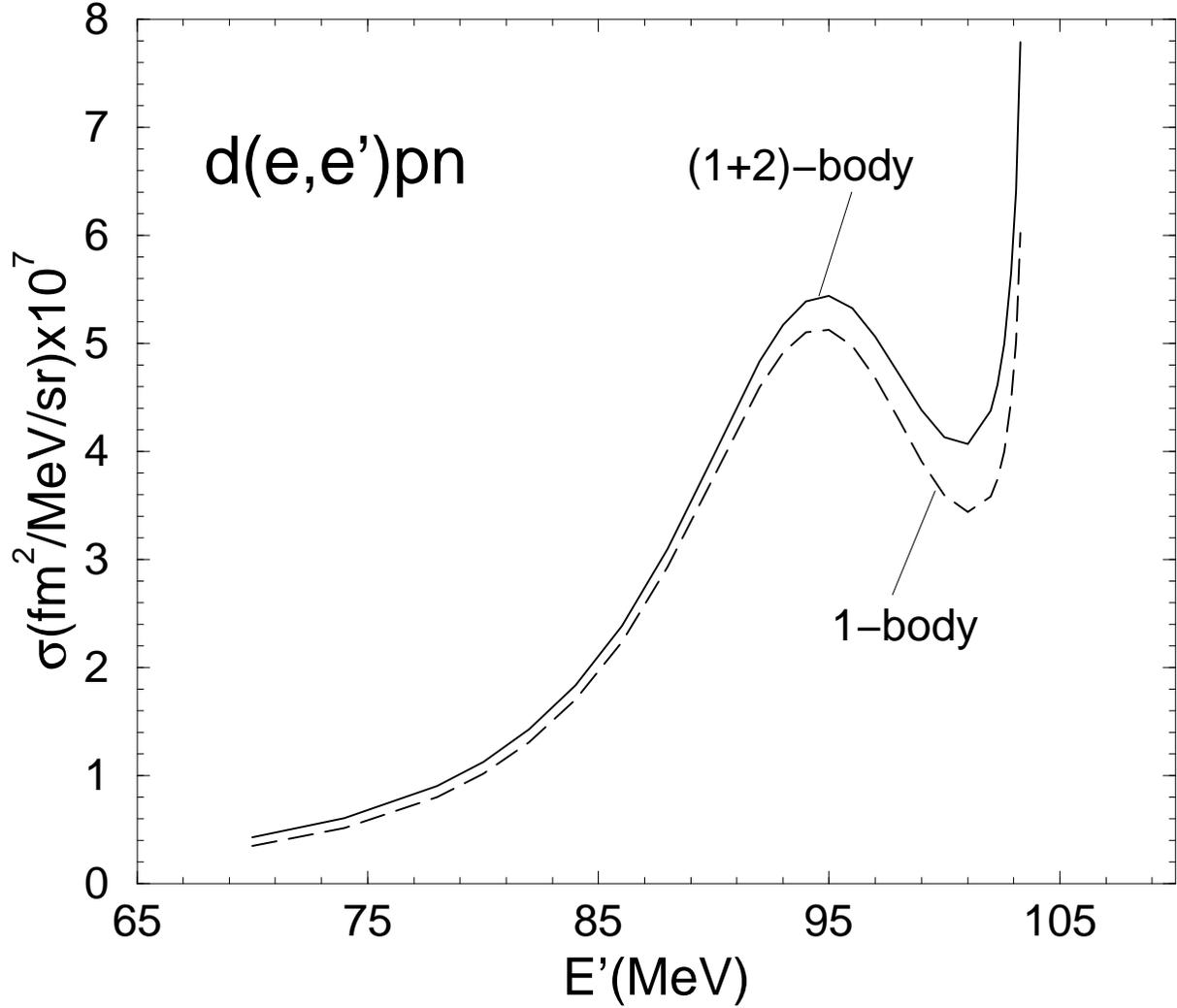}
\caption{The $d(e,e^\prime)np$ inclusive cross section calculated,
as function of the scattered electron energy $E^\prime$, with the AV18
interaction model.  The electron incident energy is 117 MeV and its
scattering angle $\theta_e$ is 138.4$^\circ$.  Predictions are shown obtained with 
one-body terms alone and both one- and two-body terms in the electromagnetic
current.}
\label{fig:xsiii}
\end{figure}

\begin{figure}[bth]
\let\picnaturalsize=N
\def\picsize{5in}
\def\picfilenamea{figs/asy_iii.eps}
\epsfbox{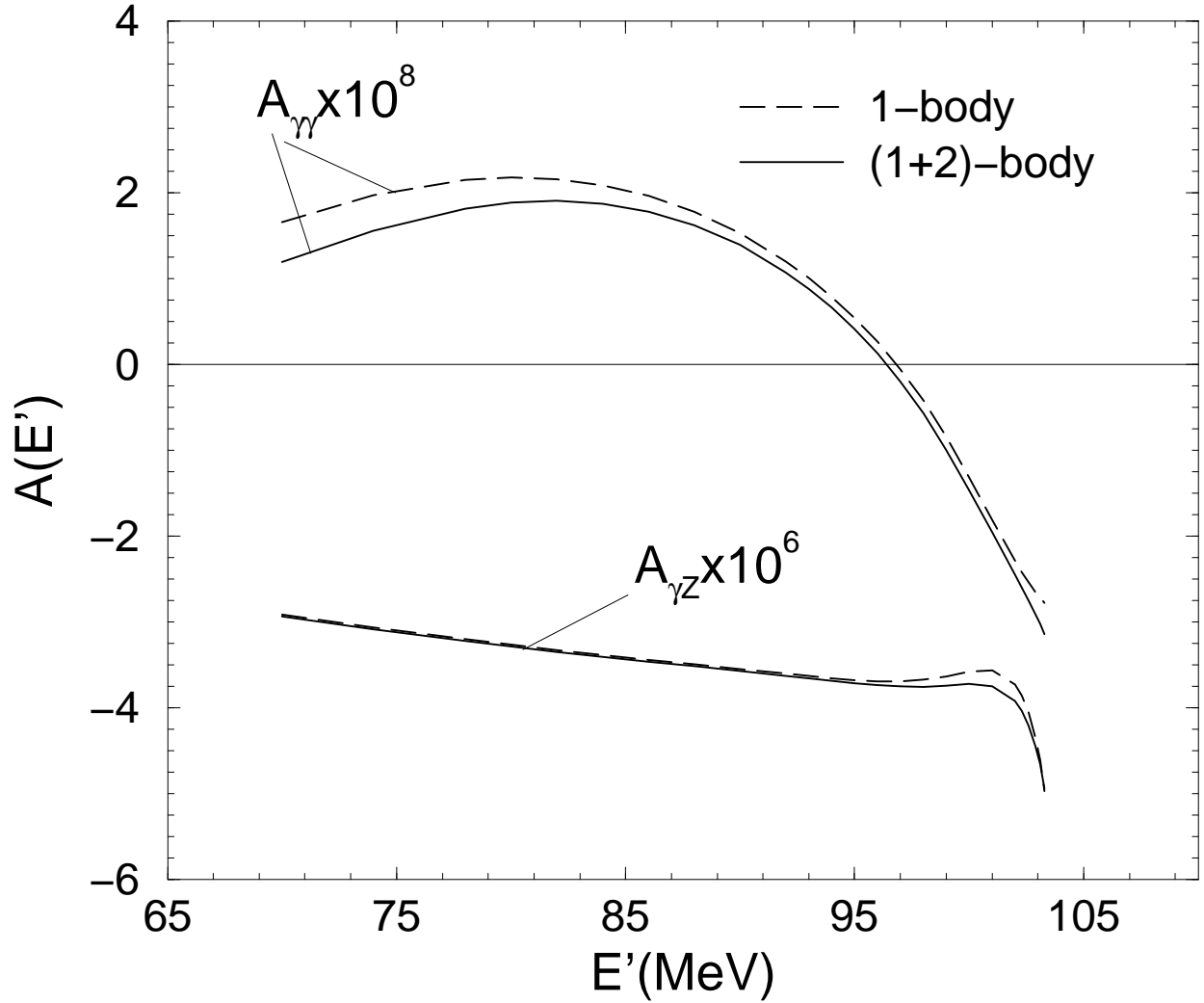}
\caption{The asymmetries $A_{\gamma\gamma}$ and $A_{\gamma Z}$ calculated,
as function of the scattered electron energy $E^\prime$, with the (PC) AV18
and (PV) DDH interaction models.  The other electron kinematical variables
are as in Fig.~\protect\ref{fig:xsiii}.  Predictions are shown obtained with 
one-body terms alone and both one- and two-body terms in the electromagnetic
and neutral weak currents.}
\label{fig:asiii}
\end{figure}

In Figs.~\ref{fig:xsiii} and~\ref{fig:asiii} we show, respectively,
the inclusive cross section and the asymmetries $A_{\gamma Z}$
and $A_{\gamma \gamma}$, obtained with the AV18 and DDH interactions,
for one of the two SAMPLE kinematics, corresponding to a three-momentum
transfer range between 176 MeV and 206 MeV at the low and high
ends of the spectrum in the scattered electron energy $E^\prime$,
the four-momentum transfer $|q_\mu^2|$ at the top of the quasi-elastic
peak is $\simeq$ 0.039 GeV$^2$.
The rise in the cross section at the high end of the $E^\prime$ spectrum---the
threshold region---is due to the $M_1$ transition connecting the deuteron
to the (quasi-bound) $n$$p$ $^1$S$_0$ state.  Note that, because of the
well-known destructive interference between the one-body current contributions
originating from the deuteron S- and D-wave components, two-body current 
contributions are relatively large in this threshold region.
However, they only amount to a $\simeq$ 5\% correction
in the quasi-elastic peak region.

In Fig.~\ref{fig:asiii} the asymmetry $A_{\gamma Z}$ labeled
(1+2)-body---$A_{\gamma Z}$ is defined in Eq.~(\ref{eq:agz})---includes,
in addition to one-body, two-body terms in the electromagnetic and neutral
weak currents (in both the vector and axial-vector components of the
latter).  These two-body contributions are negligible over the whole
$E^\prime$ spectrum.  However, the (PC and PV) two-body electromagnetic
currents play a relatively more significant role in the asymmetry
$A_{\gamma\gamma}$, Eq.~(\ref{eq:agg}).

\begin{figure}[bth]
\let\picnaturalsize=N
\def\picsize{5in}
\def\picfilenamea{figs/asy_ddh.eps}
\epsfbox{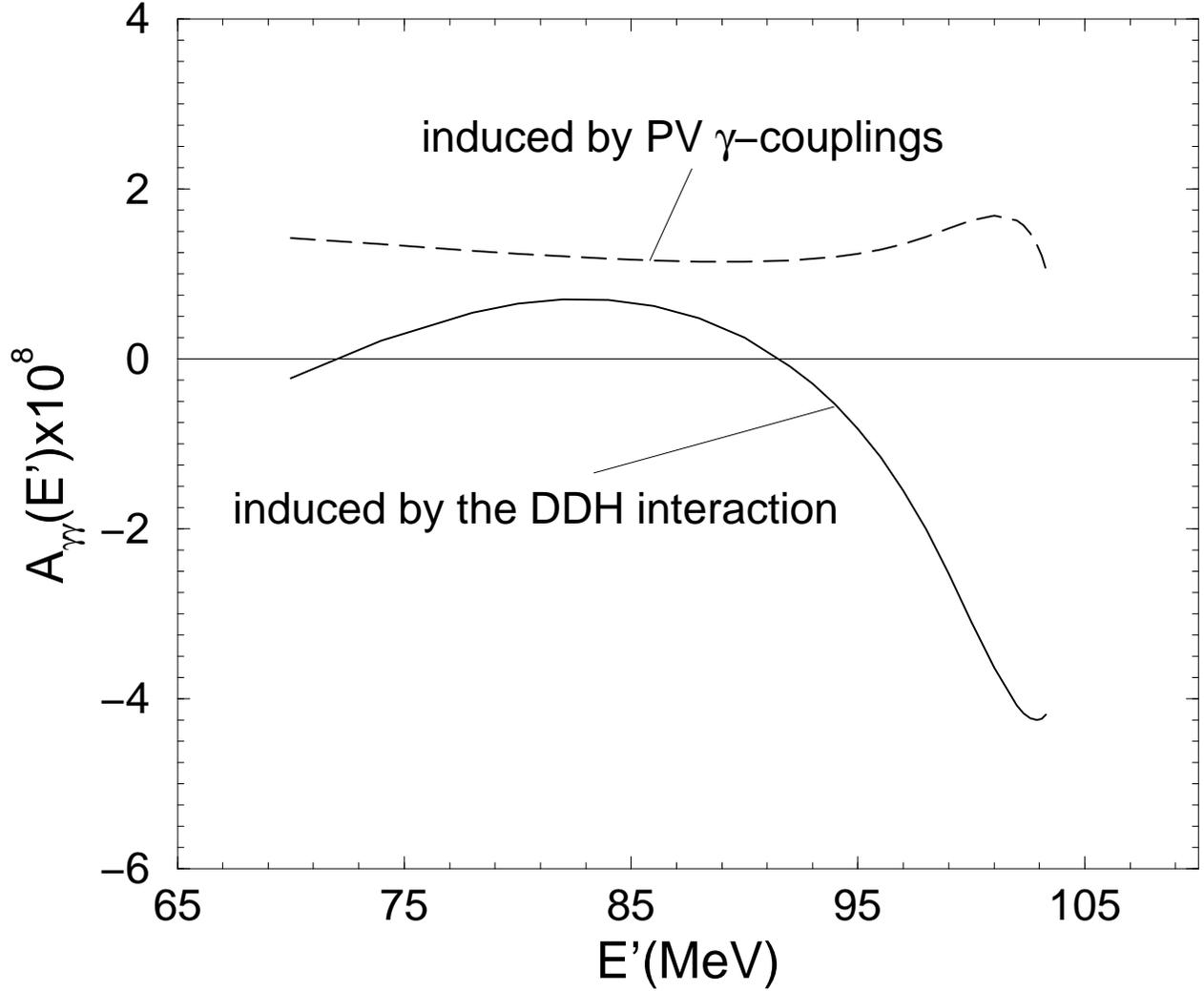}
\caption{Contributions to the asymmetry $A_{\gamma\gamma}$ calculated, as
function of the scattered electron energy $E^\prime$, with the (PC) AV18 and
(PV) DDH interaction models.  The other electron kinematical variables
are as in Fig.~\protect\ref{fig:xsiii}.  The solid line
represents the results corresponding to the presence in the wave
functions of opposite-parity components induced by the DDH interaction, while
the dashed line represents the results due to the anapole current and the PV
two-body current associated with $\pi$-exchange.  The total asymmetry
$A_{\gamma\gamma}$ shown by the solid line in Fig.~\protect\ref{fig:asiii} is
obtained as the sum of these two contributions.}
\label{fig:asddh}
\end{figure}

In Fig.~\ref{fig:asddh} we display separately, for the asymmetry $A_{\gamma\gamma}$,
the contributions originating from i) the presence in the wave
functions of opposite-parity components induced by the DDH interaction (solid curve)
and ii) the anapole current and the PV two-body
current associated with $\pi$-exchange (dashed curve).
The latter are positive and fairly constant as function of $E^\prime$, while
the former exhibit a pronounced dependence upon $E^\prime$.  Note that, up
to linear terms in the effects induced by PV interactions, the asymmetry
$A_{\gamma \gamma}$ is obtained as the sum of these two contributions. 

\begin{figure}[bth]
\let\picnaturalsize=N
\def\picsize{5in}
\def\picfilenamea{figs/asy_bonn_iii.eps}
\epsfbox{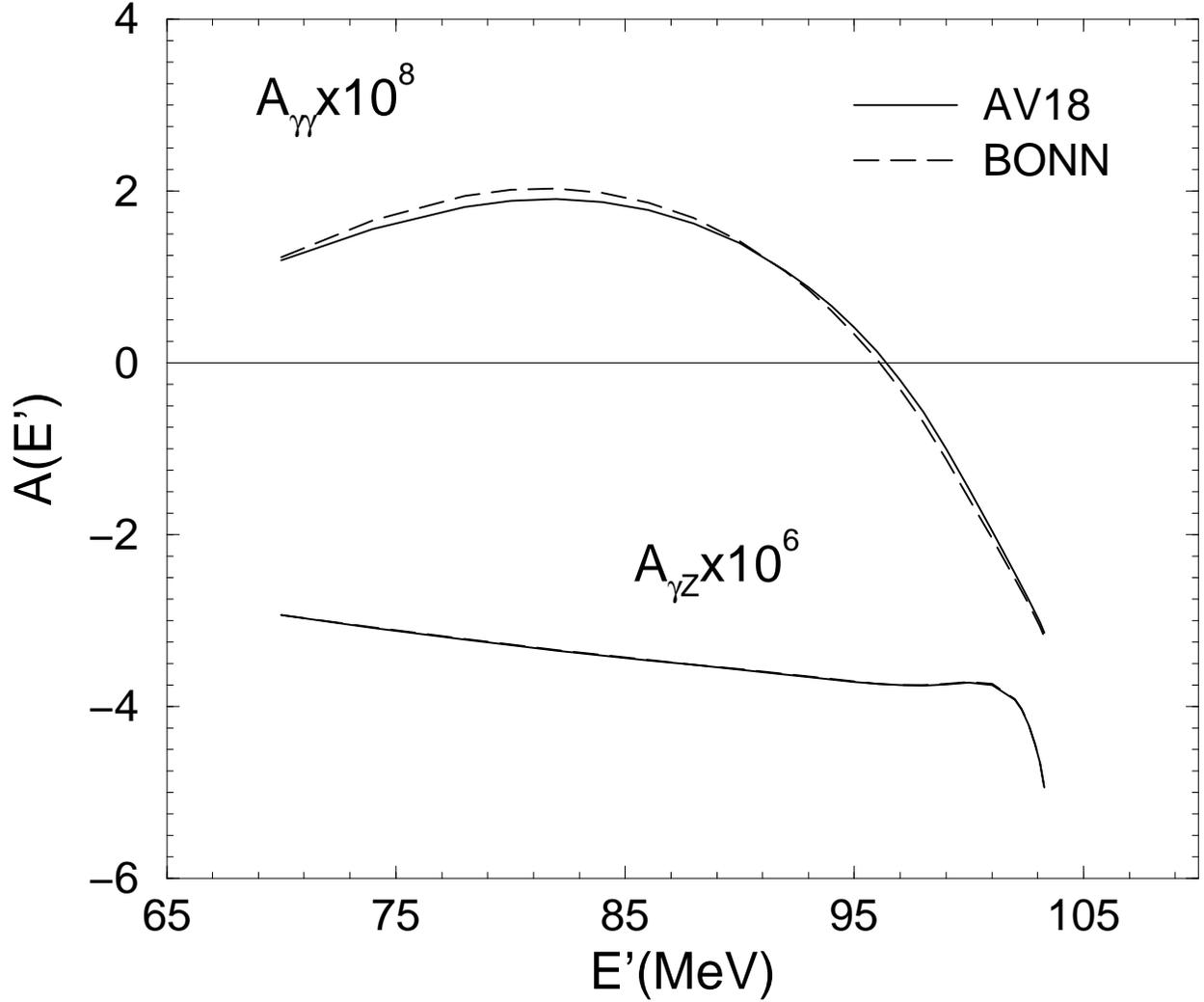}
\caption{The asymmetries $A_{\gamma\gamma}$ and $A_{\gamma Z}$ calculated,
as function of the scattered electron energy $E^\prime$, with the DDH
model in combination with either the AV18 or BONN model.  The other
electron kinematical variables are as in Fig.~\protect\ref{fig:xsiii}.
Predictions are shown obtained by including one- and two-body terms in the
electromagnetic and neutral weak currents.  For the $A_{\gamma Z}$ asymmetry
the AV18 and BONN calculated values are essentially indistinguishable.}
\label{fig:asbonn}
\end{figure}

The BONN model leads to predictions for the inclusive cross section
and asymmetries, that are very close to those obtained with the
AV18, as shown for $A_{\gamma Z}$ and $A_{\gamma\gamma}$
in Fig.~\ref{fig:asbonn}.  Thus the strong-interaction model dependence
is negligible for these observables.

\begin{figure}[bth]
\let\picnaturalsize=N
\def\picsize{5in}
\def\picfilenamea{figs/asy.eps}
\epsfbox{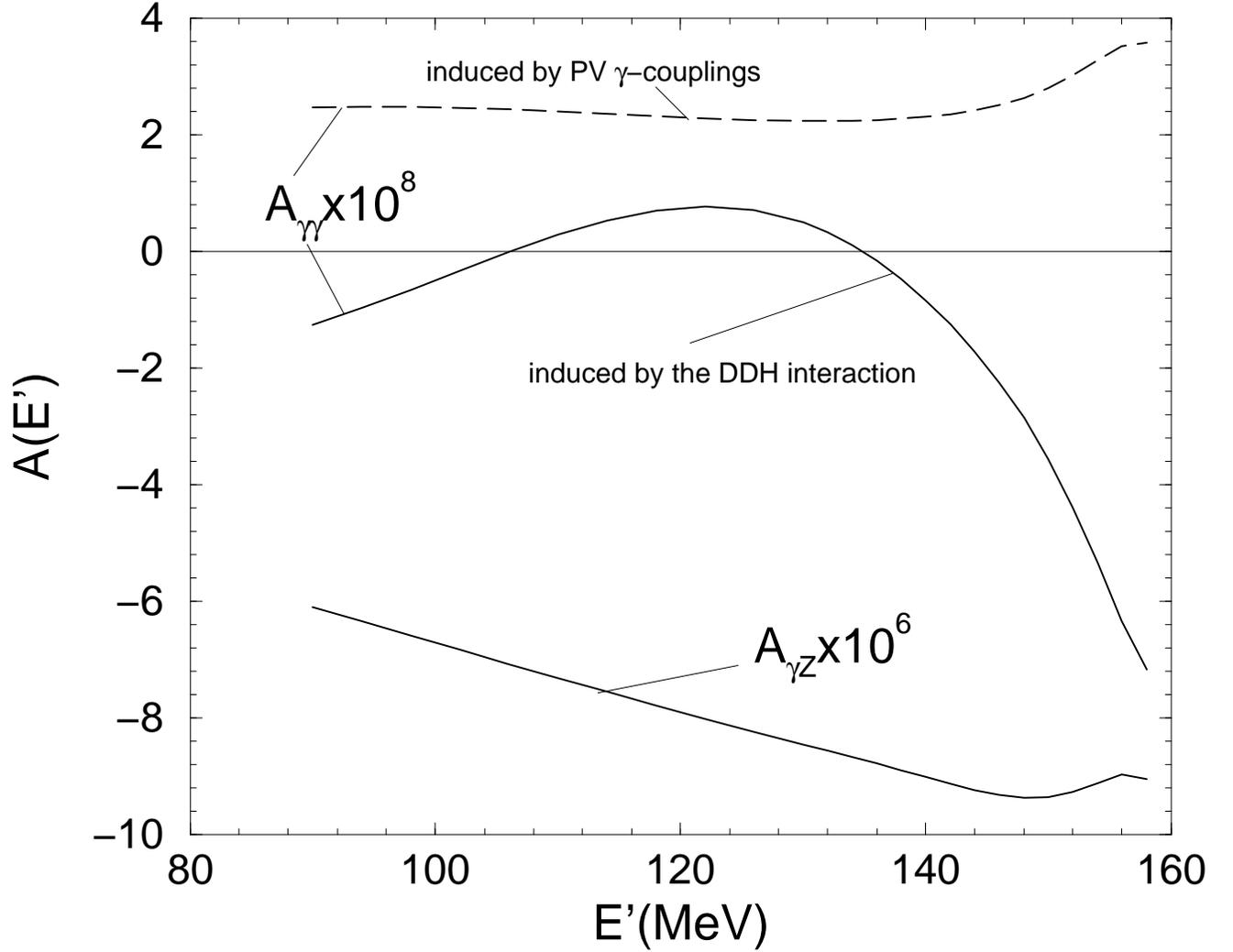}
\caption{The asymmetry $A_{\gamma Z}$ and the two contributions to the asymmetry
$A_{\gamma\gamma}$ (notation as in Fig.~\protect\ref{fig:asddh}) calculated,
as function of the scattered electron energy $E^\prime$, with the (PC) AV18
and (PV) DDH interaction models.  Predictions are shown obtained by including 
one- and two-body terms in the electromagnetic and neutral weak currents.  Note
that the electron incident energy is 192 MeV and its scattering angle $\theta_e$ is
138.4$^\circ$.}
\label{fig:as}
\end{figure}

In Fig.~\ref{fig:as} we present results for the asymmetries
corresponding to a four-momentum transfer $|q_\mu^2|$
at the top of the quasi-elastic peak of about 0.094 GeV$^2$, the
three-momentum transfer values span the range (266--327) MeV over the
$E^\prime$ spectrum shown.  The calculations are based on the AV18 model
and include one- and two-body currents.
The asymmetry from $\gamma$-$Z$ interference scales with $q_\mu^2$ and
therefore is, in magnitude, about a factor of two larger than calculated
in Fig.~\ref{fig:asiii} where $|q_\mu^2| \simeq$0.039 GeV$^2$.  The
contributions to $A_{\gamma\gamma}$ exhibit, as functions of $E^\prime$,
a behavior qualitatively similar to that obtained at the lower $|q_\mu^2|$
value, see Fig.~\ref{fig:asddh}.

\begin{figure}[bth]
\let\picnaturalsize=N
\def\picsize{5in}
\def\picfilenamea{figs/asy_ddh_pai.eps}
\epsfbox{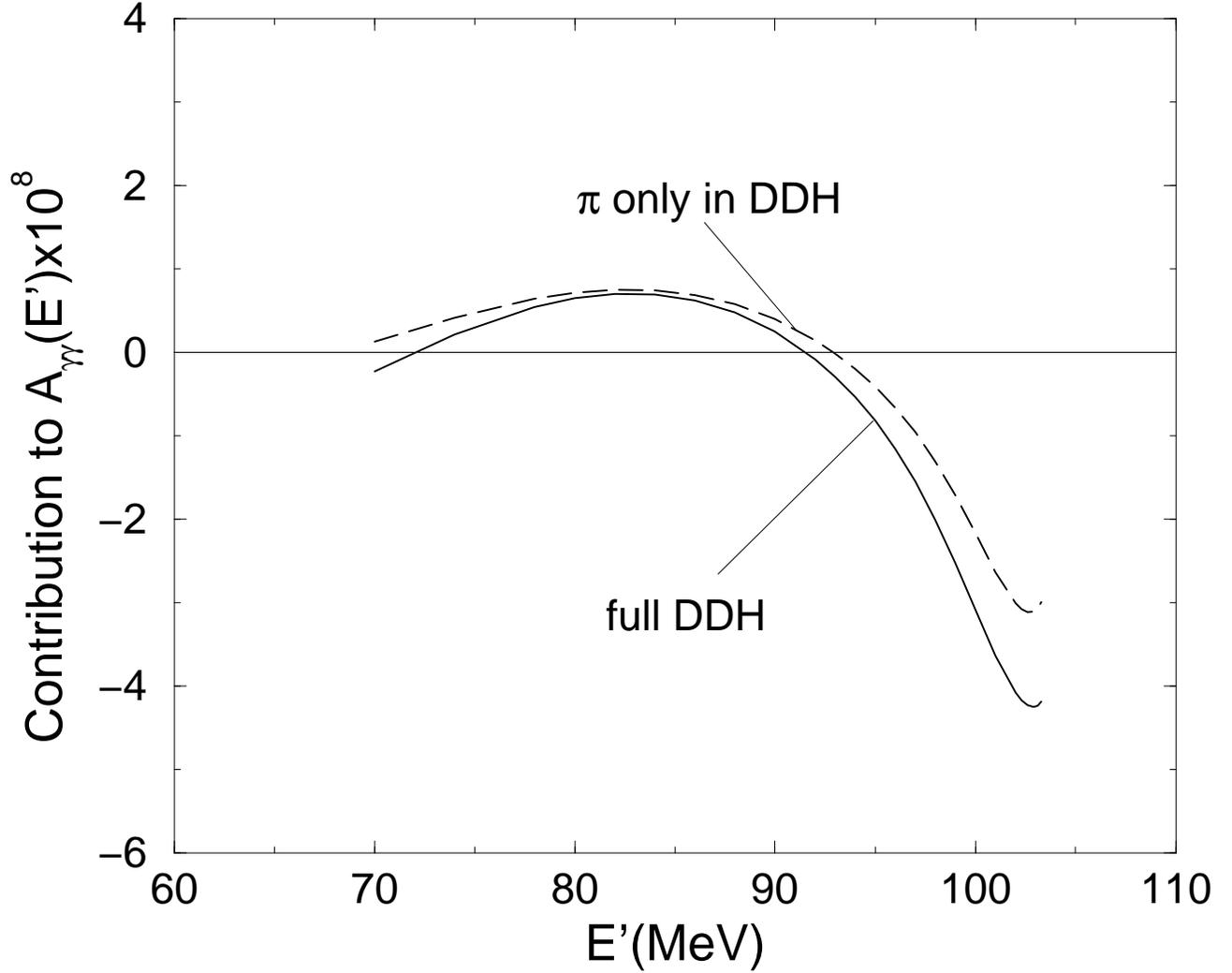}
\caption{Contribution to the asymmetry $A_{\gamma\gamma}$, calculated as function
of the scattered electron energy $E^\prime$, corresponding to the presence
in the wave functions of opposite-parity components induced by either the full
DDH or a truncated DDH model, consisting of its pion-exchange component only,
in combination with the AV18 model.  The other
electron kinematical variables are as in Fig.~\protect\ref{fig:xsiii}.  Predictions
are shown obtained by including (PC and PV) one- and two-body terms in the electromagnetic
current.}
\label{fig:aspai}
\end{figure}

In Fig.~\ref{fig:aspai} we compare results for the contribution to
$A_{\gamma\gamma}$ due to the presence in the wave function
of opposite-parity components induced by the full DDH and a truncated
version of it, including only the pion-exchange term.  Note that,
up to linear terms in the effects produced by PV interactions, the other
contribution to $A_{\gamma\gamma}$, namely that originating from (PV) one- and
two-body currents, remains the same as in Fig.~\ref{fig:asddh},
since---as mentioned earlier---only the (PV) two-body currents associated
with pion-exchange are considered in the present work.  Figure~\ref{fig:aspai}
shows that the asymmetry $A_{\gamma\gamma}$ is dominated
by the long-range pion-exchange contribution.  Hence, $A_{\gamma\gamma}$ will
scale essentially linearly with the PV $\pi$$N$$N$ coupling constant.

\begin{figure}[bth]
\let\picnaturalsize=N
\def\picsize{5in}
\def\picfilenamea{figs/asy_v6.eps}
\epsfbox{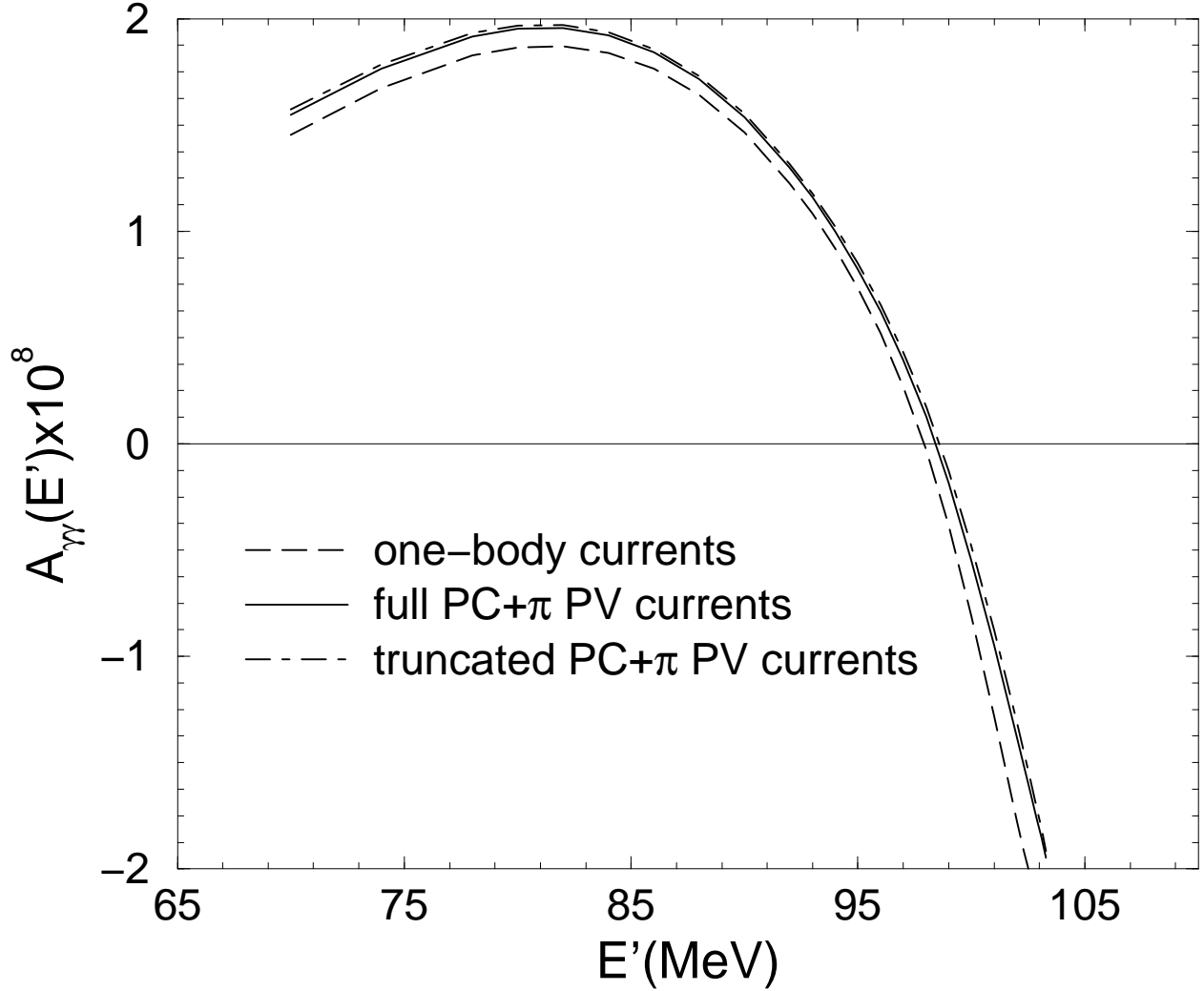}
\caption{The asymmetry $A_{\gamma\gamma}$ calculated, as function of the scattered
electron energy $E^\prime$, with the AV18 model in combination with a truncated DDH
model consisting of its pion-exchange component only.  The other electron kinematical
variables are as in Fig.~\protect\ref{fig:xsiii}.  Predictions are shown obtained by
including the one-body terms alone and both the one- and two-body terms in the PC and
PV components of the electromagnetic current (dashed and solid lines, respectively).
Also shown are the results obtained by ignoring in the PC two-body currents those
terms from the momentum-dependent components of the AV18 model (dashed-dotted line).}
\label{fig:asv6}
\end{figure}

Finally, Fig.~\ref{fig:asv6} is meant to illustrate the sensitivity of the
asymmetry $A_{\gamma \gamma}$ to those PC two-body currents derived from
the momentum-dependent interaction components of the AV18 model
(i.e., the spin-orbit, ${\bf L}^2$, and quadratic spin-orbit terms),
see Sec.~\ref{sec:cnt} and~\ref{sec:r_np}.  While these currents play
a crucial role in the photon asymmetry in the $\vec n$$p$ radiative
capture at thermal neutron energy, they give negligible contributions
to the present observable at quasi-elastic kinematics.

These results demonstrate that, in the kinematics of the SAMPLE
experiments~\cite{Hasty00,Beck01}, the asymmetry from $\gamma$-$Z$
interference is dominated by one-body currents, and that it is 
two-orders of magnitude larger than that associated with the
PV hadronic weak interaction.  Hence even the largest estimates of the
weak $\pi$$N$$N$ coupling constant will not affect extractions
of single-nucleon matrix elements.  These conclusions corroborate
those of the authors of Ref.~\cite{Liu03}, who have carried out a similar
study of the impact of hadronic weak interaction on quasi-elastic
electro-deuteron scattering.
\section{Conclusions}
\label{sec:cons}

A systematic study of parity-violating (PV) observables in the $n$$p$
system, including the asymmetries in $\vec{n}$$p$ radiative capture and
$d(\vec{\gamma},n)p$ photo-disintegration, the spin rotation and longitudinal
asymmetry in $\vec{n}$$p$ elastic scattering, and the asymmetry in electro-disintegration
of the deuteron by polarized electrons at quasi-elastic kinematics, has been
carried out by using a variety of latest-generation, strong-interaction potentials
in combination with the DDH model of the PV hadronic weak interaction.
We find that the model dependence of the $\vec{n}$$p$-capture asymmetry
is quite small, at a level similar to the expected contributions of the
short-range parts of the interaction.  This process is in fact dominated
by the long-range interaction components associated with pion exchange.
A measurement of the $\vec{n}$$p$-capture asymmetry is then a clean probe
of that physics.

Similarly, we find that the asymmetry in the $d(\vec{e},e^\prime)np$
reaction at quasi-elastic kinematics is a very clean probe of the electro-weak
properties of individual nucleons.  The processes associated with two nucleons,
including PV admixtures in the deuteron and scattering wave functions and 
electromagnetic two-body currents induced by hadronic weak interactions, play
a very small role at the values of momentum transfers explored so far~\cite{Hasty00,Beck01}.

We also find that the neutron spin rotation is sensitive to both the pion and
vector-meson PV couplings to the nucleon, while exhibiting a modest model dependence,
at the level of 5-10\%, due to the input strong-interaction potential adopted in
the calculation.  Thus a measurement of this observable~\cite{Snow03}, when combined
with measurements of the asymmetries in $\vec{n}$$p$ radiative capture~\cite{Snow00}
and $\vec{p}$$p$ elastic scattering~\cite{Berdoz01}, could provide useful constraints 
for some of these PV amplitudes. 

The asymmetry in the deuteron disintegration by circularly polarized
photons from threshold up to 20 MeV energies is dominated by the short-range
components of the DDH interaction.  However, it also displays enhanced sensitivity
to the short-range behavior in the strong-interaction potentials.  Indeed,
predictions for the asymmetry at threshold differ by almost a factor of two,
depending on whether the Argonne $v_{18}$ or Bonn 2000 interactions are
used in the calculations.  Therefore, this observable cannot provide an
unambiguous value of short-range weak meson nucleon couplings; however, they
would be valuable in placing constraints on the hadronic weak mixing angles.

Finally, the issue of electromagnetic current conservation in the presence
of parity-conserving (PC) and PV potetials has been carefully investigated.
In particular, in the case of the $p(\vec{n},\gamma)d$ and $d(\vec{\gamma},n)p$
processes dramatic cancellations occur betweeen the contributions associated
with the two-body currents induced, respectively, by the PC and PV potentials.

\section*{Acknowledgments}
The authors wish to thank L.E.\ Marcucci and M.\ Viviani 
for interesting discussions and illuminating correspondence,
and G.\ Hale for making available to them experimental data sets of the
deuteron photo-disintegration.  The work of J.C.\ and M.P.\ was supported
by the U.S. Department of Energy under contract W-7405-ENG-36, and
the work of R.S.\ was supported by DOE contract DE-AC05-84ER40150
under which the Southeastern Universities Research Association (SURA)
operates the Thomas Jefferson National Accelerator Facility.
Finally, some of the calculations were made possible by grants
of computing time from the National Energy Research Supercomputer
Center.
%
%
%

%
%
%   Tables
%
%
\begin{table}
\caption{Values used for the strong- and weak-interaction coupling
constants and short-range cutoff parameters of the $\pi$-, $\rho$-,
and $\omega$-meson to the nucleon in the DDH potential.}
\begin{tabular}{cdddddd}
        & $g^2_\alpha/4\pi$  & $\kappa_\alpha$  & $10^7\times h_\alpha^0$  & $10^7\times h_\alpha^1$ & $10^7\times h_\alpha^2$ & $\Lambda_\alpha$ (GeV/c)  \\
\tableline
$\pi$    & 13.9 &     &        &   4.56 &        &  1.72  \\
$\rho$   & 0.84 & 6.1 & --16.4 & --2.77 & --13.7 &  1.31  \\
$\omega$ & 20.  & 0.  &   3.23 &   1.94 &        &  1.50
\end{tabular}
\label{tb:ddh}
\end{table}
\begin{table}
\caption{Values used for the strong- and weak-interaction coupling
constants and short-range cutoff parameters of the $\pi$-, $\rho$-,
and $\omega$-meson to the nucleon in the DDH ``best values'' potential,
from Ref.~\protect\cite{Desplanques80}.}
\begin{tabular}{cddddddd}
        & $g^2_\alpha/4\pi$  & $\kappa_\alpha$  & $10^7\times h_\alpha^0$  & $10^7\times h_\alpha^1$ & $10^7\times h_\alpha^2$ & $\Lambda_\alpha$ (GeV/c)  \\
\tableline
$\pi$    & 13.9 &     &        &   4.56 &       & 2.4 \\
$\rho$   & 0.84 & 3.7 & --11.4 & --0.19 & --9.5 & 2.4 \\
$\omega$ & 20.  & 0.  & --1.90 & --1.14 &       & 2.4
\end{tabular}
\label{tb:ddhb}
\end{table}
\begin{table}
\caption{Labeling of channels.}
\begin{tabular}{ccccc}
         & \multicolumn{4}{c} {$\alpha$} \\
$J$      & 1  & 2 & 3  & 4 \\
\tableline
0        & $^1$S$_0$ & $^3$P$_0$ &           &           \\
1        & $^3$S$_1$ & $^3$D$_1$ & $^1$P$_1$ & $^3$P$_1$ \\
2        & $^3$P$_2$ & $^3$F$_2$ & $^1$D$_2$ & $^3$D$_2$ \\
3        & $^3$D$_3$ & $^3$G$_3$ & $^1$F$_3$ & $^3$F$_3$ \\
$\dots$  & $\dots$   & $\dots$   & $\dots$   & $\dots$   
\end{tabular}
\label{tb:chan}
\end{table}
\begin{table}
\caption{Classification of channel mixings for a given $J$: PC or PV
if induced by $v^{\rm PC}$ or $v^{\rm PV}$, respectively.  Note that
no coupling is allowed between channels 3 and 4.}
\begin{tabular}{cccccc}
         & \multicolumn{5}{c} {coupling} \\
$J$      & 12  & 13 & 14 & 23 & 24 \\
\tableline
0         & PV &    &    &    &    \\
1         & PC & PV & PV & PV & PV \\
2         & PC & PV & PV & PV & PV \\
$\dots$   & PC & PV & PV & PV & PV    
\end{tabular}
\label{tb:mixing}
\end{table}
\begin{table}[bthp]
\caption{Neutron spin-rotation angle per unit length, in units of
$10^{-9}$ rad cm$^{-1}$, in the limit of vanishing incident neutron energy.
Various combinations of strong- and weak interaction potentials are
used.  Also listed are the results obtained by ignoring strong-interaction
effects, row labeled ``plane waves''.}
\begin{tabular}{cddd}
               & DDH       & DDH$\pi$ &  DDHb   \\
\hline
AV18           &  5.09     &  5.21    &  7.19  \\
NIJM-I         &  4.94     &  5.35    &  7.64  \\
BONN           &  4.63     &  5.18    &  7.35  \\
Plane waves    &--5.67     &--6.87    &--5.85  \\
\end{tabular}
\label{tb:nspin}
\end{table}
\begin{table}[bthp]
\caption{Total cross-section $\sigma^\gamma$ and parity-violating asymmetry $a^\gamma$
in the $\vec n$$p$ radiative capture at thermal neutron energies, obtained in various
models.  The asymmetry is reported for pion-exchange only (DDH$\pi$) and full DDH (DDH)
interactions.}
\begin{tabular}{cdddd}
&\multicolumn{2}{c}{$\sigma^\gamma$(mb)} & \multicolumn{2}{c}{$a^\gamma \times 10^8$} \\
\hline
Interaction &Impulse Current&Full Current& DDH$\pi$  & DDH \\
AV18        & 304.6            & 334.2 & --4.98 & --4.92 \\
NIJM-I      & 305.4            & 332.5 & --5.11 & --5.02 \\
BONN        & 306.5            & 331.6 & --4.97 & --4.89 \\
\end{tabular}
\label{tb:np_x}
\end{table}
\begin{table}[bthp]
\caption{Cumulative contributions (in units of 10$^{-8}$) to the PV asymmetry $a^\gamma$
in the $\vec n$$p$ radiative capture at thermal neutron energies for the AV18 interaction
and pion-exchange-only DDH$\pi$ interaction.  See text for explanation.}
\begin{tabular}{cdd}
          & AV18 (PC) Currents & DDH$\pi$ (PV) Currents \\
\hline
Impulse                          & --15.3 &      \\
$+\pi$                           & --48.3 & 44.2 \\
$+\rho$                          & --40.4 & 44.0 \\
$+p$-dependent                   & --43.8 & 44.0 \\
\end{tabular}
\label{tb:cntre}
\end{table}
\begin{table}[bthp]
\caption{Cumulative contributions to the total cross section
$\sigma^\gamma$ and PV asymmetry $a^\gamma$ in
the $\vec n$$p$ radiative capture at thermal neutron energies for the AV8 and
pion-exchange-only DDH$\pi$ interactions.  Also listed is the asymmetry obtained with the Siegert
form of the $E_1$ operator.  See text for explanation.}
\begin{tabular}{cdddd}
&\multicolumn{1}{c}{$\sigma^\gamma$(mb)} & \multicolumn{3}{c}{$a^\gamma \times 10^8$} \\
\hline
                           &        & AV8 (PC) Currents & DDH$\pi$ (PV) Currents & Total \\
\hline
Impulse                    &  226.4      & --17.7  &        & --17.7  \\
$+\pi$                     &  239.2      & --57.9  &  51.3  & --6.60  \\
$+\rho$                    &  241.7      & --50.3  &  51.1  &  +0.790 \\
$+{\rm SO}$                &  241.7      & --57.0  &  51.1  & --5.89  \\
$+\Delta+\omega\pi\gamma$  &  247.4      & --56.3  &  50.5  & --5.82  \\
\hline
Siegert $E_1$              &              &         &        & --5.76 \\
\end{tabular}
\label{tb:cnte1}
\end{table}
\begin{table}[bthp]
\caption{Photon helicity-dependent asymmetries (in units of $10^{-8}$)
calculated with various combinations of strong- and weak-interaction
potentials at an incident photon energy of 2.2259 MeV, about 1.3 keV
above threshold.  Predictions are listed obtained by including only
one-body terms (impulse) and both one- and two-body terms (full) in
the electromagnetic current, right-hand-side of Eq.~(\protect\ref{eq:j_s}).}
\begin{tabular}{cddd}
                   & AV18+DDH (BONN+DDH)    & AV18+DDH$\pi$  &  AV18+DDHb   \\
\hline
Impulse            & 5.44  (9.41)  & --0.035    & 2.49   \\
Full               & 5.19  (9.05)  & --0.037    & 2.38    
\end{tabular}
\label{tb:tadg}
\end{table}
%
%
%
% Figures
%
%
\end{document}